\documentclass{article}

\usepackage{arxiv}

\renewcommand{\undertitle}{Published in Computer Methods in Applied Mechanics and Engineering 464 (2027) 119466\\[2pt]
  {\normalfont\small \url{https://doi.org/10.1016/j.cma.2026.119466}}}
\renewcommand{\headeright}{Published in CMAME 464 (2027) 119466}
\usepackage[utf8]{inputenc}
\usepackage[T1]{fontenc}
\usepackage{graphicx}
\usepackage{amssymb}
\usepackage{xcolor}
\usepackage{tikz}
\usepackage{comment}
\usepackage{pgfplots}
\usepackage{subcaption}
\usepackage{float}
\usepackage[numbers,sort&compress]{natbib}
\usepackage{hyperref}
\usepackage{xurl}
\pgfplotsset{compat=1.18}
\usepackage{placeins}
\usepackage{caption}
\usepackage{booktabs}
\usepackage{multirow}
\usepackage[ruled,vlined]{algorithm2e}

\graphicspath{{./}}

\newcommand{\periodicModelName}{P-GNN}
\newcommand{\proposedModelName}{P-DivGNN}

\usepackage{amsmath,bm}

\def\vv{{\bm{v}}}
\def\ee{{\bm{e}}}

\begin{document}

\title{Non-linear mechanical field reconstruction coupling recurrent neural networks with physics-informed graph neural networks}

\author{%
  Manuel Ricardo Guevara Garban$^{1,2,3}$\thanks{Corresponding author: \texttt{manuel.guevara-garban@u-bordeaux.fr}}\qquad
  Yves Chemisky$^{4}$\qquad
  Michaël Clément$^{2}$\\[4pt]
  \bfseries
  Étienne Prulière$^{5}$\qquad
  Martin Abendroth$^{3}$\qquad
  Bjoern Kiefer$^{3}$\\[10pt]
  \mdseries\small
  $^{1}$Univ. Bordeaux, CNRS, Bordeaux INP, I2M, UMR 5295, Talence, France\\
  $^{2}$Univ. Bordeaux, CNRS, Bordeaux INP, LaBRI, UMR 5800, Talence, France\\
  $^{3}$Institute of Mechanics and Fluid Dynamics, TU Bergakademie Freiberg, Freiberg, 09599, Germany\\
  $^{4}$Univ. Grenoble Alpes, CNRS, UMR 5525, La Tronche, 38000, France\\
  $^{5}$Arts et Metiers Institute of Technology, CNRS, Bordeaux INP, I2M, UMR 5295, Talence, 33400, France
}
\date{}

\maketitle

\begin{abstract}
Reconstructing local stress fields in heterogeneous microstructures under non-linear, history-dependent loading remains a major computational bottleneck in multi-scale simulations. We propose a coupled LSTM-GNN framework that links the temporal and spatial aspects of local stress field reconstruction. A Long Short-Term Memory network encodes macroscopic stress-strain sequences into a compact hidden state that captures the path-dependent constitutive response, while a physics-informed Graph Neural Network reconstructs the spatially-resolved stress field at each time step. We introduce a relative weighting strategy with linear warm-up to balance the data-driven reconstruction loss and a discrete divergence-based equilibrium penalty. This resolves the scale mismatch that prevents fixed-weight formulations from converging in the elasto-plastic regime.
The model is trained on $10{,}000$ non-proportional loading paths applied to a periodic plate-with-a-hole microstructure and von Mises elasto-plasticity. The model achieves three orders of magnitude speedup over finite element simulations and generalizes to loading sequences twice the training length, with $3.0\%$ cumulative error. Because the graph relies on mesh connectivity instead of the specific element type, one trained surrogate can be applied directly without retraining to meshes with different element types and to both coarser and finer resolutions, while in all cases reproducing the high-fidelity quad-element FE field used during training. Indeed, the message passing characteristics inherent to GNN  and MeshGraphNet architecture render the model mesh-agnostic. Analysis of the LSTM hidden states suggests a low-dimensional structure related to the internal state variables of the constitutive model.
\end{abstract}


\section{Introduction}\label{sec:intro}

Recent manufacturing processes especially additive manufacturing make it possible to design architectured materials with far more complex geometries than before, opening new possibilities for lightweight, multi-functional components.
Designing these structures still requires repeated full-field simulations under changing loads, which quickly becomes expensive.
Multi-scale optimization is especially costly because it requires repeated iterations over microstructural geometries, macroscale geometries, and material parameters.
The cost even becomes prohibitive once plasticity or kinematic non-linearities are included.

\subsection{Multi-scale simulation}
A computationally efficient way to simulate mechanical problems that have two (or more) characteristic length scales  lies in the use of homogenization theory.
If the scales are well separated, we can split the problem into microscopic and macroscopic parts.

These two boundary value sub-problems are coupled in the context of a linearized two-scale finite element (FE) simulation:
considering appropriate boundary conditions, the microscopic problem provides at each integration point the homogenized properties (mean stress, tangent stiffness matrix) that are required to solve the macroscopic problem.

The resolution of the macroscopic PDE system, in turn, provides the predicted strain that allows defining the appropriate local boundary condition for the microscopic problem.
This method is referred to as the FE$^2$ method \cite{ref:fe2_feyel1, ref:fe2_feyel2}.
In linear elasticity, the microscopic problem, which consists of six linear simulations in 3D (and three in 2D) for the determination of the homogenized stiffness matrix, needs only to be solved once since the stiffness matrix is time independent, whereas for non-linear problems, like elasto-plasticity, a new homogenization of the microstructure needs to be done at each time step, at each integration point and considering an implicit solver, for each iteration inside the time step (e.g. Newton-Raphson iterations).

In many cases, homogenized properties are not sufficient in the design process, for instance considering the elasticity limit or the fatigue durability, and the evaluation of mechanical strength under a specific loading scenario that involves a local non-linear response of the material.
In these cases, a local criterion based on the stress components is required, so the full local stress field needs to be computed.
Note that the computation of full fields comes naturally considering FE$^2$ methods, which suffer from poor computational efficiency, especially considering material non-linearities that require a linearized, incremental scheme.

Prior to the development of FE$^2$ methods, micro-mechanical approaches have been developed~\cite{quFundamentalsMicromechanicsSolids2006} to link the average local fields of different phases with the global average.
These relations are restricted to a specific geometry (i.e., ellipsoidal inhomogeneities), while approximations were developed for other geometries, the field being described as a piecewise field per phase~\cite{Dvorak.1992}.

\subsection{Reduced order models}
To mitigate the high costs associated with multi-scale simulations and following micro-mechanical approaches, reduced order models (ROMs) have been proposed to simplify simulations at the micro-scale level while retaining essential microstructure features and
behaviors, such as mean stress values.
Such ROMs, which have added plastic modes to better represent the local fields of non-linear heterogeneous microstructures, have been developed in~\cite{michelNonuniformTransformationField2003} and extended in~\cite{roussetteNonuniformTransformationField2009}, considering the use of Proper Orthogonal Decomposition (POD) to identify those plastic modes.
The plastic modes are therefore used to better represent strain localization that appears in phases subjected to the development of plastic strains.

AI-based approaches were also developed to drastically cut down computational times~\cite{ref:aymen_fe_lstm}, at the cost of retaining only the average stress field and tangent modulus.

Hybrid approaches embedding neural networks into elasto-plastic constitutive models have also been developed~\cite{ref:malik_hybrid_NN_foams_2022}, alongside ROM-FE$^2$ methods that reduce the cost of non-linear multi-scale simulations~\cite{ref:lange_monolithic, ref:lange_hyperROM_2024}. A recent study compared these surrogate families and reports trade-offs in efficiency, accuracy, and flexibility~\cite{ref:lange_comparison_foams_2025, ref:abendroth_foam_dataset_2025}.

Another local field estimation has been introduced in~\cite{LiuBessaLiu.2016}, referred to as the self-consistent clustering analysis.
In this approach, local fields are approximated as a group of clusters with similar mechanical features.
This method is very relevant to determine the homogenized response of complex heterogeneous structures without the determination of specific local non-linear deformation modes that are microstructure-dependent.

Nonetheless, all these methods are not dedicated to the precise definition of local fields, but rather focus on satisfying, computationally efficient estimation of the global response of non-linear, multi-scale structures.

\subsection{Deep learning methods}
Machine-learning (ML) methods are now widely used for surrogate modeling \cite{ref:ml_stochastic, ref:probabilistic_response, ref:pinn_surrogate_modeling} and, in some cases, for full-field prediction beyond the specific case of multi-scale modeling.
Deep Learning methods, such as Convolutional Neural Networks (CNNs) \cite{ref:cnn_legacy}, have been employed to predict full stress fields on 2D geometries, as demonstrated by the framework "StressNet," which uses CNNs to estimate stress fields in cantilevered structures \cite{ref:cnn_cantilevered_structures}.
In addition, \cite{ref:gupta_cnn_multiscale_mechanics} proposed to use a U-Net \cite{ref:cnn_unet} architecture as an end-to-end ML-driven framework to accelerate multi-scale mechanics simulations of heterogeneous macro-structures.
The U-Net architecture has also been employed for predicting stress fields in additively manufactured metals with complex defect networks \cite{ref:cnn_stress_fields_additive}.

Other CNN-based methods have explored the use of generative models, such as Generative Adversarial Networks (GANs) \cite{ref:cnn_gan_legacy} and Diffusion Models \cite{ref:diffusion_models_legacy}.
For example, \cite{ref:cnn_gan_strain_stress_tensors} proposed a GAN model for predicting strain and stress tensors in hierarchical composite microstructures.
Similarly, \cite{ref:cnn_stress_estimation_diffusion_model} used Diffusion Models to estimate stress fields in 2D geometries.

CNNs work well on regular grids but are awkward on unstructured meshes with varying refinement, where local features at different resolutions matter for accuracy.
Point-cloud methods such as \cite{ref:point_cloud_deep_onet} avoid rasterization by using 1D convolutional layers for field predictions on parameterized geometries, yet they still miss explicit connectivity, which matters for stress localization on mesh data.

To overcome the limitations of CNNs in simulating unstructured geometries, Graph Neural Networks (GNNs) \cite{ref:gnn_legacy} have gained increasing attention.
By operating directly on graph-structured data, such as mesh data,
GNNs can effectively capture spatial relationships and multi-scale dependencies inherent in unstructured domains, offering a more flexible approach to stress field prediction.
One notable contribution is \textit{MeshGraphNet}, a GNN architecture proposed by \cite{ref:mesh_graph_net}.
This framework adopts an Encode-Message Passing-Decode architecture, enabling accurate representations of mesh-based simulations.
\textit{MeshGraphNet} has become a state-of-the-art approach for
predicting scalar fields on mesh data \cite{ref:gnn_nature_stress_strain, ref:gnn_multiscale_periodic, ref:gnn_multiscale_mesh_graph_net}.

Despite these advances, integrating physical constraints such as equilibrium conditions and periodic boundary conditions into GNN frameworks remains a challenge.
Some exploratory works aim to incorporate physics-informed priors,
also known as Physics-Informed Neural Networks (PINNs) \cite{ref:PINN_legacy}, into GNNs.
For instance, \cite{ref:gnn_pde_gnn_embedded} proposed a GNN-based
solver for partial differential equations (PDEs) considering boundary conditions, while \cite{ref:gnn_thermo_informed} introduced a thermodynamics-informed GNN to predict the temporal evolution of dissipative dynamic systems.
However, enforcing equilibrium constraints on predicted stress fields using GNNs remains an unresolved issue.
\cite{ref:divergence_free_nn} explored the concept of divergence-free neural networks for fluid mechanics, using automatic differentiation, even so, this approach remains experimental and difficult to scale on larger meshes due to the expensive use of
automatic differentiation.
In a previous work \cite{ref:guevara_ijnme_2025}, we addressed this gap by introducing a discrete divergence operator into the GNN loss function for linear elastic and hyperelastic materials.
The present work extends this approach to the history-dependent setting.

\par
A parallel line of research has explored recurrent neural networks (RNNs) for constitutive modeling and multi-scale surrogates.
Mozaffar et al.~\cite{ref:mozaffar2019deep_learning_plasticity} demonstrated that RNNs can learn path-dependent plasticity directly from stress-strain data, while Gorji et al.~\cite{ref:gorji2020rnn_path_dependent_plasticity} systematically evaluated RNN architectures for modeling path-dependent plasticity under complex loading.
Ghavamian and Simone~\cite{ref:ghavamian2019accelerating_multiscale_rnn} used RNNs to accelerate FE$^2$ simulations of history-dependent materials by replacing microscale solves.
Wu et al.~\cite{ref:wu2020rnn_multiscale_elastoplastic} extended this to non-proportional loading with an RNN-accelerated multi-scale model for elasto-plastic composites.
Bonatti and Mohr~\cite{ref:bonatti2022self_consistency_rnn_plasticity} highlighted the importance of self-consistency constraints in RNN-based constitutive models.

These RNN-based approaches predict only the \textit{mean} (homogenized) constitutive response and do not provide spatially resolved local fields.
Meanwhile, GNN-based methods for local field reconstruction have been limited to the elastic and hyperelastic regimes \cite{ref:guevara_ijnme_2025, ref:gnn_multiscale_periodic}.

A recent physics-informed framework does combine the two architectures: a GNN extracts spatial features of the polycrystal microstructure that feed an LSTM predicting microvoid growth over the loading history~\cite{ref:gnn_lstm_polycristals}. There the GNN serves as a feature extractor and the prediction is a homogenized scalar, the void volume fraction. A coupling in which the GNN instead performs spatial localization to reconstruct the full local stress field in the history-dependent regime has, to our knowledge, not been addressed.

\par
Beyond classical surrogate modeling, data-driven computational mechanics \cite{ref:kirchdoerfer2016data_driven_mechanics} and thermodynamics-based neural networks \cite{ref:masi2021thermodynamics_ann_constitutive} have emerged as principled alternatives to handcrafted constitutive laws.
Im et al.~\cite{ref:im2021pod_lstm_elastoplastic_surrogate} combined POD with LSTM (Long Short-Term Memory) for elasto-plastic surrogate modeling, while Maia et al.~\cite{ref:maia2023physically_recurrent_nn} proposed physically recurrent neural networks that embed constitutive models within data-driven surrogates for path-dependent heterogeneous materials.
These approaches illustrate a growing trend toward architectures that respect physical structure while learning from data.

\par
Temporal extensions of graph-based architectures have been proposed for physics simulation. Sanchez-Gonzalez et al.~\cite{ref:sanchez2020learning_simulate_physics} introduced Graph Network Simulators for learning complex particle-based dynamics, and Brandstetter et al.~\cite{ref:brandstetter2022message_passing_pde} developed message-passing PDE solvers.
Hu et al.~\cite{ref:hu2024temporal_gnn_polycrystal} proposed a temporal GNN for cross-scale modeling of polycrystals.
However, these approaches usually unfold time steps autoregressively over the graph, which leads to an accumulation of errors in the graph domain where many node values evolve simultaneously. Although this strategy is efficient for processing temporal data, it is not ideally suited for reconstructing local fields when the goal is to recover a specific snapshot of interest. This limitation continues to pose a significant challenge in this area.

Neural operators such as the Fourier Neural Operator \cite{ref:li2021fourier_neural_operator} and DeepONet \cite{ref:lu2021deeponet} learn mappings between function spaces and have been extended to time-dependent PDEs \cite{ref:he2024sequential_deeponet, ref:kovachki2023neural_operator}.
While these approaches offer mesh-independent representations, they require re-training for new geometries and do not naturally incorporate the periodic boundary conditions and mesh connectivity that are central to computational homogenization.

\subsection{Contributions}\label{sec:contributions}
In this work, we extend the physics-informed GNN localization framework introduced in our previous work \cite{ref:guevara_ijnme_2025} to non-linear, history-dependent constitutive behaviors.
The main contributions are:
\begin{enumerate}
\item A coupled LSTM-GNN architecture that decouples homogenized temporal sequence learning from spatial field reconstruction, enabling local stress field prediction in elasto-plastic microstructures under non-proportional loading;
\item A relative weighting strategy with linear warm-up for the divergence-based physics-informed loss, which automatically balances the reconstruction and equilibrium objectives across material regimes without problem-specific tuning;
\item A systematic evaluation of the model capabilities including out-of-distribution generalization to longer loading sequences, cross-mesh transfer from quadrilateral to triangular elements, and mesh refinement robustness;
\item An interpretability analysis of the LSTM hidden states, revealing a low-dimensional manifold structure aligned with the physical state variables of the constitutive model.
\end{enumerate}

\section{Background}
\subsection{Mechanical problem}\label{sec:mechanical_problem}
The mechanical problem considered in this work is the microscale boundary value problem arising in periodic computational homogenization. We consider a representative unit cell (RUC) occupying a domain $\Omega$, with volume $V$, subject to periodic boundary conditions. The local static equilibrium, in the absence of body forces, reads
\begin{equation}\label{eq:div_sigma_eq_0}
\operatorname{div} \boldsymbol{\sigma} (\mathbf{x}) = \boldsymbol{0}\ ,\qquad \forall \;\mathbf{x}\in\Omega\ ,
\end{equation}
which in 2D reduces to
\begin{equation}\label{eq:div_sigma_continu}
\operatorname{div} \boldsymbol{\sigma}
    =  \begin{bmatrix} \frac{\partial \sigma_{xx}}{\partial x} + \frac{\partial \sigma_{xy}}{\partial y} \\[1ex] \frac{\partial \sigma_{xy}}{\partial x} + \frac{\partial \sigma_{yy}}{\partial y}\end{bmatrix} = \boldsymbol{0}\ .
\end{equation}

Considering a linearized incremental scheme, the increment of displacement $\eth \mathbf{u}$ is decomposed using periodic boundary conditions \cite{ref:suquet_periodic, ref:chatzigeorgiou.etal.2016} as
\begin{equation}
\eth \mathbf{u}(\mathbf{x}) = \eth \bar{\boldsymbol{\varepsilon}} \cdot \mathbf{x} + \eth \mathbf{\tilde{u}}(\mathbf{x})\ ,
\end{equation}
where $\eth \bar{\boldsymbol{\varepsilon}} = \frac{1}{V} \int_{\Omega} \eth \boldsymbol{\varepsilon} \, dV$ is the increment of prescribed mean strain tensor and $\eth \mathbf{\tilde{u}}$ is a periodic fluctuation satisfying $\eth \mathbf{\tilde{u}}(\mathbf{x}^+) = \eth \mathbf{\tilde{u}}(\mathbf{x}^-)$ on opposite boundary faces.
The linearized increment of strain tensor is defined as
\begin{equation}\label{eq:strain_def}
    \eth \boldsymbol{\varepsilon} = \frac{1}{2} \left[ \nabla \eth \mathbf{u} + \nabla^T \eth \mathbf{u} \right].
\end{equation}

\subsubsection{Elasto-plastic constitutive model}
The constitutive behavior is described within the framework of small-strain, rate-independent elasto-plasticity \cite{ref:chaboche_book,ref:simo_hughes_1998}. The total strain tensor is decomposed additively into elastic and plastic contributions
\begin{equation}\label{eq:strain_decomposition}
\boldsymbol{\varepsilon} = \boldsymbol{\varepsilon}^e + \boldsymbol{\varepsilon}^p\ .
\end{equation}
The elastic part follows Hooke's law for linear-elastic materials
\begin{equation}\label{eq:constitutive_law}
\boldsymbol{\sigma} = \mathbb{C} : \boldsymbol{\varepsilon}^e\ .
\end{equation}
Here, $\mathbb{C}$ is the fourth-order elasticity tensor, which in the case of isotropy can be expressed in terms of the Young's modulus $E$ and the Poisson's ratio $\nu$.

Plastic yielding is assumed to be governed by the von Mises criterion, based on the
equivalent stress measure
\begin{equation}
	\sigma_\text{vm} = \sqrt{\tfrac{3}{2}\,\boldsymbol{s}:\boldsymbol{s}} \ .
\end{equation}
Therein, $\boldsymbol{s} = \boldsymbol{\sigma} - \tfrac{1}{3} \, \text{tr}(\boldsymbol{\sigma}) \, \boldsymbol{I}$ is the deviatoric stress tensor.
The yield function then reads
\begin{equation}
f(\boldsymbol{\sigma}, \bar{\varepsilon}^p) = \sigma_\text{vm}(\boldsymbol{\sigma}) - \sigma_y(\bar{\varepsilon}^p) \leq 0 ,
\end{equation}
where $\bar{\varepsilon}^p$ is the accumulated equivalent plastic strain.
Isotropic hardening is modeled through the power-law rule
\begin{equation}
\sigma_y(\bar{\varepsilon}^p) = \sigma_{y_{0}} + K \, (\bar{\varepsilon}^p)^n ,
\end{equation}
with $\sigma_{y_{0}}$ the initial yield stress, $K$ a hardening parameter, and $n$ the hardening exponent.
The associative flow rule governs the evolution of plastic strain according to
\begin{equation}
\dot{\boldsymbol{\varepsilon}}^p = \dot{\lambda}\, \frac{\partial f}{\partial \boldsymbol{\sigma}} ,
\end{equation}
subject to the Karush-Kuhn-Tucker loading-unloading conditions,
\begin{equation}
\dot{\lambda} \geq 0, \quad f \leq 0, \quad \dot{\lambda}\, f = 0 \ .
\end{equation}
where $\lambda$ is a plastic Lagrange multiplier. The accumulated plastic strain thus evolves as $\dot{\bar{\varepsilon}}^p = \sqrt{\tfrac{2}{3} \, \dot{\boldsymbol{\varepsilon}}^p : \dot{\boldsymbol{\varepsilon}}^p }$.

This constitutive model introduces intrinsic path dependence: the stress response at any time depends on the entire history of plastic deformation through the internal variable $\bar{\varepsilon}^p$.
This path dependence, absent in the elastic setting treated in our previous work \cite{ref:guevara_ijnme_2025}, motivates coupling a recurrent architecture \cite{ref:aymen_fe_lstm} with the GNN localization framework.

\subsubsection{Stress localization in periodic homogenization}\label{sec:localization}
In periodic homogenization, the relationship between macroscopic and microscopic fields is mediated by localization tensors.
Given a macroscopic strain increment $\delta\bar{\boldsymbol{\varepsilon}}$, the local strain increment at position $\mathbf{x}$ within the unit cell is obtained through the fourth-order strain localization tensor $\mathbb{A}$,
\begin{equation}\label{eq:strain_localization}
\eth \boldsymbol{\varepsilon}(\mathbf{x}, t)  = \mathbb{A}(\mathbf{x}, t) : \eth \bar{\boldsymbol{\varepsilon}} (t) .
\end{equation}
The corresponding stress localization tensor $\mathbb{B}$ connects the local stress increment to the macroscopic stress increment,
\begin{equation}\label{eq:stress_localization}
\eth \boldsymbol{\sigma}(\mathbf{x}, t) = \mathbb{B}(\mathbf{x}, t) : \eth \bar{\boldsymbol{\sigma}} (t) \ .
\end{equation}
Note that the kinematic and kinetic localization tensors are related via $\mathbb{B} = \mathbb{D} : \mathbb{A} : \bar{\mathbb{D}}^{-1}$,
where $\mathbb{D}(\mathbf{x})$ is the local tangent modulus and $\bar{\mathbb{D}} = \langle \mathbb{D} : \mathbb{A} \rangle$ the homogenized tangent modulus \cite{ref:suquet_periodic}.

In linear elasticity, the tangent modulus reduces to the time-independent elasticity tensor $\mathbb{D}\! = \!\mathbb{C}$, so that $\mathbb{B}$ is constant and the stress takes the total form $\boldsymbol{\sigma}(\mathbf{x}) = \mathbb{B}(\mathbf{x}) : \bar{\boldsymbol{\sigma}}$.

In elasto-plasticity, $\mathbb{D}$ evolves with the accumulated plastic strain among other internal variables, making the localization tensor loading path-dependent, i.e.,
\begin{equation}\label{eq:localization_history}
\mathbb{B}_t = \mathbb{B}\!\left(\mathbf{x},\, \{\bar{\boldsymbol{\varepsilon}}_\tau\}_{\tau \leq t}\right) .
\end{equation}
This history dependence motivates the coupled architecture proposed in this work, where an LSTM encodes the loading history into a hidden state $\boldsymbol{h}_t$, while a GNN performs spatial localization conditioned on $\boldsymbol{h}_t$.

The macroscopic stress $\bar{\boldsymbol{\sigma}}$ is defined as volume average of the
local stress field over the RUC,
\begin{subequations}\label{eq:mean_continuous}
\begin{align}
  \bar{\boldsymbol{\sigma}} &= \frac{1}{|V|} \int_\Omega \boldsymbol{\sigma} \, \mathrm{d}\Omega \ .
\end{align}
\end{subequations}
In the FE setting the integrals are evaluated by numerical integration using Gauss quadrature on the
nodal stress and strain fields. The reference volume $|V|$ is the RUC
bounding box.

\subsection{Long Short-Term Memory (LSTM)}\label{sec:lstm}
Long Short-Term Memory (LSTM) networks \cite{ref:lstm} are recurrent architectures designed to capture long-range dependencies in sequential data while mitigating the vanishing gradient problem \cite{ref:vanishing_grad}.
Each LSTM unit maintains a cell state $\boldsymbol{c}_t \in \mathbb{R}^h$ regulated by three gates---forget, input, and output---that control information flow, where $h$ denotes the dimension of the hidden state $\boldsymbol{h}$.
\begin{figure}[!htbp]
	\centering
	\includegraphics[width=0.7\textwidth]{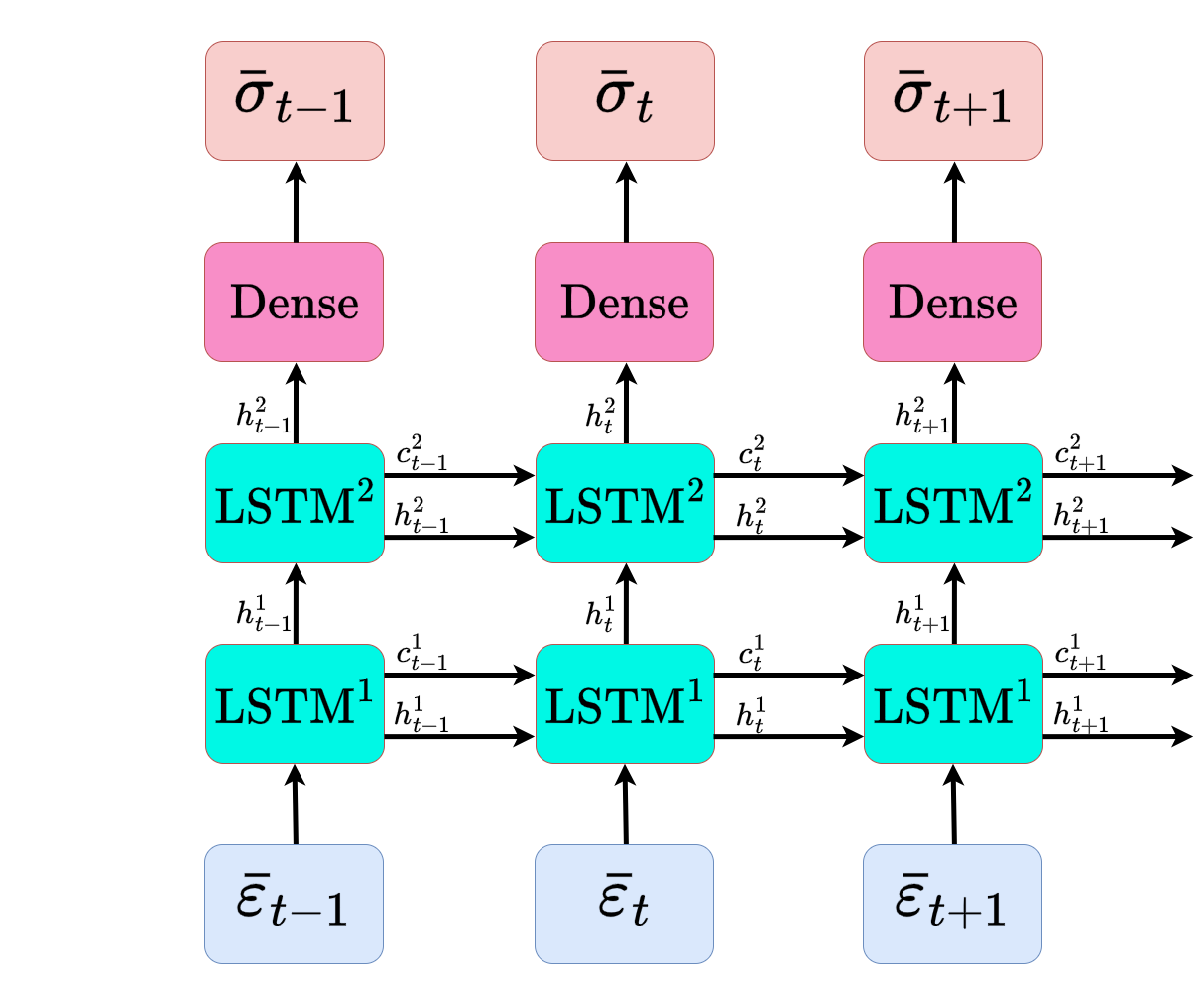}
	\caption{Implemented two-layer, stacked LSTM architecture, unrolled through time. At each step $t$, the mean strain $\bar{\boldsymbol{\varepsilon}}_t$ is processed through two LSTM layers, and the final hidden state $\boldsymbol{h}^2_t$ is mapped to the mean stress prediction $\bar{\boldsymbol{\sigma}}_t$ via a dense layer.}
	\label{fig:stress_lstm_schema}
\end{figure}

Given an input $\boldsymbol{x}_t \in \mathbb{R}^d$ and the previous hidden state $\boldsymbol{h}_{t-1} \in \mathbb{R}^h$, the update equations are:
\begin{subequations}
\begin{align}
\boldsymbol{f}_t &= \tilde{\sigma}(\boldsymbol{W}_f \boldsymbol{x}_t + \boldsymbol{U}_f \boldsymbol{h}_{t-1} + \boldsymbol{b}_f) \\[.3ex]
\boldsymbol{i}_t &= \tilde{\sigma}(\boldsymbol{W}_i \boldsymbol{x}_t + \boldsymbol{U}_i \boldsymbol{h}_{t-1} + \boldsymbol{b}_i) \\[.3ex]
\boldsymbol{o}_t &= \tilde{\sigma}(\boldsymbol{W}_o \boldsymbol{x}_t + \boldsymbol{U}_o \boldsymbol{h}_{t-1} + \boldsymbol{b}_o) \\[.3ex]
\tilde{\boldsymbol{c}}_t &= \tanh(\boldsymbol{W}_c \boldsymbol{x}_t + \boldsymbol{U}_c \boldsymbol{h}_{t-1} + \boldsymbol{b}_c) \\[.3ex]
\boldsymbol{c}_t &= \boldsymbol{f}_t \odot \boldsymbol{c}_{t-1} + \boldsymbol{i}_t \odot \tilde{\boldsymbol{c}}_t \\[.3ex]
\boldsymbol{h}_t &= \boldsymbol{o}_t \odot \tanh(\boldsymbol{c}_t)
\end{align}\end{subequations}
where $\tilde{\sigma}$ denotes the sigmoid activation, $\odot$ the element-wise product, $\boldsymbol{W} \in \mathbb{R}^{h \times d}$, $\boldsymbol{U} \in \mathbb{R}^{h \times h}$, and $\boldsymbol{b} \in \mathbb{R}^h$ are learnable parameters, with $\boldsymbol{c}_0 = \boldsymbol{h}_0 = \boldsymbol{0}$.

In this work, a two-layer, stacked LSTM is used, following the architecture validated in \cite{ref:these_aymen} for mean stress prediction (Figure~\ref{fig:stress_lstm_schema}).
The stacked configuration enables the second layer to operate on the hidden representations of the first, capturing higher-order temporal abstractions.

The model takes as input the mean strain
$\bar{\boldsymbol{\varepsilon}}_t \in \mathbb{R}^d$
and predicts the mean stress
$\bar{\boldsymbol{\sigma}}_t \in \mathbb{R}^d$
together with its hidden state:
\begin{equation}\label{eq:lstm_for_stress}
\textrm{LSTM}(\bar{\boldsymbol{\varepsilon}}_t, \boldsymbol{h}^2_{t-1}, \boldsymbol{c}^2_{t-1}) \mapsto (\bar{\boldsymbol{\sigma}}_t,
\boldsymbol{h}^2_t, \boldsymbol{c}^2_t)
\end{equation}
with $d$ the number of independent components of the symmetric
stress and strain tensors: $d = 3$ in 2D and $d = 6$
in 3D.

The hidden state $\boldsymbol{h}^2_t$ encodes the accumulated loading history and will serve as additional input to the GNN localization model (Section~\ref{sec:method}).
In the following sections, we will write $\boldsymbol{h}_t = \boldsymbol{h}_t^2$, the index 2 being dropped for the sake of readability.

\subsection{Graph Neural Networks}\label{sec:gnn_mp}
Graph Neural Networks (GNNs) \cite{ref:gnn_legacy} operate on graph-structured data. A finite element mesh is naturally a graph: the mesh nodes act as vertices and the element edges define the connections between them. We denote the resulting undirected mesh graph by $\mathcal{M} = (\mathcal{V}, \mathcal{E})$, where $\mathcal{V} = \{v_1, \ldots, v_n\}$ is a set of $n$ mesh nodes and $\mathcal{E}$ is a set of unordered edges $\{v_i, v_j\}\ (v_i, v_j \in \mathcal{V}, i \neq j)$. Each node is associated with a feature vector $\vv_i \in \mathbb{R}^{d_v}$ and each edge with a feature vector $\ee_{ij} \in \mathbb{R}^{d_e}$.
GNNs update node representations through message passing: at each layer, every node aggregates information from its neighbors $\mathcal{N}(i)$, enabling the network to learn spatial correlations across the graph topology.
This mechanism makes GNNs particularly well-suited for FE meshes, where nodes and edges naturally encode the geometric connectivity and local refinement that grid-based methods cannot represent.

\begin{figure}[!htbp]
    \centering
    \includegraphics[width=0.9\textwidth]{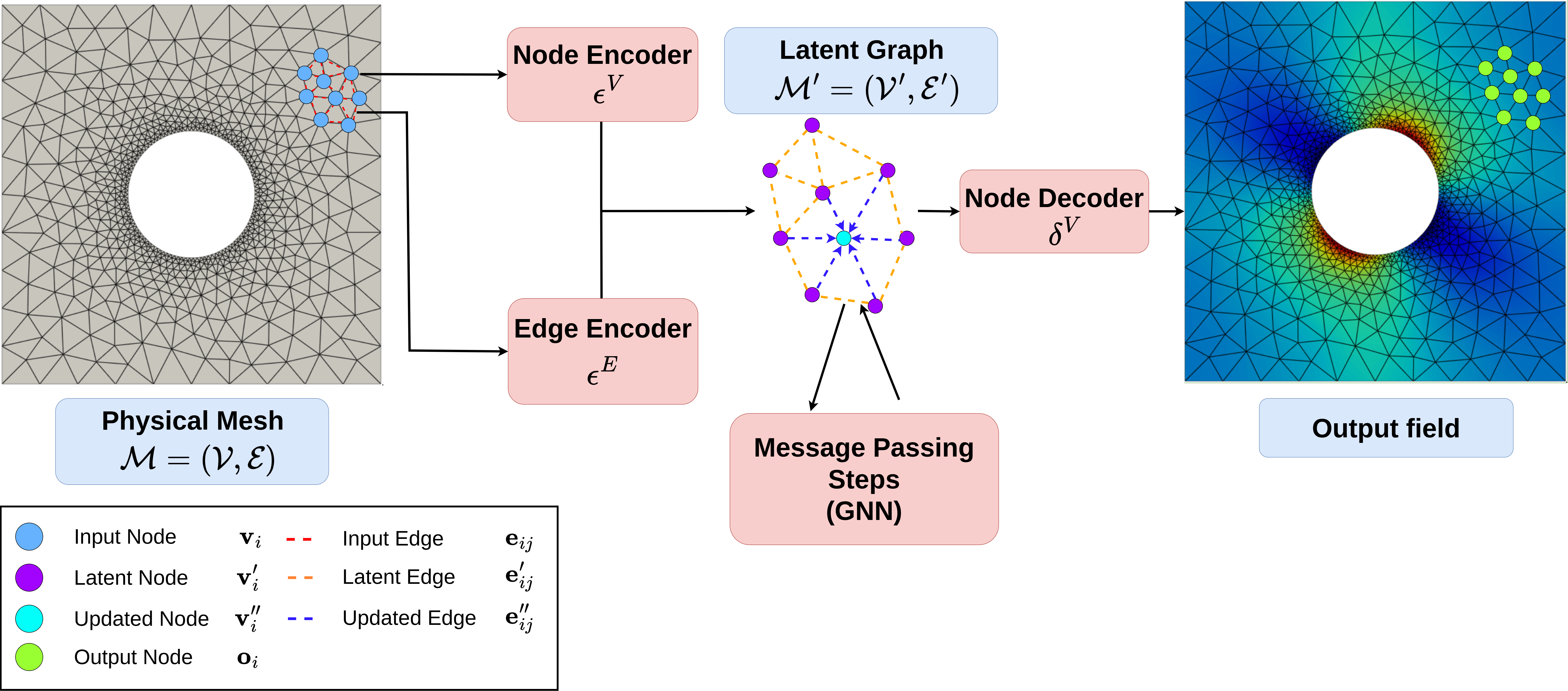}
    \caption{GNN architecture based on the encode-process-decode paradigm. The physical mesh $\mathcal{M}$ is mapped to a latent graph $\mathcal{M}'$ through node and edge encoders. Message passing steps update node and edge features in the latent space, and a node decoder produces the output stress field. \cite{ref:guevara_ijnme_2025}.}
    \label{fig:gnn_architecture}
\end{figure}

In our previous work \cite{ref:guevara_ijnme_2025}, we introduced a GNN-based stress localization framework (Equation~\eqref{eq:stress_localization}) following the \textit{MeshGraphNet} architecture \cite{ref:mesh_graph_net} for the linear elastic and hyperelastic cases. The implemented GNN architecture is presented in Figure \ref{fig:gnn_architecture}.
The architecture employs an Encode-Message Passing-Decode pipeline: MLP encoders project node and edge features into a latent space of dimension $H$; $L$ message-passing steps propagate information across the graph with shared-weight MLPs and residual connections; and an MLP decoder maps the final latent node representations to the target output.

The main adaptation is the enrichment of node features with temporal information from the LSTM. In the elastic formulation \cite{ref:guevara_ijnme_2025}, each node carried only $\vv_i = (\bar{\boldsymbol{\sigma}}, \mathbf{x}_i, \alpha_i)$, where $\bar{\boldsymbol{\sigma}}$ is the mean stress, $\mathbf{x}_i$ the spatial coordinates, and $\alpha_i \in \{-1, 0, 1\}$ a boundary classification label.
In the present work, the feature vector is augmented with the LSTM hidden state:
\begin{equation}\label{eq:node_features}
\vv_{i,t} = (\bar{\boldsymbol{\sigma}}_t, \boldsymbol{h}_t, \mathbf{x}_i, \alpha_i),
\end{equation}
where $\boldsymbol{h}_t \in \mathbb{R}^{64}$ and $\bar{\boldsymbol{\sigma}}_t \in \mathbb{R}^{d}$ are respectively the hidden state and the mean stress tensor produced by the LSTM at time step $t$ (Equation~\eqref{eq:lstm_for_stress}).
This hidden state, broadcast uniformly to all nodes, provides the GNN with a compact encoding of the loading history up to the current increment.
The GNN mapping thus becomes time-dependent:
\begin{equation}\label{eq:gnn_mapping}
\textrm{GNN}(\bar{\boldsymbol{\sigma}}_t, \boldsymbol{h}_t, \vv_{i,t}, \mathcal{M}) \mapsto (\sigma_{xx_{i,t}}, \sigma_{yy_{i,t}}, \sigma_{xy_{i,t}})
\end{equation}
Each edge $\{v_i, v_j\}$ carries the Euclidean distance between nodes as its feature. To enforce periodicity, additional edges with zero-valued features connect boundary node pairs on opposite faces, enhancing message passing across the periodic domain.
The model incorporating these periodic edges is denoted \textit{\periodicModelName}.

The implemented model uses a latent space of dimension $H = 128$ and $L = 10$ message-passing steps, yielding a total of $175{,}491$ trainable
parameters. Every block is a multilayer perceptron of hidden width $H$ with ReLU activations: the node and edge encoders and the per-step message-passing networks use two hidden layers followed by a layer normalization, as in \cite{ref:mesh_graph_net}, while the decoder is a single hidden layer that maps $H$ to the three stress components, without normalization. These hyper-parameters were selected in our previous work \cite{ref:guevara_ijnme_2025} for a finite-strain hyperelastic case and are reused here without re-tuning. The underlying hypothesis is that the GNN's task, spatial localization given a homogenized state, is of comparable difficulty in the present setting, since the path-dependence is factored out into the LSTM.

A variant incorporating a divergence-based physics-informed loss, originally proposed in \cite{ref:guevara_ijnme_2025} and referred to as \textit{\proposedModelName}, will be considered in Section~\ref{sec:divergence_loss}.

\section{Coupled LSTM-GNN Framework}\label{sec:method}
\begin{figure}[!htbp]
    \centering
    \includegraphics[width=0.8\textwidth]{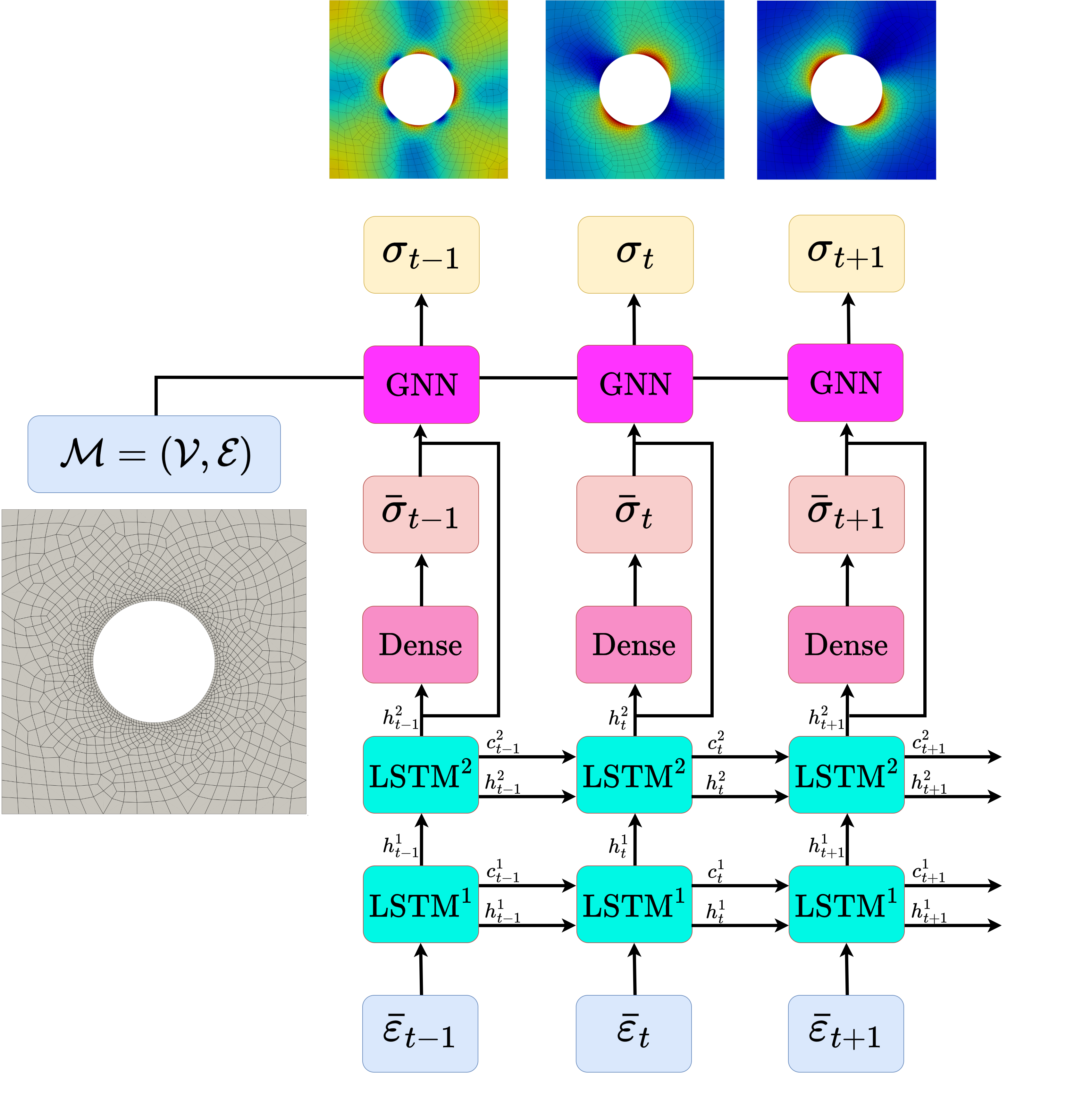}
    \caption{Architecture of the coupled LSTM-GNN framework. The LSTM processes the mean strain history to produce hidden states $\boldsymbol{h}_t$ and mean stress predictions $\bar{\boldsymbol{\sigma}}_t$.
    These outputs serve as enriched node features for the GNN, which performs message passing on the periodic mesh graph to reconstruct the local stress field at each time step.}
    \label{fig:schema_lsr}
\end{figure}
The LSTM encodes the loading history into a compact hidden state that captures path-dependent behavior, but by itself it cannot reconstruct spatially resolved fields.
To recover local stress distributions, the LSTM hidden state is coupled with the GNN localization model.

At each time step, the LSTM processes the input mean strain $\bar{\boldsymbol{\varepsilon}}_t$ and produces two outputs: the predicted mean stress $\bar{\boldsymbol{\sigma}}_t$ and a hidden representation $\boldsymbol{h}_t$ encoding the loading history up to time $t$.
These outputs are broadcast as uniform node features to the mesh graph, where the GNN performs message passing to reconstruct the full local stress distribution (Figure~\ref{fig:schema_lsr}).
The complete pipeline reads:
\begin{equation}\label{eq:coupled_pipeline}
\textrm{GNN}\bigl(\textrm{LSTM}(\bar{\boldsymbol{\varepsilon}}_t, \boldsymbol{h}_{t-1}), \vv_{i,t}, \mathcal{M}\bigr) \mapsto (\sigma_{xx_{i,t}}, \sigma_{yy_{i,t}}, \sigma_{xy_{i,t}})\,,
\end{equation}
where $\vv_{i,t}$ contains the per-node spatial coordinates $\mathbf{x}_i$ and boundary label $\alpha_i$, along with the time-dependent global features $\bar{\boldsymbol{\sigma}}_t$ and $\boldsymbol{h}_t$ that are broadcast to all nodes (Equation~\eqref{eq:node_features}).

The LSTM is a two-layer network with input size $3$, hidden size $64$, and output size $3$, totaling $51{,}139$ parameters.
The GNN contains $175{,}491$ parameters.
These configurations are fixed for all experiments.

\subsection{Database generation}\label{sec:database}
\begin{figure}[!htbp]
	\centering
	\includegraphics[width=\textwidth]{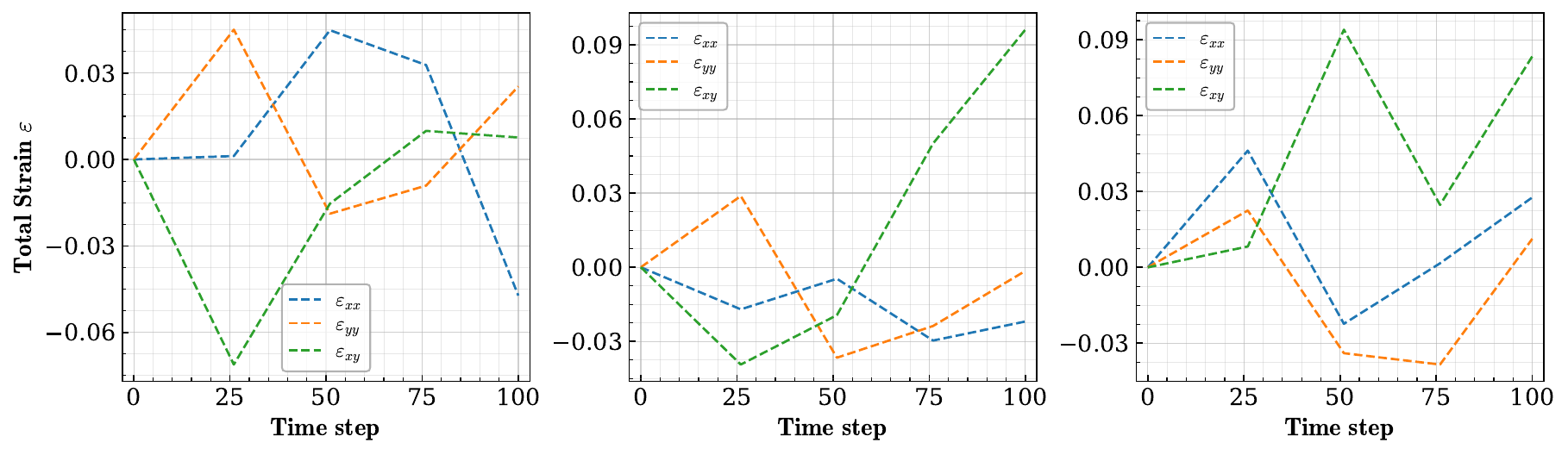}
	\caption{Examples of non-proportional strain paths randomly sampled from the test database. Each path consists of four segments with linearly interpolated increments.}
	\label{fig:strain_time_paths}
\end{figure}

A finite element database containing $10{,}000$ distinct, non-proportional loading cases was generated using the open-source FE library \textit{Fedoo} \cite{ref:3mah} which we develop and maintain.
Each loading case consists of four strain segments, with target strain components ($\bar{\varepsilon}_{xx}, \bar{\varepsilon}_{yy}, \bar{\varepsilon}_{xy}$) randomly sampled within $[-0.05, 0.05]$ and $25$ intermediate steps between successive targets, yielding $101$ time steps per case (Figure~\ref{fig:strain_time_paths}).
The dataset is split into $70\%$ training and $30\%$ testing subsets.

\begin{figure}[!htbp]
	\centering
	\begin{minipage}[c]{0.5\textwidth}
		\centering
		\vspace{0.5em}
		\begin{tabular}{l c}
			\hline
			\textbf{Parameter} & \textbf{Value} \\
			\hline
			Young’s modulus $E$ & $10^5$ MPa \\
			Poisson’s ratio $\nu$ & $0.3$ \\
			Yield stress $\sigma_{y0}$ & $300$ MPa \\
			Hardening parameter $K$  & $1000$ MPa \\
			Hardening exponent $n$ & $0.3$ \\
			Plate width/height $L$ & $1$ \\
			Hole radius $r$ & $0.2$ \\
			\hline
		\end{tabular}
		\captionof{table}{Material and geometric parameters of the hole-plate microstructure.}
		\label{tab:material_props}
	\end{minipage}
	\hfill
	\begin{minipage}[c]{0.4\textwidth}
		\centering
		\includegraphics[width=\linewidth]{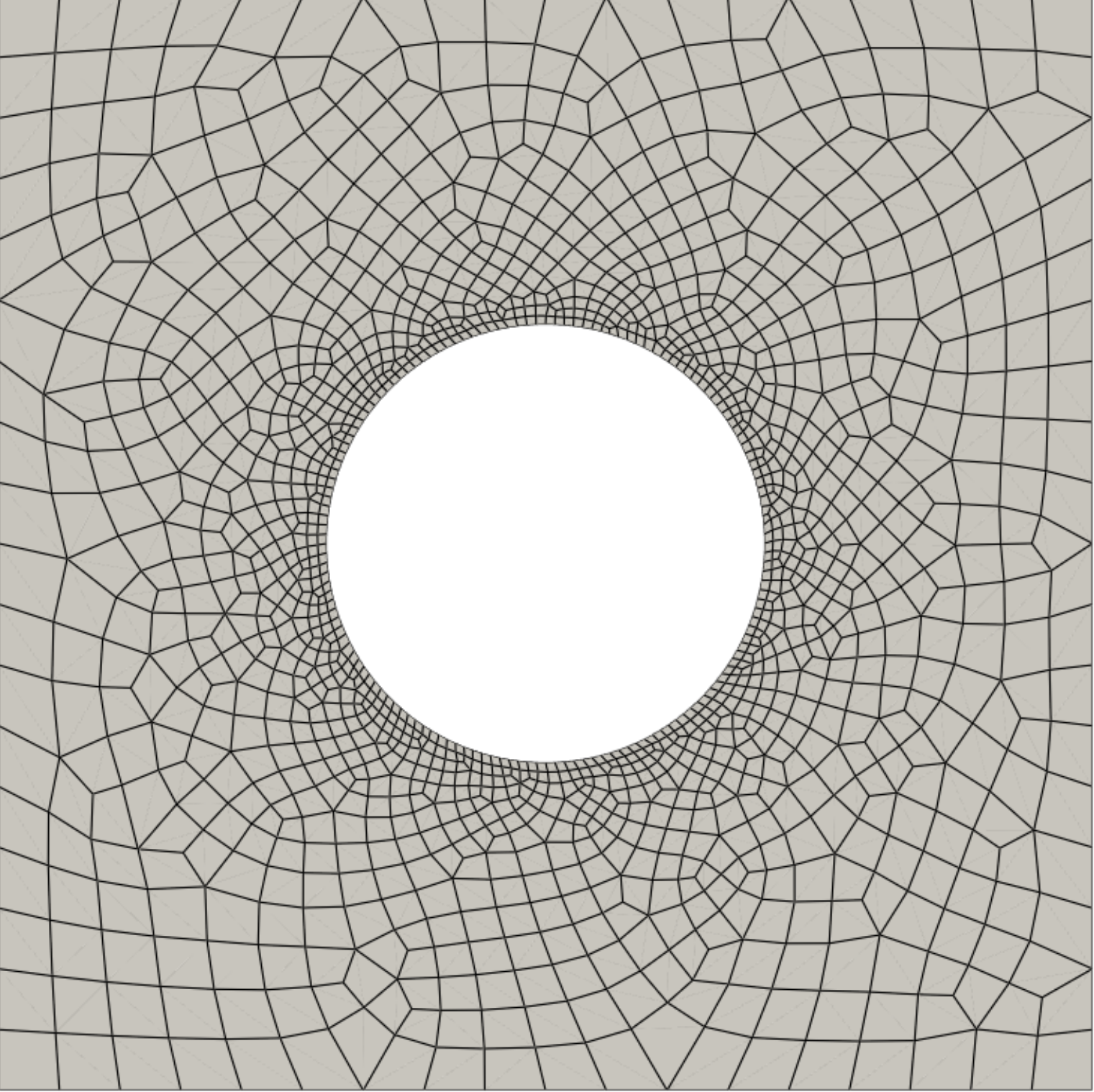}
		\captionof{figure}{FE mesh of the plate with central hole ($1{,}512$ nodes, $1{,}402$ quad elements).}
		\label{fig:quad_mesh}
	\end{minipage}%
\end{figure}

\begin{figure}[!htbp]
	\centering
	\begin{subfigure}[t]{0.48\textwidth}
		\centering
		\includegraphics[width=\textwidth]{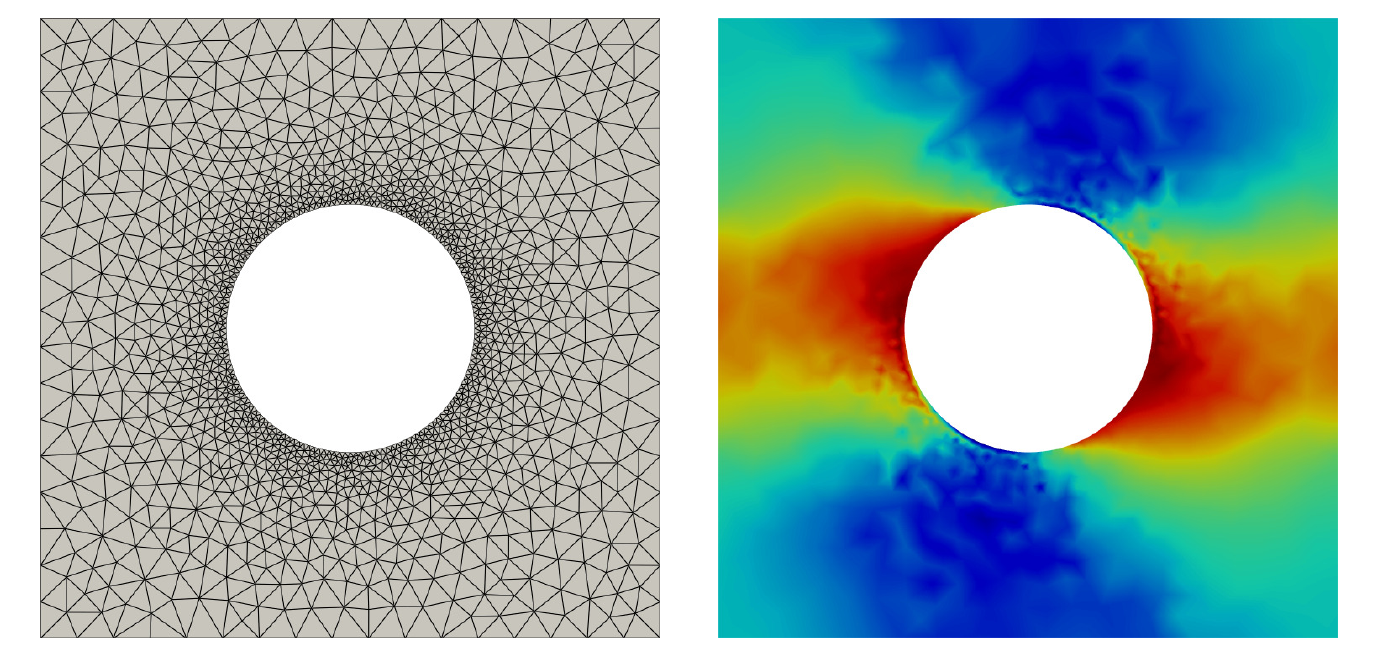}
		\caption{Triangular mesh ($1{,}487$ nodes, $2{,}758$ CST elements): visible locking artifacts.}
		\label{fig:fem_tri_mesh}
	\end{subfigure}
	\hfill
	\begin{subfigure}[t]{0.48\textwidth}
		\centering
		\includegraphics[width=\textwidth]{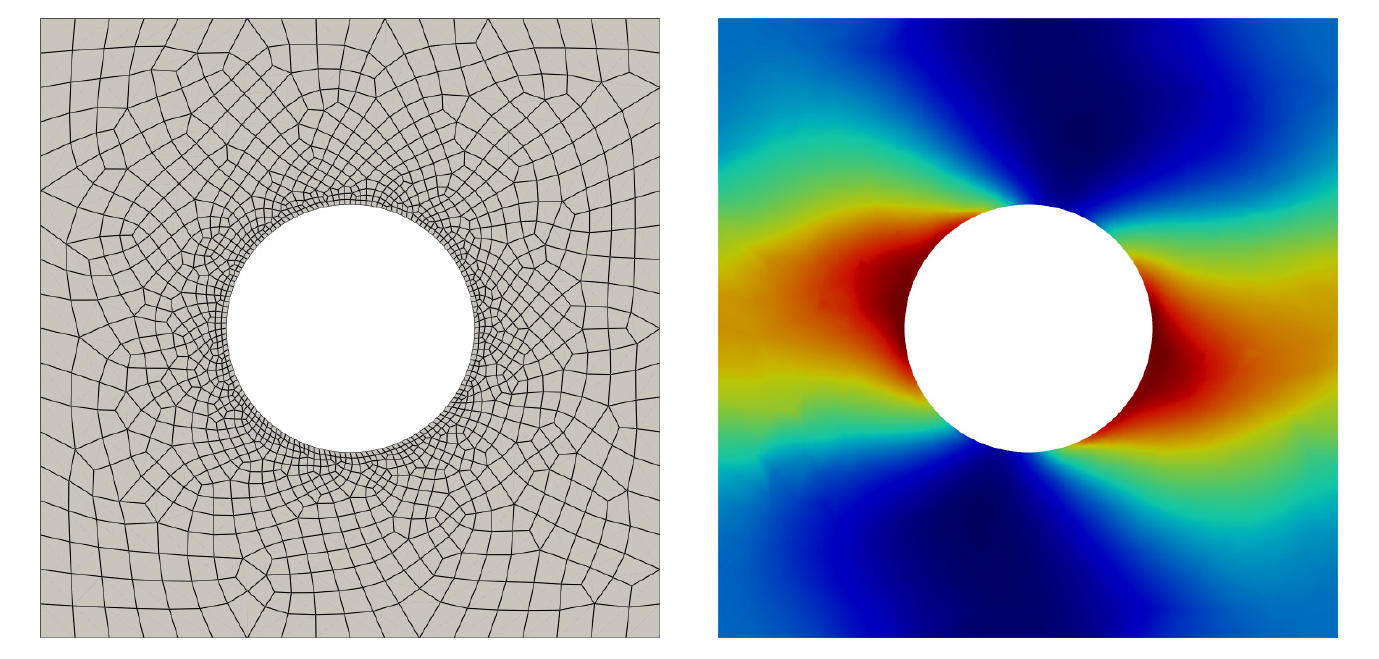}
		\caption{Quadrilateral mesh ($1{,}512$ nodes, $1{,}402$ elements): smooth stress field.}
		\label{fig:fem_quad_mesh}
	\end{subfigure}
	\caption{Comparison of constant strain triangular (CST) and quadrilateral discretizations for the same elasto-plastic loading case.}
\end{figure}

Simulations use plane strain conditions with periodic boundary conditions and the elasto-plastic constitutive model described in Section~\ref{sec:mechanical_problem}.
The material parameters correspond to a titanium alloy (Table~\ref{tab:material_props}).
The geometry is a square plate with a central hole, discretized with $1{,}402$ quadrilateral elements, totaling $1{,}512$ nodes (Figure~\ref{fig:quad_mesh}).
Quadrilateral elements were preferred over triangular ones because linear triangular elements, also called constant strain triangles (CST), suffer from volumetric locking under the incompressibility constraint of plastic flow \cite{ref:belytschko2000nonlinear}, producing artificially stiff responses.
Figures~\ref{fig:fem_tri_mesh} and~\ref{fig:fem_quad_mesh} illustrate this contrast.
The stored outputs are the in-plane stress components $(\sigma_{xx}, \sigma_{yy}, \sigma_{xy})$, which are the only ones entering the two-dimensional weak form and therefore the only ones appearing in the equilibrium residual penalized in Section~\ref{sec:divergence_loss}. The out-of-plane stress $\sigma_{zz}$ is not zero under plane strain and is present in the finite element solutions, but it is not part of the model output.

\begin{figure}[!htbp]
	\centering
	\includegraphics[width=\textwidth]{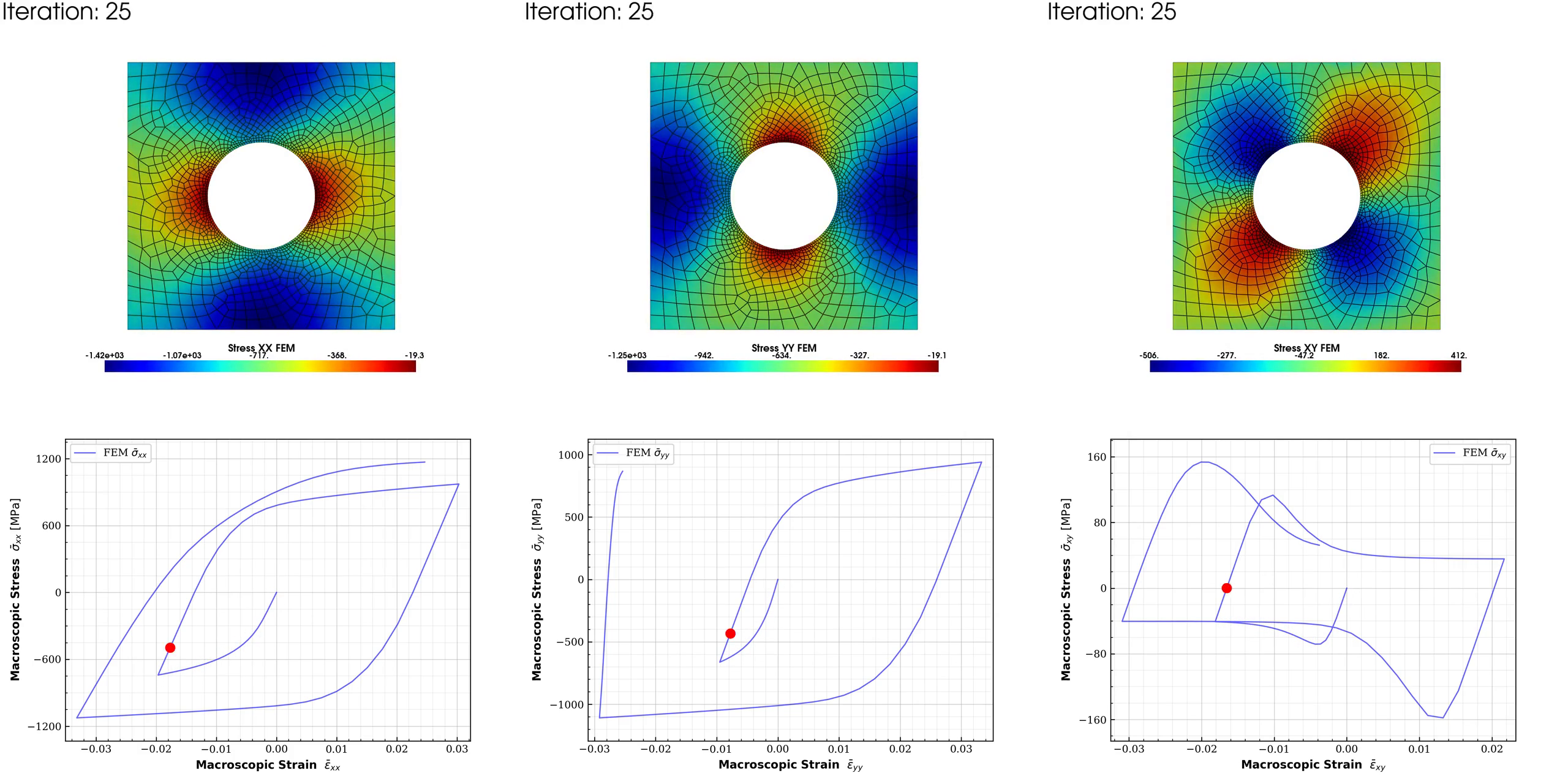}
	\caption{Example of a local stress field snapshot (MPa) with its corresponding mean stress-strain curve. The red point indicates the current loading step.}
	\label{fig:local_mean_curve}
\end{figure}

From each simulation, both macroscopic (volume-averaged, equation~\eqref{eq:mean_continuous}) and local (nodal) stress fields are stored at every time step (Figure~\ref{fig:local_mean_curve}).
Within each element, the Gauss-point stresses are projected onto the element nodes by a least-squares inversion of the shape functions, and the values from all elements sharing a node are averaged to give a single nodal stress.
Before training, the inputs and outputs of each model are
standardized to zero mean and unit variance, with separate statistics for inputs and outputs estimated on the training set.

The LSTM operates on the $d$ macroscopic strain and stress
components, each standardized independently:
\begin{equation}\label{eq:standardization_lstm}
\hat{x}_c = \frac{x_c - \mu_c}{s_c}\ , \qquad c = 1, \ldots, d,
\end{equation}
where $\mu_c$ and $s_c$ are the mean and standard deviation of
component $c$ over all time steps and loading cases. This per-component
scaling lets components of different magnitude contribute comparably
during optimization.

The GNN instead uses a single global mean and standard deviation,
computed over all components and all mesh nodes at once, following the normalization used in the hyperelastic version of the model~\cite{ref:guevara_ijnme_2025}:
\begin{equation}\label{eq:standardization_gnn}
\hat{x} = \frac{x - \mu}{s}\ .
\end{equation}
This global scaling is applied to the mean stress and hidden states
broadcast to the nodes and to the local stress targets, while the
spatial coordinates and boundary label are passed without scaling.

\subsection{LSTM training}\label{sec:lstm_training}
\begin{figure}[!htbp]
	\centering
	\includegraphics[width=0.7\textwidth]{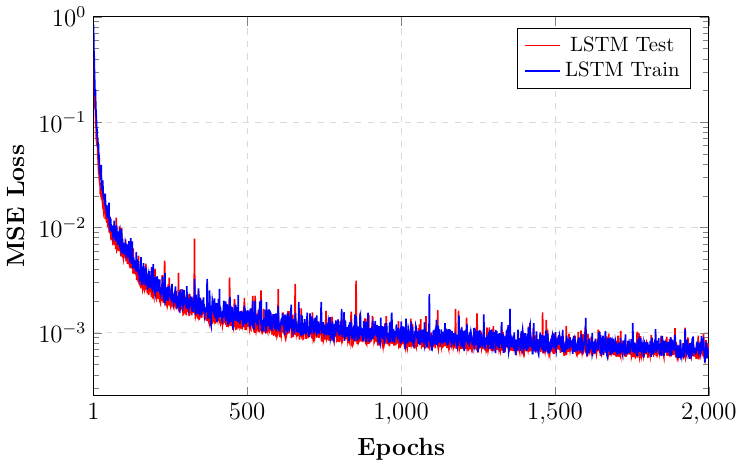}
	\caption{LSTM training and validation loss convergence over $2{,}000$ epochs.}
	\label{fig:lstm_loss}
\end{figure}

The LSTM is trained first in isolation to predict the evolution of the mean stress tensor components from prescribed mean strain sequences.
The model is optimized using Adam \cite{ref:adam} with a learning rate of $10^{-3}$ and a mean squared error (MSE) loss.
Training runs for $2{,}000$ epochs with a mini-batch size of $64$; the loss convergence is shown in Figure~\ref{fig:lstm_loss}. The total training time is approximately 10~minutes on an NVIDIA RTX A1000 GPU (4~GB VRAM), a substantial improvement over the $\sim$7~hours reported in \cite{ref:aymen_fe_lstm} for a comparable dataset, owing to the batched GPU-oriented implementation.
To improve generalization and avoid a possible learning bias related to the discretization resolution used during training, every mini-batch is resampled in time to a random number of steps drawn log-uniformly between $25$ and $1001$ (the native discretization has $100$ steps): the strain and stress sequences are interpolated linearly between their original time points onto the new step count. The network therefore observes the loading paths at many increment sizes rather than at one fixed discretization.

\subsection{Macro-scale model stability}\label{sec:lstm_stability}

A recurrent surrogate of a rate-independent elasto-plastic response should satisfy two requirements. \emph{Stationarity} means that the predicted stress and the internal state remain constant when the applied strain is held constant. \emph{Self-consistency} means that the predictions do not change when the same strain path is described with a finer time discretization. Both requirements were identified by Bonatti and Mohr~\cite{ref:bonatti2022self_consistency_rnn_plasticity} as necessary conditions for a recurrent constitutive surrogate. The random time resampling of the training batches (Section~\ref{sec:lstm_training}) prevents the network from associating one increment size with the constitutive response, within the range of increment sizes it was trained on. Robustness beyond that range is hard to obtained by training alone: since the recurrence of Equation~\eqref{eq:lstm_for_stress} performs one state update per presented step, the model is evaluated behind a committed-state rule (Algorithm~\ref{alg:committed_state}) that fixes the resolution at which the state evolves, independently of the caller's discretization: the recurrent state advances only when the input strain has moved by at least a tolerance $\delta_{\mathrm{tol}}$, in the Euclidean norm of the strain components, since the last committed strain. Smaller moves produce a provisional stress evaluated from the committed state without modifying it, and an input identical to the committed strain returns the committed output unchanged.

\begin{algorithm}[!htbp]
\caption{Committed-state inference of the macro-scale LSTM. The recurrent state advances once per accepted strain increment; sub-tolerance moves are answered with a trial stress that leaves the state untouched. Superscript $\mathrm{c}$ marks committed values, superscript $\mathrm{tr}$ trial ones.}
\label{alg:committed_state}
\KwIn{strain sequence $\{\bar{\boldsymbol{\varepsilon}}_t\}_{t=1}^{T}$, tolerance $\delta_{\mathrm{tol}}$}
\KwOut{stress sequence $\{\bar{\boldsymbol{\sigma}}_t\}_{t=1}^{T}$}
$(\boldsymbol{h}^{\mathrm{c}}, \boldsymbol{c}^{\mathrm{c}}) \leftarrow (\boldsymbol{0}, \boldsymbol{0})$;\quad
$\bar{\boldsymbol{\varepsilon}}^{\mathrm{c}} \leftarrow \boldsymbol{0}$;\quad
$\bar{\boldsymbol{\sigma}}^{\mathrm{c}} \leftarrow \boldsymbol{0}$\;
\For{$t = 1, \ldots, T$}{
  \eIf{$\bar{\boldsymbol{\varepsilon}}_t = \bar{\boldsymbol{\varepsilon}}^{\mathrm{c}}$}{
    $\bar{\boldsymbol{\sigma}}_t \leftarrow \bar{\boldsymbol{\sigma}}^{\mathrm{c}}$\tcp*{repeated input}
  }{
    $(\bar{\boldsymbol{\sigma}}^{\mathrm{tr}}, \boldsymbol{h}^{\mathrm{tr}}, \boldsymbol{c}^{\mathrm{tr}}) \leftarrow \textrm{LSTM}(\bar{\boldsymbol{\varepsilon}}_t, \boldsymbol{h}^{\mathrm{c}}, \boldsymbol{c}^{\mathrm{c}})$\tcp*{trial step}
    $\bar{\boldsymbol{\sigma}}_t \leftarrow \bar{\boldsymbol{\sigma}}^{\mathrm{tr}}$\;
    \If{$\lVert \bar{\boldsymbol{\varepsilon}}_t - \bar{\boldsymbol{\varepsilon}}^{\mathrm{c}} \rVert \geq \delta_{\mathrm{tol}}$}{
      $(\boldsymbol{h}^{\mathrm{c}}, \boldsymbol{c}^{\mathrm{c}}) \leftarrow (\boldsymbol{h}^{\mathrm{tr}}, \boldsymbol{c}^{\mathrm{tr}})$\tcp*{commit}
      $\bar{\boldsymbol{\varepsilon}}^{\mathrm{c}} \leftarrow \bar{\boldsymbol{\varepsilon}}_t$;\quad
      $\bar{\boldsymbol{\sigma}}^{\mathrm{c}} \leftarrow \bar{\boldsymbol{\sigma}}^{\mathrm{tr}}$\;
    }
  }
}
\end{algorithm}

The rule reproduces the trial-state contract of an implicit finite element material routine. Within a load increment the routine is called once per equilibrium iteration, and each call recomputes the response from the state converged at the end of the previous increment, so that repeated probing leaves the stored internal variables untouched; they are overwritten only when the increment is accepted. The sub-tolerance branch of Algorithm~\ref{alg:committed_state} plays the role of an elastic step in a return-mapping algorithm, which returns a stress evaluated from the committed plastic state and leaves the internal variables unchanged. The two rules differ in what triggers the update: a finite element routine commits when the global equilibrium iteration converges, whereas Algorithm~\ref{alg:committed_state} commits when the applied strain has moved by $\delta_{\mathrm{tol}}$. What the analogy transfers is the guarantee that a provisional evaluation cannot corrupt the stored state, not the criterion by which an increment is accepted. Because the distance is measured from the last committed strain, sub-tolerance moves accumulate: any net strain displacement is committed as soon as it reaches the tolerance, so no loading is lost. The tolerance is set to $\delta_{\mathrm{tol}} = 0.1\,\Delta\bar{\varepsilon}_{\mathrm{med}}$, where
\begin{equation}\label{eq:ds50}
\Delta\bar{\varepsilon}_{\mathrm{med}} = \underset{p,\,t}{\mathrm{median}}\;\bigl\lVert \bar{\boldsymbol{\varepsilon}}^{(p)}_{t+1} - \bar{\boldsymbol{\varepsilon}}^{(p)}_{t} \bigr\rVert
\end{equation}
is the median Euclidean norm of the increment between consecutive stored strain states, taken over the three components and pooled over all paths $p$ and all $100$ native increments $t$ of each ($\Delta\bar{\varepsilon}_{\mathrm{med}} = 3.2\times10^{-3}$); the tests of Appendix~\ref{app:tolerance} show that the results are insensitive to this choice over the band $[0.05,\,0.1]\,\Delta\bar{\varepsilon}_{\mathrm{med}}$.

\subsection{GNN training}\label{sec:gnn_training}
\begin{figure}[!htbp]
	\centering
	\includegraphics[width=0.7\textwidth]{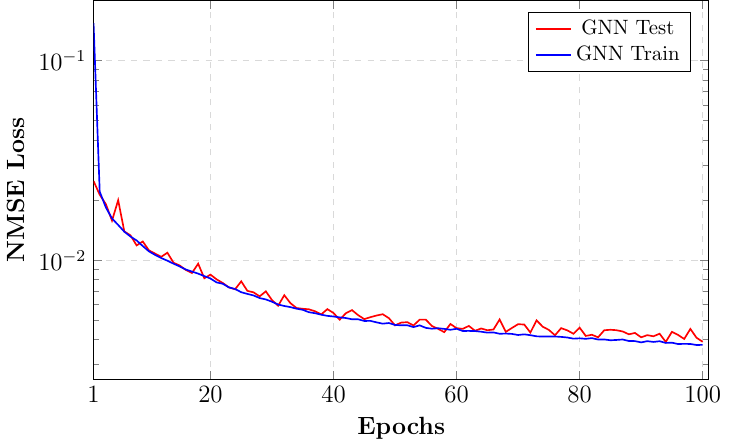}
	\caption{GNN training and validation NMSE convergence over $100$ epochs.}
	\label{fig:gnn_loss}
\end{figure}

Once the LSTM is trained, it is run in inference mode to extract the hidden state sequences for each loading path. These hidden states are incorporated as additional node features in the graphs used for GNN training (Equation~\eqref{eq:node_features}).

To reduce memory cost, the GNN is trained only on snapshots at the four main strain steps per loading case, not on the full step trajectories. Despite this, the results presented in Section \ref{sec:results} show that the GNN model is capable of reconstructing the whole loading history even for intermediate increments not seen during training. Each training batch is formed from a contiguous subsequence of a loading path. This is an implementation choice of the data loader and not a requirement: the GNN is time-agnostic given its macro-scale inputs, and a control run with a random snapshot sampler converges to the same accuracy.

The GNN is optimized, again using Adam, at a learning rate of $10^{-3}$, using the normalized mean squared error (NMSE) loss over $100$ epochs with batch size of $50$.
Training takes approximately twelve hours on an NVIDIA RTX A4000 GPU (16~GB); the NMSE convergence is shown in Figure~\ref{fig:gnn_loss}.
The LSTM and GNN are implemented in PyTorch~\cite{ref:pytorch}, with the graph operations built on PyTorch Geometric~\cite{ref:pyg}.
The NMSE ensures equal contribution from each stress component
\begin{equation}\label{eq:nmse}
    \textrm{NMSE}\, (\boldsymbol{\sigma}, \hat{\boldsymbol{\sigma}}) = \frac{1}{d} \sum_{c=1}^{d} \frac{\sum_{i=1}^{n} (\sigma_{c,i} - \hat{\sigma}_{c,i})^2}{\sum_{i=1}^{n} (\sigma_{c,i} - \frac{1}{n} \sum_{j=1}^n \sigma_{c,j})^2}\ ,
\end{equation}
where $\sigma_{c,i}$ is the ground truth local stress computed from FEA at node $i$, $\hat{\sigma}_{c,i}$ the GNN predicted local stress for the stress component $c \in \{ xx, yy, xy \}$.

\subsection{Divergence-regularized training}\label{sec:divergence_loss}

\par
A central challenge in physics-informed learning is balancing data-driven and physics-based loss terms.
Wang et al.~\cite{ref:wang2021gradient_pathologies_pinn, ref:wang2022ntk_pinn_failure} analyzed gradient flow pathologies in PINNs from a Neural Tangent Kernel perspective, showing that naive multi-objective losses lead to training failure.
GradNorm \cite{ref:chen2018gradnorm} and more recent approaches \cite{ref:bischof2025relobralo_loss_balancing} propose adaptive weighting strategies based on gradient magnitudes or loss ratios.
We propose a relative weighting strategy that addresses a similar challenge in the GNN context, where the scale mismatch between the NMSE and divergence terms is particularly severe for elasto-plastic materials.

In our previous work \cite{ref:guevara_ijnme_2025}, we introduced the \textit{\proposedModelName} architecture with a physics-informed loss that guides the predicted stress field toward the local equilibrium condition $\operatorname{div} \boldsymbol{\sigma} = \boldsymbol{0}$.
The penalty relies on a discrete divergence operator $\hat{\mathbf{D}}_{\sigma}$, assembled from the FE shape-function derivatives and averaged across adjacent elements at shared nodes, that maps the stacked vector of nodal stress components to its nodal divergence. Its full construction is given in \cite{ref:guevara_ijnme_2025}.
The divergence penalty reads
\begin{equation}\label{eq:div_term}
\mathcal{L}_{\mathrm{div}} = (\mathbf{div}(\boldsymbol{\hat{\sigma}}))^2 =
\frac{1}{n} \sum_{i=1}^n \left\| \left. \mathbf{div}(\boldsymbol{\hat{\sigma}}) \right|_{\mathbf{x}=\mathbf{x}_i}\right\|^2,
\end{equation}
evaluated only at interior nodes, since boundary nodes exhibit artificially high divergence values in both the FE reference and the predicted fields.

A direct application of the fixed-weight formulation $\mathcal{L} = \textrm{NMSE} + \lambda \, \mathcal{L}_{\mathrm{div}}$ from \cite{ref:guevara_ijnme_2025} is problematic in the elasto-plastic setting.
The NMSE is a dimensionless ratio of order $10^{-2}$, while the divergence term can reach $10^{6}$ during early training.
Any fixed $\lambda$ large enough to influence equilibrium dominates the gradient signal and prevents NMSE convergence.

\begin{figure}[!htbp]
    \centering
    \includegraphics[width=0.7\textwidth]{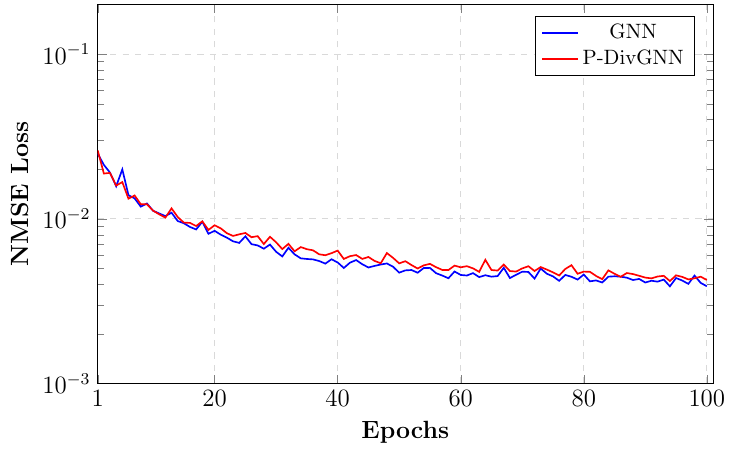}
    \caption{Validation NMSE convergence for the baseline GNN and \textit{\proposedModelName} over $100$ training epochs. Both models reach comparable final NMSE, confirming the divergence penalty does not hinder reconstruction.}
    \label{fig:gnn_loss_comparison}
\end{figure}

To resolve this scale mismatch, we introduce a relative weighting strategy with linear warm-up
\begin{equation}\label{eq:relative_weighting}
    \mathcal{L}_{\mathrm{total}} = \textrm{NMSE} + \lambda_{\mathrm{eff}} \, \mathcal{L}_{\mathrm{div}}\ , \qquad
    \lambda_{\mathrm{eff}} = \lambda_{\mathrm{rel}} \cdot w(e) \cdot \frac{\overline{\textrm{NMSE}}}{\overline{\mathcal{L}}_{\mathrm{div}} + \varepsilon}\ ,
\end{equation}
where $\overline{\textrm{NMSE}}$ and $\overline{\mathcal{L}}_{\mathrm{div}}$ are the current batch values treated as constants, and $w(e) \!=\! \min(1, e/K)$ is a linear warm-up over $K$ epochs.
By construction, $\lambda_{\mathrm{eff}} \cdot \mathcal{L}_{\mathrm{div}} \approx \lambda_{\mathrm{rel}} \cdot w(e) \cdot \textrm{NMSE}$, so the divergence term contributes a controlled fraction of the total gradient regardless of absolute scales.
We use $\lambda_{\mathrm{rel}} \!= \!0.1$ and $K \!= \!20$, so that the physics penalty reaches approximately $10\%$ of the total loss after warm-up. Figure~\ref{fig:gnn_loss_comparison} compares the validation NMSE of the baseline GNN and the proposed model over the 100 training epochs; both converge to comparable final values, confirming that the divergence penalty does not degrade reconstruction accuracy. A preliminary study showing the sensitivity of the reconstruction accuracy and of the equilibrium residual to $\lambda_{\mathrm{rel}}$ and to the warm-up length $K$ is reported in Appendix~\ref{app:lambda_k}.

\section{Results}\label{sec:results}

The following sections evaluate the coupled LSTM-GNN framework against reference FE solutions. Macroscopic accuracy is quantified using the Mean Squared Error (MSE), Mean Absolute Error (MAE), and weighted Mean Absolute Percentage Error (wMAPE):
\begin{align}
	\text{MSE}\,(\mathbf{y},\hat{\mathbf{y}}) &= \frac{1}{N} \sum_{i=1}^{N} (\hat{y}_i - y_i)^{2}, \\[0.3em]
	\text{MAE}\,(\mathbf{y},\hat{\mathbf{y}}) &= \frac{1}{N} \sum_{i=1}^{N} \lvert \hat{y}_i - y_i \rvert\ , \\[0.3em]
	\text{wMAPE}\,(\mathbf{y},\hat{\mathbf{y}}) &= \frac{\sum_{i=1}^{N} \lvert y_i - \hat{y}_i \rvert}{\sum_{i=1}^{N} \lvert y_i \rvert}\ .
\end{align}
Microscopic field accuracy is assessed using the NMSE (Equation~\eqref{eq:nmse}) and the discrete stress divergence (Equation~\eqref{eq:div_term}).

\subsection{Macroscopic stress predictions}\label{sec:macro_results}

\begin{figure}[!htbp]
	\centering
	\includegraphics[width=\textwidth]{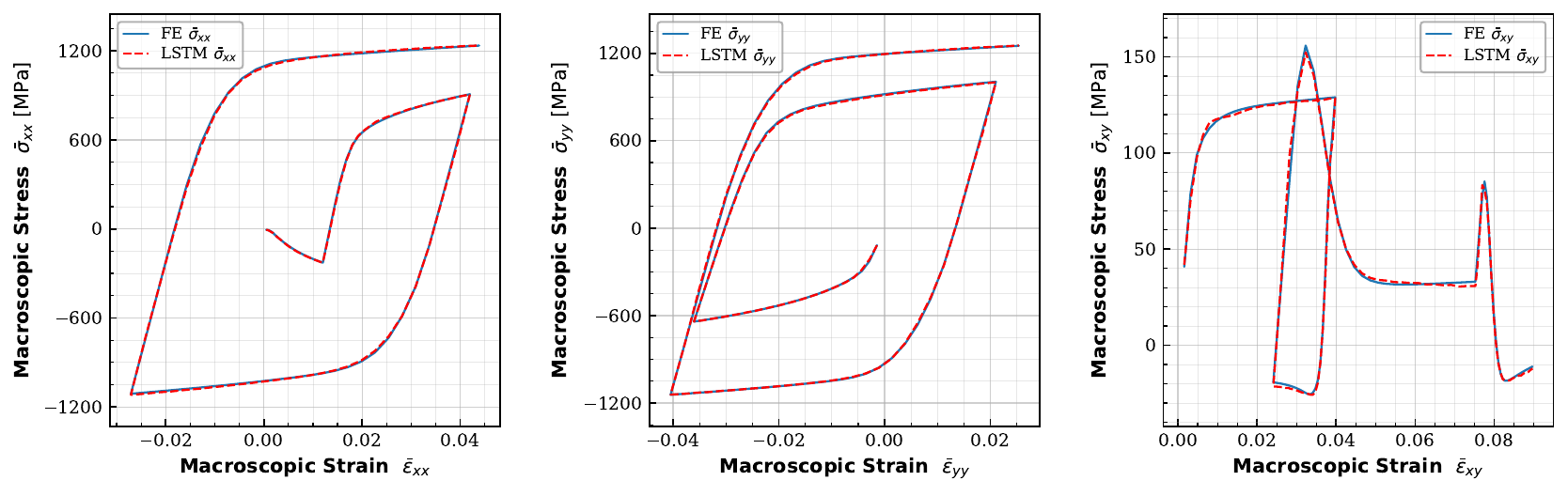}\\[0.8em]
	\includegraphics[width=\textwidth]{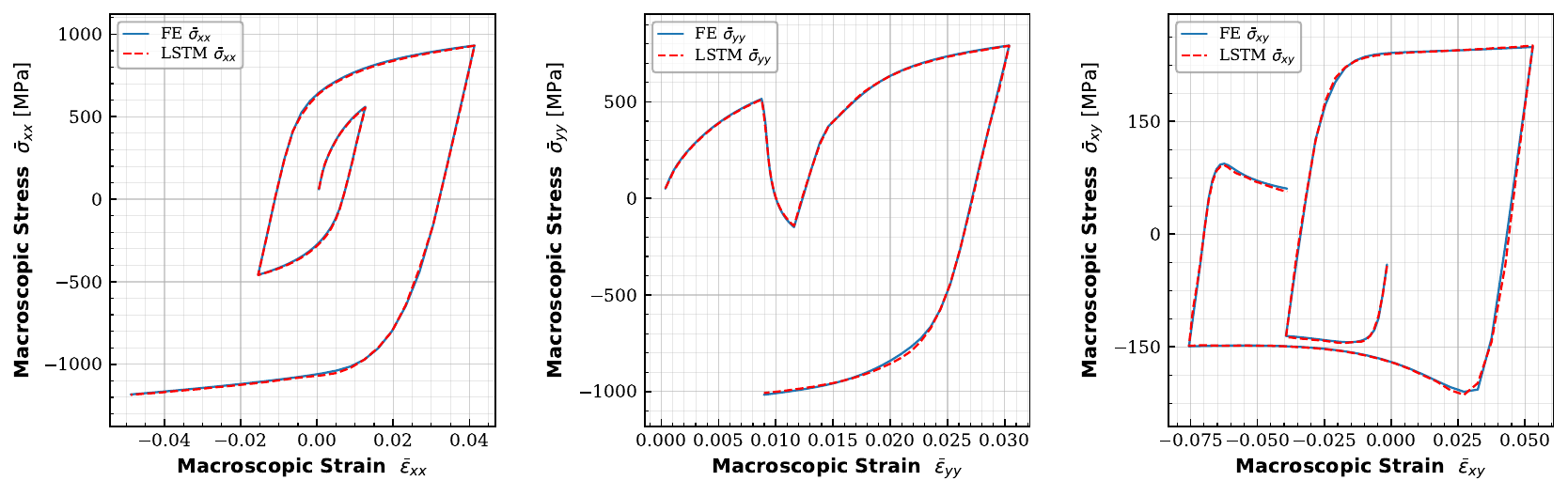}
	\caption{Macroscopic stress-strain responses for the two best test predictions, ranked by mean absolute error: FE reference (solid blue) vs.\ LSTM prediction (dashed red).}
	\label{fig:lstm_best}
\end{figure}

\begin{figure}[!htbp]
	\centering
	\includegraphics[width=\textwidth]{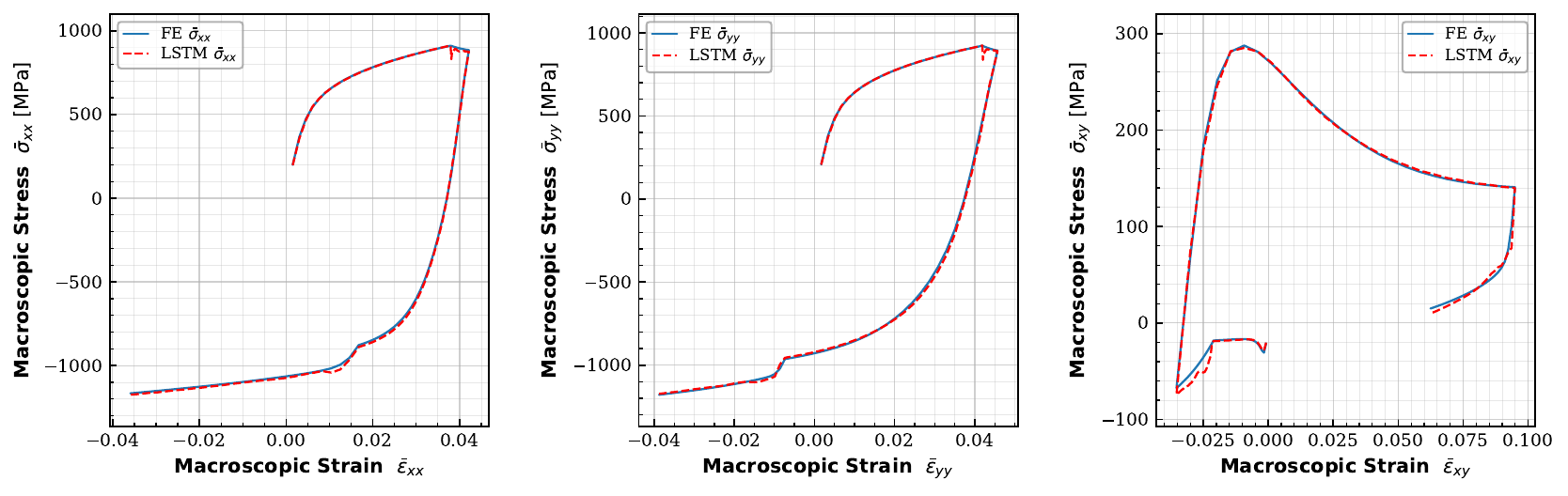}\\[0.8em]
	\includegraphics[width=\textwidth]{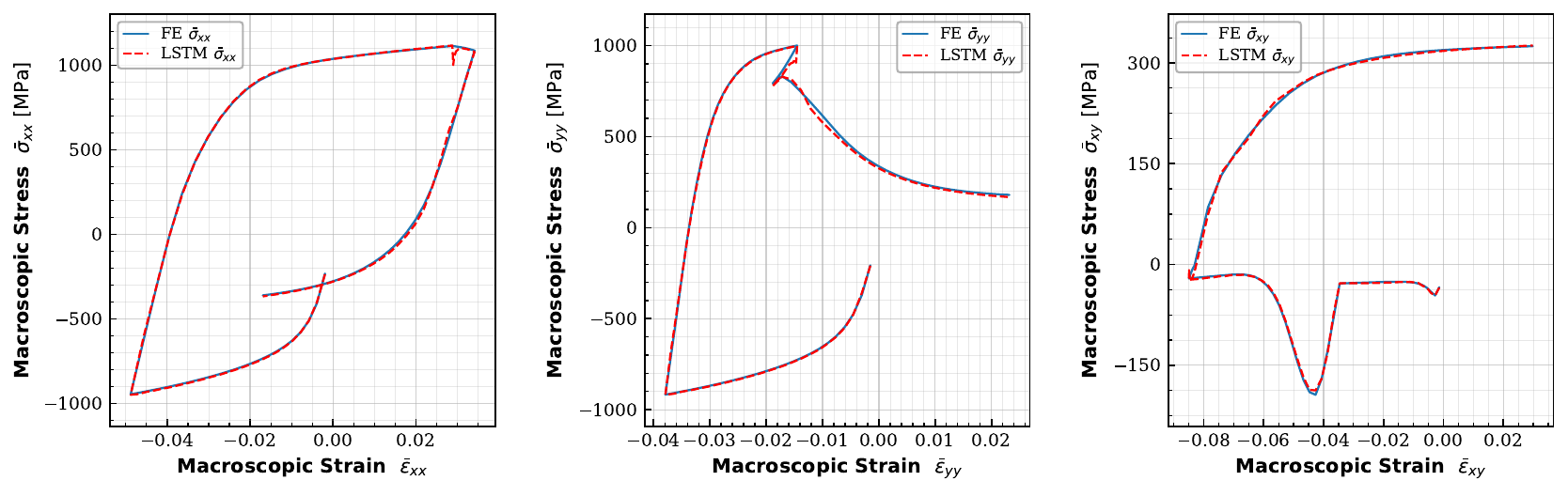}
	\caption{Macroscopic stress-strain responses for the two worst test predictions, ranked by mean absolute error: FE reference (solid blue) vs.\ LSTM prediction (dashed red).}
	\label{fig:lstm_worst}
\end{figure}

The $3{,}000$ test simulations are ranked by their MAE averaged over the three stress components. Figure~\ref{fig:lstm_best} presents the two best predictions and Figure~\ref{fig:lstm_worst} the two worst, illustrating the range of the model's accuracy.

Even in the worst cases, the LSTM closely tracks the reference solution throughout the non-proportional loading path. The most notable discrepancies occur during rapid transitions between loading segments, where elastic-plastic regime changes coincide with multiaxial stress rotations.

Table~\ref{tab:lstm_metrics} reports the global error metrics over the full test set.
Despite the localized discrepancies in the worst cases, the low MSE and MAE values confirm accurate predictions across the database, with an overall wMAPE of $1.65\%$.
The shear component $\bar{\sigma}_{xy}$ exhibits the largest errors due to its sensitivity to non-proportional loading and its lower magnitudes, which amplify percentage-based metrics.

\begin{table}[!htbp]
\centering
\caption{LSTM prediction error metrics on the full test set ($3{,}000$ simulations).}
\label{tab:lstm_metrics}
\begin{tabular}{lcccc}
\hline
\textbf{Metric} & \textbf{Overall} & $\bar{\sigma}_{xx}$ & $\bar{\sigma}_{yy}$ & $\bar{\sigma}_{xy}$ \\
\hline
MSE   & $5.55 \times 10^{-4}$ & $3.91 \times 10^{-4}$ & $4.10 \times 10^{-4}$ & $8.64 \times 10^{-4}$ \\
MAE   & $1.50 \times 10^{-2}$ & $1.32 \times 10^{-2}$ & $1.33 \times 10^{-2}$ & $1.85 \times 10^{-2}$ \\
wMAPE & $1.65\%$ & $1.56\%$ & $1.57\%$ & $2.24\%$ \\
\hline
\end{tabular}
\end{table}

\FloatBarrier
\subsection{LSTM hidden state analysis}\label{sec:hidden_state}

\begin{figure}[!htbp]
	\centering
	\includegraphics[width=0.55\textwidth]{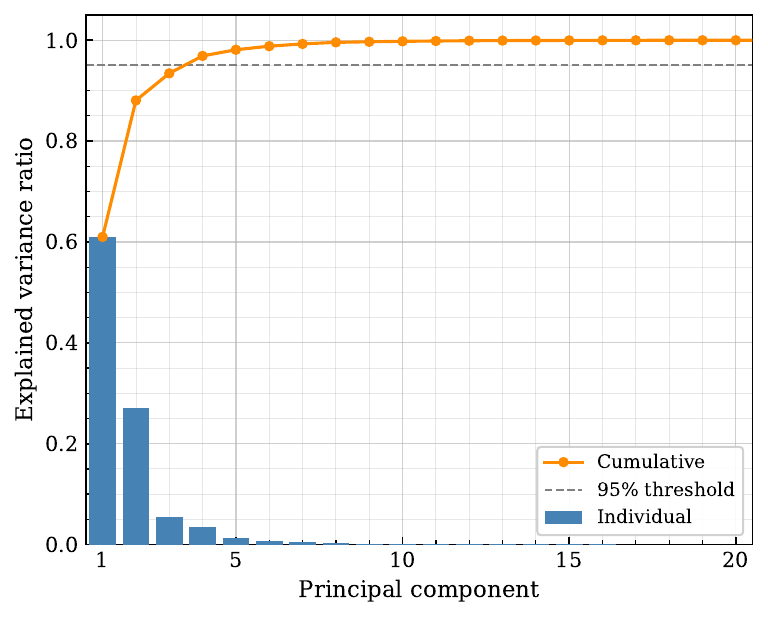}
	\caption{Explained variance ratio of the principal components of the LSTM hidden states. The dashed line marks the $95\%$ cumulative threshold, reached at the fourth component; the first two components alone capture about $88\%$ of the total variance.}
	\label{fig:pca_explained_variance}
\end{figure}

\begin{figure}[!htbp]
	\centering
	\begin{subfigure}[t]{0.48\textwidth}
		\centering
		\includegraphics[width=\textwidth]{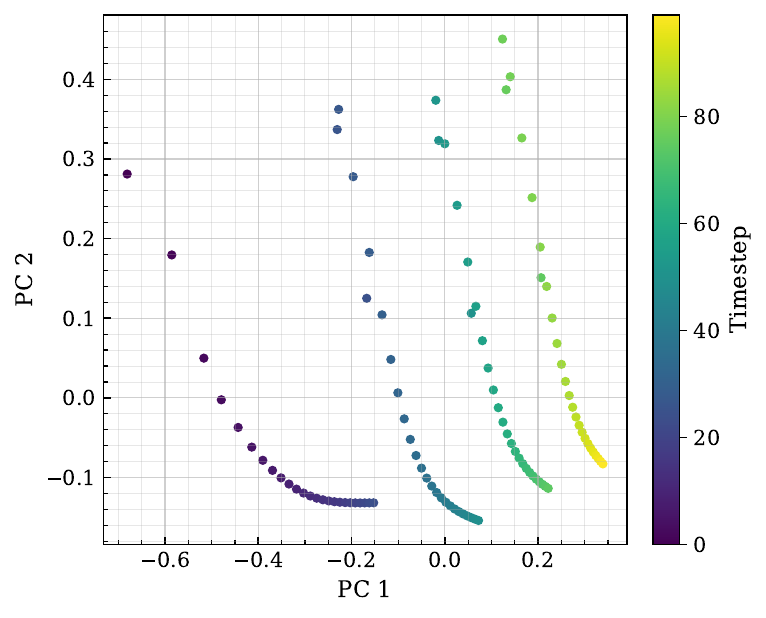}
		\caption{PCA projection onto the two leading principal components. The left-to-right progression along PC~1 reflects the accumulation of loading history, while repeating arc patterns correspond to the four strain segments.}
		\label{fig:pca_hidden_states}
	\end{subfigure}
	\hfill
	\begin{subfigure}[t]{0.48\textwidth}
		\centering
		\includegraphics[width=\textwidth]{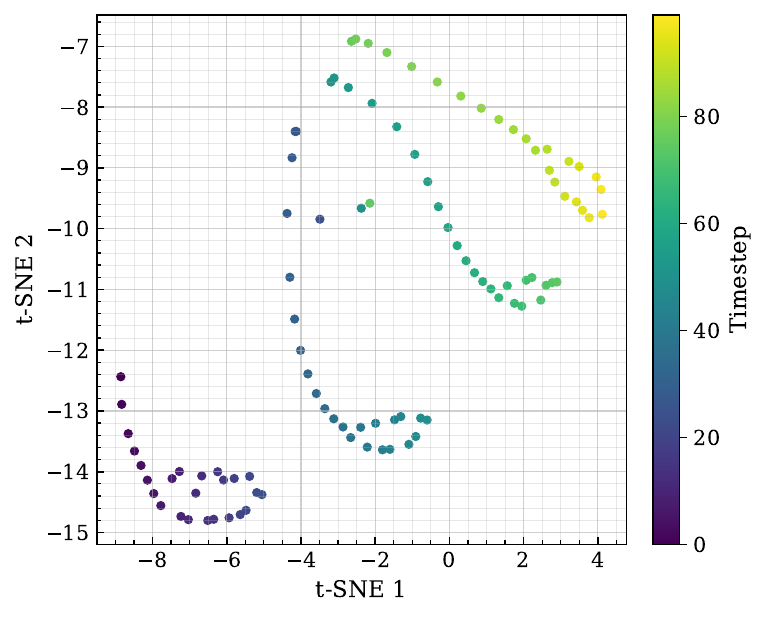}
		\caption{t-SNE embedding. Early time steps (dark, bottom-left cluster) occupy a compact region, while later steps spread progressively as plastic deformation accumulates.}
		\label{fig:tsne_hidden_states}
	\end{subfigure}
	\caption{Dimensionality reduction of the LSTM hidden states averaged across all test simulations. Each point corresponds to one time step, colored by its index in the loading sequence.}
	\label{fig:hidden_states_dr}
\end{figure}

To gain insight into the information encoded by the LSTM, the $64$-dimensional hidden state vectors $\boldsymbol{h}_t$ are analyzed using two dimensionality reduction techniques.
Principal Component Analysis (PCA) \cite{ref:jolliffe_pca} finds orthogonal directions that capture the most variance in the data. t-distributed Stochastic Neighbor Embedding (t-SNE) \cite{ref:tsne} provides a visual representation of the hidden-state variables; it is a non-linear method that maps nearby points in high-dimensional space to nearby points in a two-dimensional embedding, making clusters and local groupings visible.
PCA gives a global, linear view of the hidden state geometry; t-SNE resolves finer structure that a linear projection would flatten.
The explained variance ratio (Figure~\ref{fig:pca_explained_variance}) shows that the first two components alone capture about $88\%$ of the total variance, and the first three exceed $93\%$, which justifies inspecting $\boldsymbol{h}_t$ in a two-dimensional projection.
Figure~\ref{fig:hidden_states_dr} then shows the PCA (Figure~\ref{fig:pca_hidden_states}) and t-SNE (Figure~\ref{fig:tsne_hidden_states}) embeddings of the hidden states averaged across all test simulations.

\begin{figure}[!htbp]
	\centering
	\includegraphics[width=0.8\textwidth]{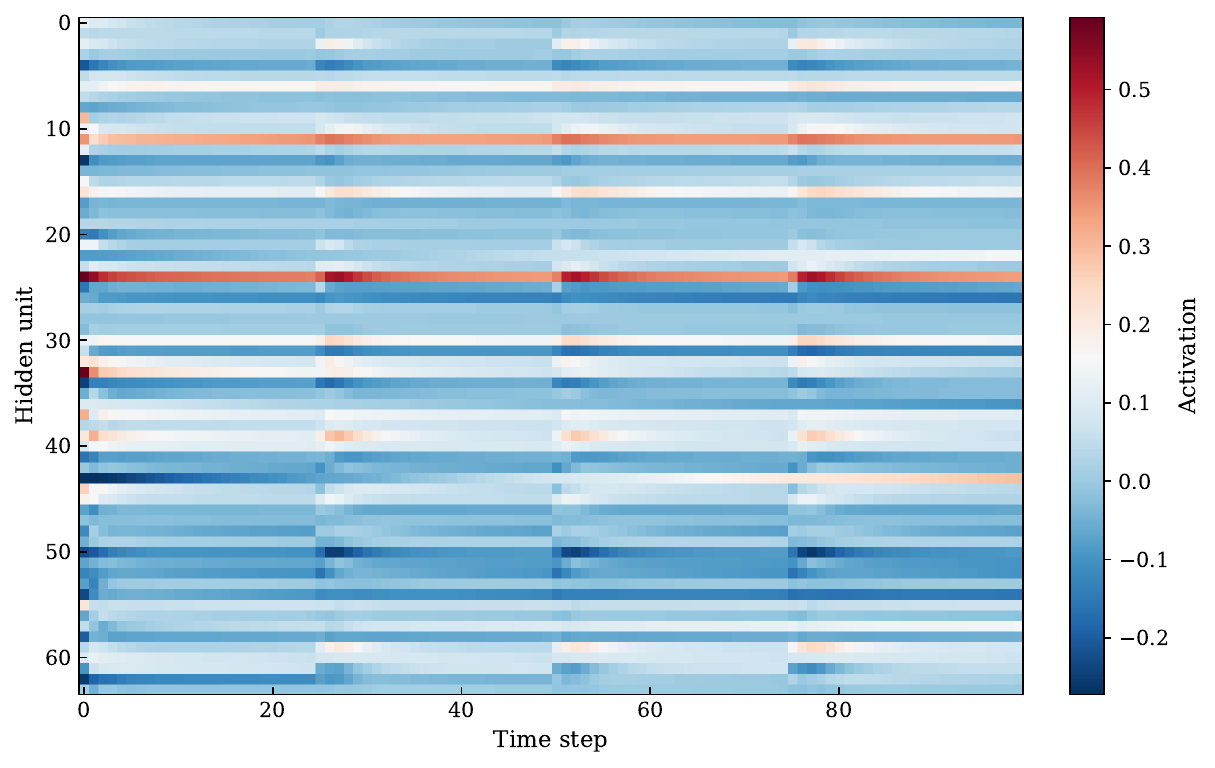}
	\caption{Activation heatmap of the $64$ LSTM hidden units, averaged over the test set. Persistent units encode accumulated plastic state while transient units respond to loading changes.}
	\label{fig:hidden_state_heatmap}
\end{figure}

In the PCA projection, four distinct clusters are clearly visible; each one corresponds to the change in direction following a macro strain increment. The low effective dimensionality of $\boldsymbol{h}_t$ suggests that the hidden states lie on a low-dimensional structure related to the main constitutive variables.

The activation heatmap (Figure~\ref{fig:hidden_state_heatmap}) displays the $64$ hidden unit activations across the loading sequence.
Each plotted activation is the output-gated cell state $\boldsymbol{h}_t \in (-1,1)$; since the output gate is non-negative, a negative value reflects a negative cell-state component rather than a physical sign. Vertical band transitions at steps $\approx 25, 50, 75$ correspond to segment boundaries.

Some units maintain persistent activation throughout the sequence, consistent with encoding accumulated plastic history, while others respond transiently to loading direction changes---suggesting a functional specialization between ``memory'' and ``event'' neurons.

The hidden state variables do not behave like the classical internal variables in thermodynamics of irreversible processes. The second law is not enforced in the LSTM, and even if it were, the constraint would be hard to express, since the macroscopic dissipation rate of a heterogeneous unit cell has no closed form~\cite{ref:these_aymen}. Classical internal variables grow monotonically during dissipative loading and stay constant in the transient elastic regime; most of the hidden units shown here do neither. The LSTM appears to encode a mixture: fields that evolve rapidly in the elastic regime, such as stress, together with variables that track the change of microstructural state once a dissipative process is active. The present results do not characterize that mixture, but the physical meaning of the hidden state is an open direction for data-driven models of dissipative materials.

\FloatBarrier
\subsection{Generalization to longer loading sequences}\label{sec:generalization}

\begin{figure}[!htbp]
	\centering
	\includegraphics[width=0.7\textwidth]{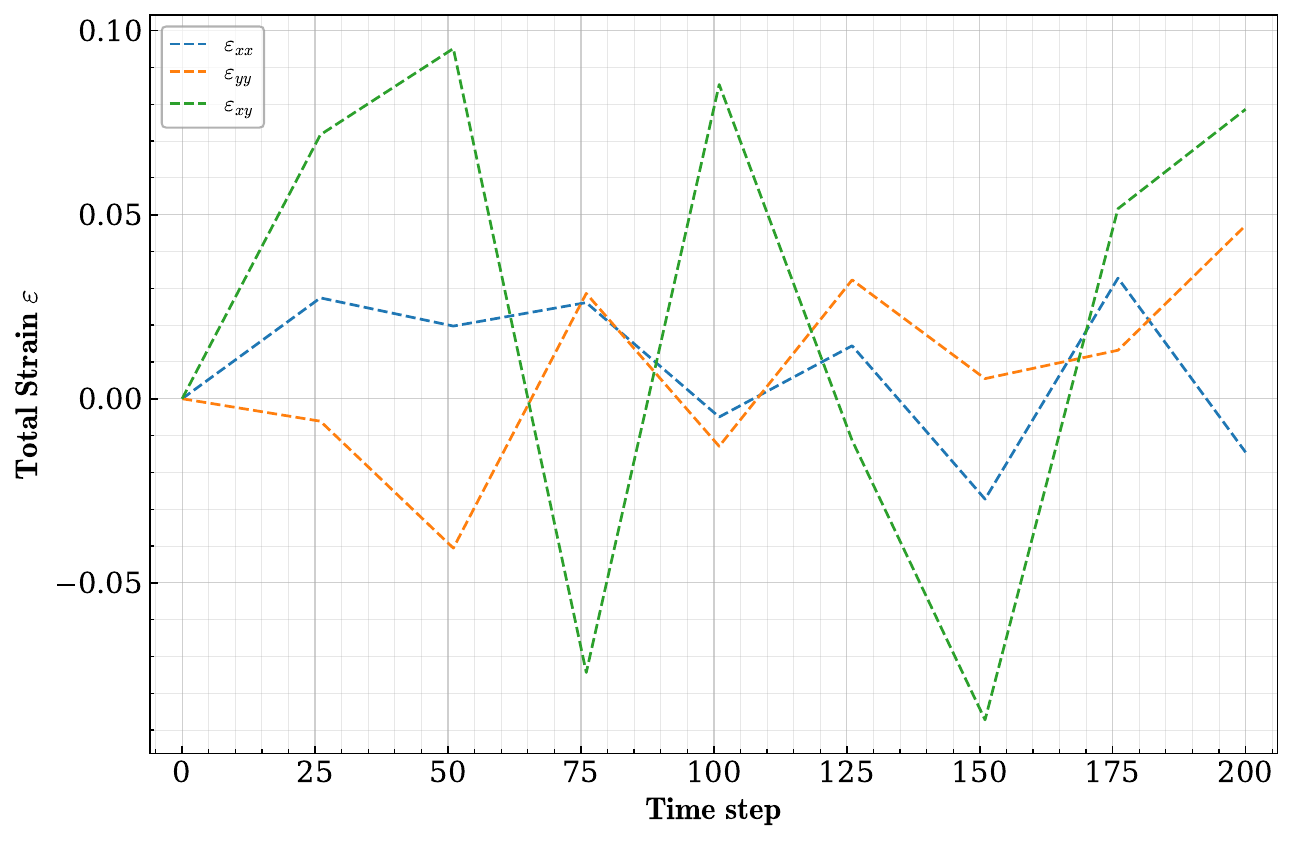}
	\caption{Eight-segment, non-proportional strain loading path used for out-of-distribution testing ($201$ time steps, twice the training length).}
	\label{fig:8seg_strain_path}
\end{figure}

An important question is whether the LSTM generalizes to loading histories longer than those seen during training.
The model, trained exclusively on four-segment paths ($\sim$101 time steps), is evaluated on an eight-segment path with $201$ time steps---approximately double the training horizon (Figure~\ref{fig:8seg_strain_path}).

\begin{figure}[!htbp]
	\centering
	\includegraphics[width=\textwidth]{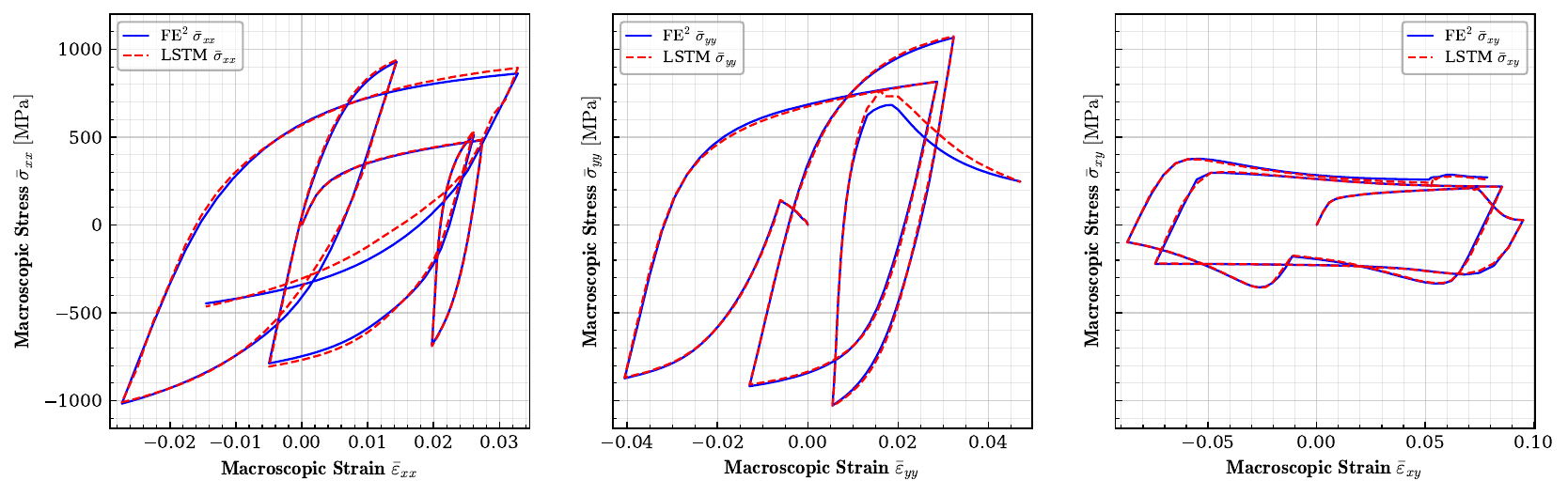}
	\caption{Macroscopic stress-strain response for the eight-segment path: FE reference (solid blue) vs.\ LSTM prediction (dashed red). Cumulative relative error: $\sim$3.0\%.}
	\label{fig:8seg_stress_strain}
\end{figure}

Figure~\ref{fig:8seg_stress_strain} compares the corresponding macroscopic stress-strain response. The LSTM closely reproduces the reference over all eight segments, achieving a cumulative relative error of approximately $3.0\%$.
In this eight-segment test, the error in segments 5--8 stays comparable to that in the first four, which suggests that the recurrent model can extend beyond the training horizon without obvious drift.

\begin{figure}[!htbp]
	\centering
	\begin{subfigure}[t]{0.48\textwidth}
		\centering
		\includegraphics[width=\textwidth]{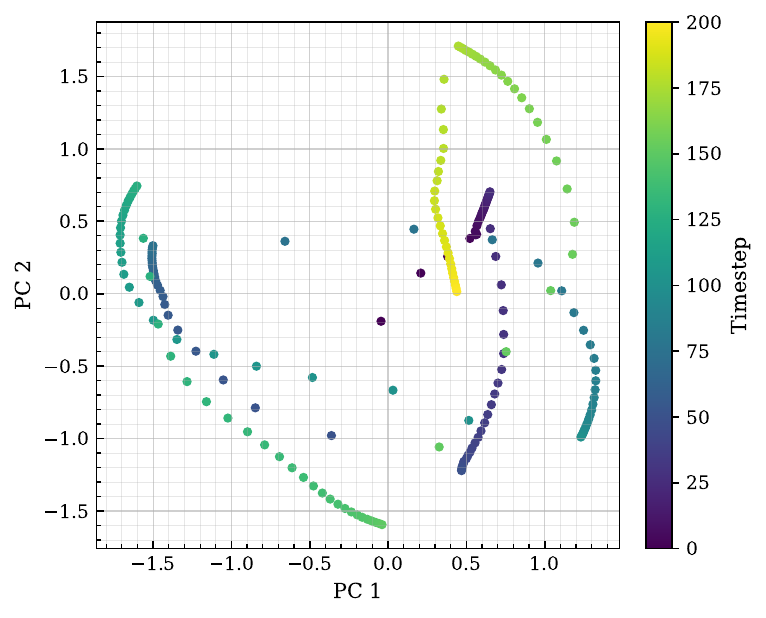}
		\caption{PCA of hidden states.}
	\end{subfigure}
	\hfill
	\begin{subfigure}[t]{0.48\textwidth}
		\centering
		\includegraphics[width=\textwidth]{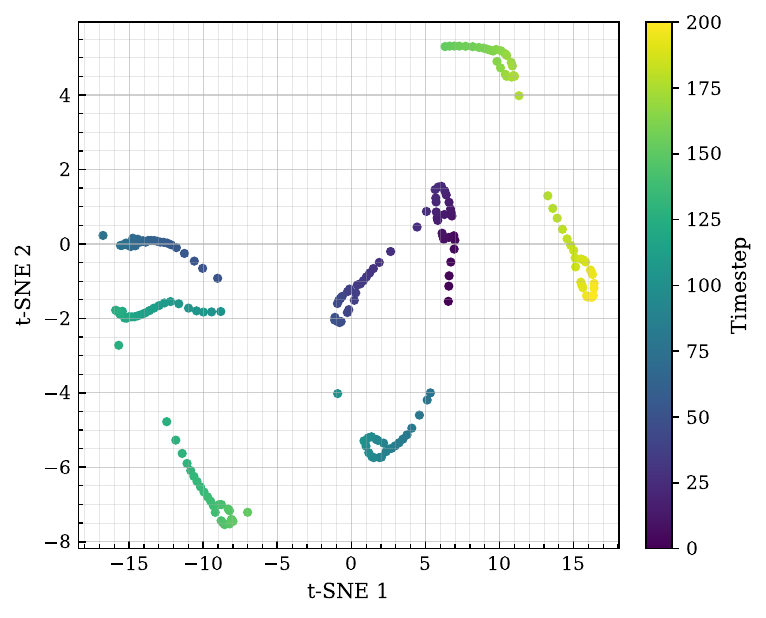}
		\caption{t-SNE of hidden states.}
	\end{subfigure}
	\caption{Dimensionality reduction of the LSTM hidden states for the eight-segment path.
    Colors indicate time step (dark: early, bright: late).
    Both projections reveal eight distinct segment-level clusters.}
	\label{fig:8seg_dr}
\end{figure}

\begin{figure}[!htbp]
	\centering
	\includegraphics[width=0.8\textwidth]{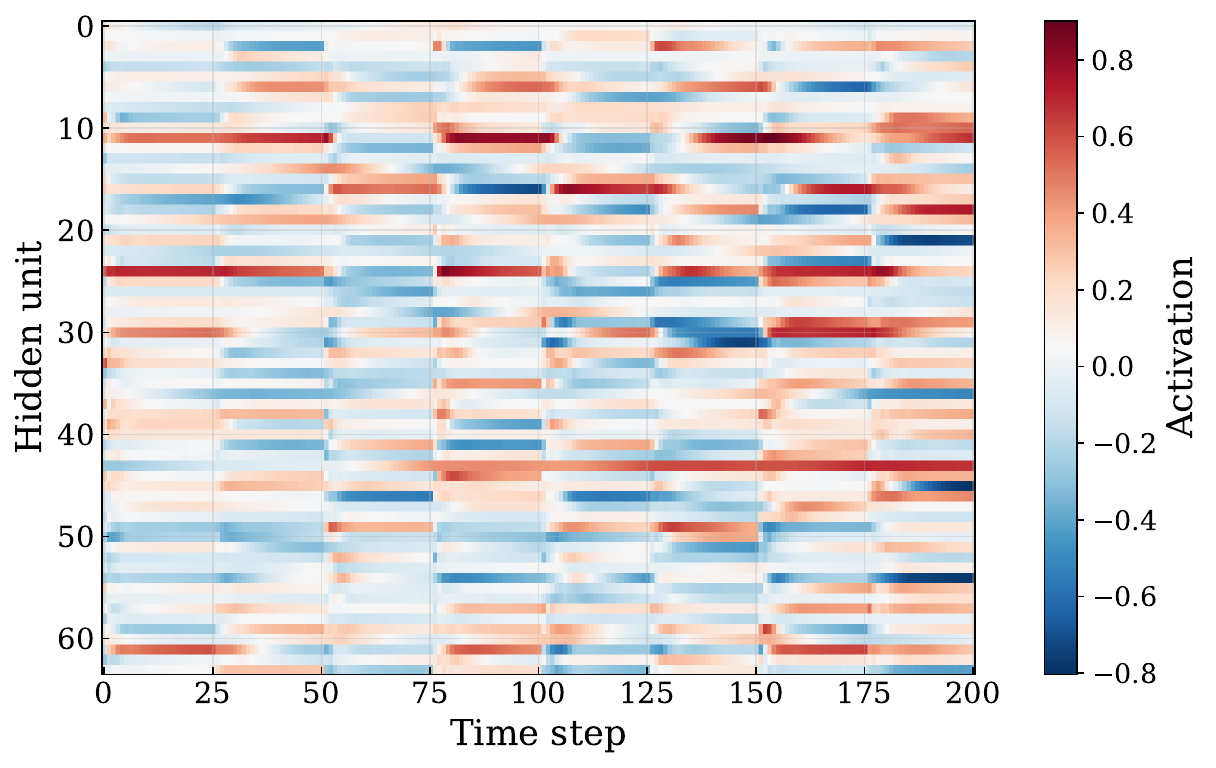}
	\caption{Activation heatmap of the $64$ LSTM hidden units across $201$ time steps.
    Eight vertical band transitions are visible.
    Persistent units encode accumulated plastic history, while transient units respond to loading reversals.}
	\label{fig:8seg_heatmap}
\end{figure}

The hidden-state plots are consistent with this behavior.
The PCA projection (Figure~\ref{fig:8seg_dr}, left) reveals eight distinct arc-shaped trajectories, one per segment, while t-SNE (right) separates them into well-defined clusters.
The activation heatmap (Figure~\ref{fig:8seg_heatmap}) shows eight clear band transitions with no sign of saturation or divergence in the later segments, explaining the sustained prediction accuracy.
A few points in the PCA projection sit away from the eight arcs. These are the first one to four steps after each change of strain direction, where the per-step plastic increment briefly drops and the response is close to elastic before plastic flow resumes.

\FloatBarrier
\subsection{Time-discretization invariance and stationarity tests}\label{sec:stability_tests}

The two requirements introduced in Section~\ref{sec:lstm_stability} are verified on the full $3{,}000$-path validation split, with the macro-scale model in its deployed configuration: random time resampling during training (Section~\ref{sec:lstm_training}) and committed-state inference (Algorithm~\ref{alg:committed_state}) with $\delta_{\mathrm{tol}} = 0.1\,\Delta\bar{\varepsilon}_{\mathrm{med}}$.

Self-consistency is tested by rewriting each validation strain path on a finer time grid and replaying it. Every native increment is subdivided into equal sub-increments, so the stored states are preserved exactly on the refined grid and the corners between the four loading segments remain represented at every refinement level. The strain is interpolated linearly between the stored states, the FE reference is interpolated onto the same grid, and the error is accumulated over all refined steps. Stationarity is tested by appending $100$ steps to each path during which the strain is held at its final value.

\begin{table}[!htbp]
	\centering
	\caption{Self-consistency test: macroscopic prediction error of the deployed macro-scale model when identical validation strain paths are described with an increasing number of time steps. The last column is the number of recurrent state updates per path, averaged over the split.}
	\label{tab:refinement_invariance}
	\begin{tabular}{lccc}
		\toprule
		Discretization & Steps per path & wMAPE & State updates per path \\
		\midrule
		Native                & $101$            & $1.68\%$ & $101$ \\
		$10\times$ finer       & $1{,}001$        & $1.73\%$ & $729$ \\
		$100\times$ finer      & $10{,}001$       & $1.92\%$ & $989$ \\
		$1000\times$ finer     & $100{,}001$      & $1.95\%$ & $1{,}043$ \\
		$10^{4}\times$ finer   & $1{,}000{,}001$  & $1.95\%$ & $1{,}049$ \\
		$10^{5}\times$ finer   & $10{,}000{,}001$ & $1.95\%$ & $1{,}049$ \\
		\bottomrule
	\end{tabular}
\end{table}

\begin{figure}[!htbp]
    \centering
    \includegraphics[width=0.82\textwidth]{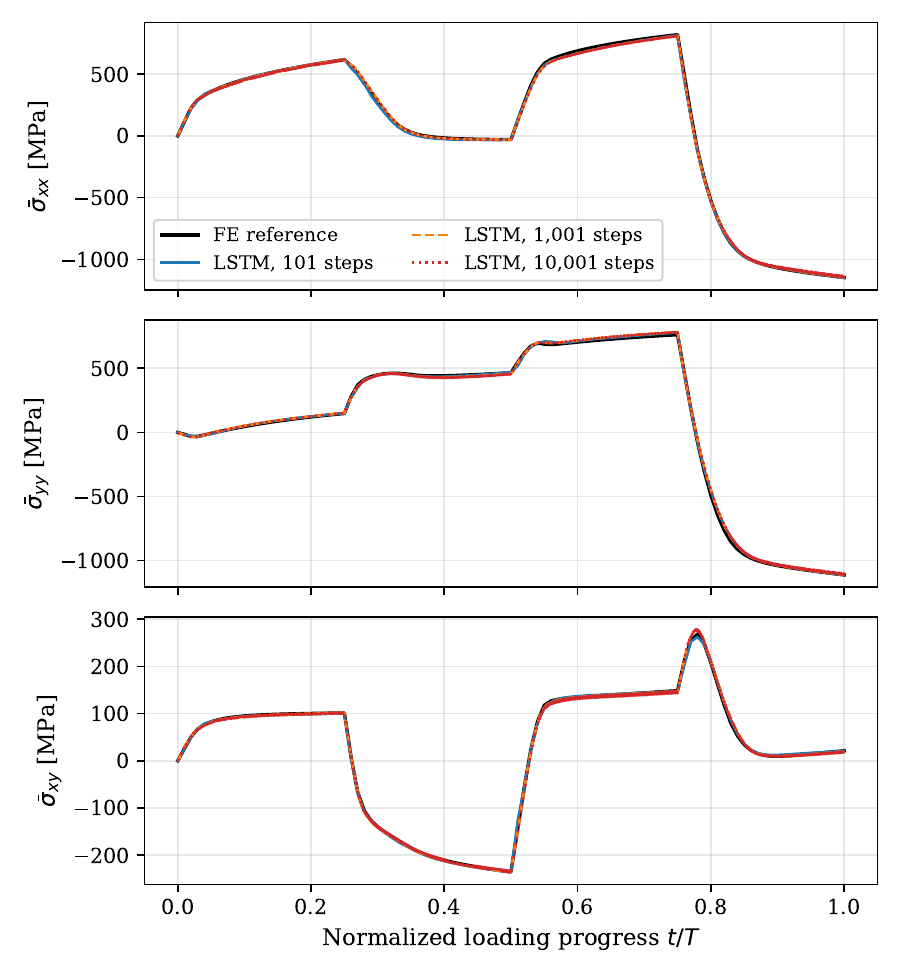}
    \caption{Macroscopic stress components for one validation case: FE reference (black solid) and predictions of the macro-scale model with the same strain path described by $101$ steps (blue solid), $1{,}001$ steps (orange dashed) and $10{,}001$ steps (red dotted).}
    \label{fig:stability_refinement}
\end{figure}

Table~\ref{tab:refinement_invariance} shows that the error remains close to its native value at every refinement level tested. The committed-state rule advances the recurrent state once per $\delta_{\mathrm{tol}}$ of strain arc length, so it is set by the length of the path and not by the number of steps used to describe it. It saturates near $1{,}050$ updates, the arc length of a typical validation path divided by $\delta_{\mathrm{tol}}$, while the number of presented steps grows by a factor of $10^{5}$. Figure~\ref{fig:stability_refinement} shows the three finest of these descriptions replayed through the model for one validation case.

\begin{figure}[!htbp]
    \centering
    \includegraphics[width=0.82\textwidth]{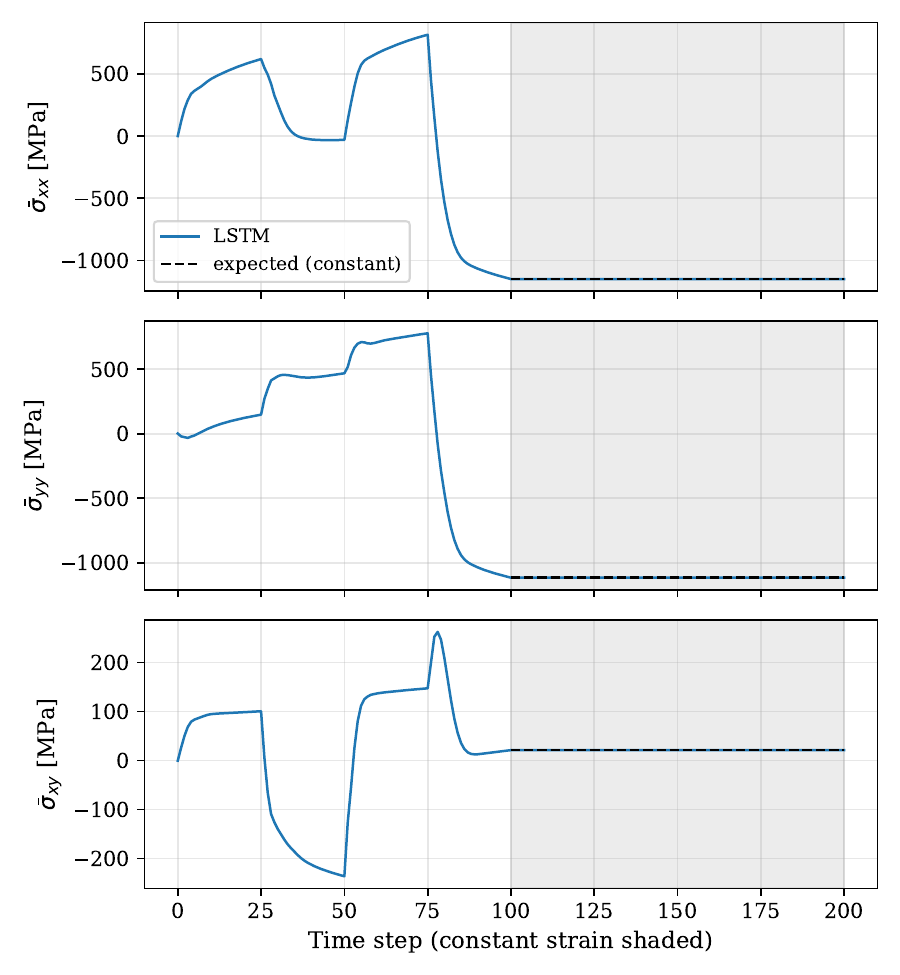}
    \caption{Macroscopic stress components for one validation case extended by $100$ steps at constant strain (shaded): prediction of the macro-scale model (blue solid) and the value held at the last loading step (black dashed).}
    \label{fig:stability_stationarity}
\end{figure}

Stationarity is guaranteed by construction: an input equal to the committed strain returns the committed output without evaluating the recurrence, so the property follows from the inference rule rather than from the trained weights. Over the constant-strain tails the drift of the predicted stress is therefore identically zero on every validation path. Figure~\ref{fig:stability_stationarity} shows one such extended path.

Self-consistency holds within bounds rather than exactly. Two mechanisms set those bounds: the random resampling of the step count during training, which exposes the network to many increment sizes, and the committed-state rule, which fixes the resolution at which the state advances. Together they keep the error between $1.68\%$ and $1.95\%$ over the whole refinement range, without making the prediction strictly independent of the discretization. The sensitivity of this behavior to the value of $\delta_{\mathrm{tol}}$ is reported in Appendix~\ref{app:tolerance}. Below the trained range, at fewer than $25$ steps per path, the accuracy degrades instead, and such inputs should be subdivided before evaluation. An architecture that embeds the property in the recurrent update, such as the Minimal State Cells of Bonatti and Mohr~\cite{ref:bonatti2022self_consistency_rnn_plasticity}, gives the guarantee by construction, by restricting how the state may respond to the size of the increment. The localization stage consumes a fixed-size hidden state vector regardless of the cell that produced it, so such a model can replace the present one without modifying the GNN. Such cells are therefore a direct alternative to gated recurrent networks in this framework. Their state vector can play the same role as the LSTM state does here, carrying the history dependence to the GNN, which makes them a natural candidate for the macro-scale stage of a future version of the model.

\FloatBarrier
\subsection{Microscopic local stress fields}\label{sec:micro_results}
To evaluate both the baseline GNN and \textit{\proposedModelName} under non-linear, history-dependent conditions, an arbitrary loading path is generated for simultaneous comparison against the FE reference.

\begin{figure}[!htbp]
    \centering
    \begin{subfigure}[t]{0.49\textwidth}
        \centering
        \includegraphics[width=\textwidth]{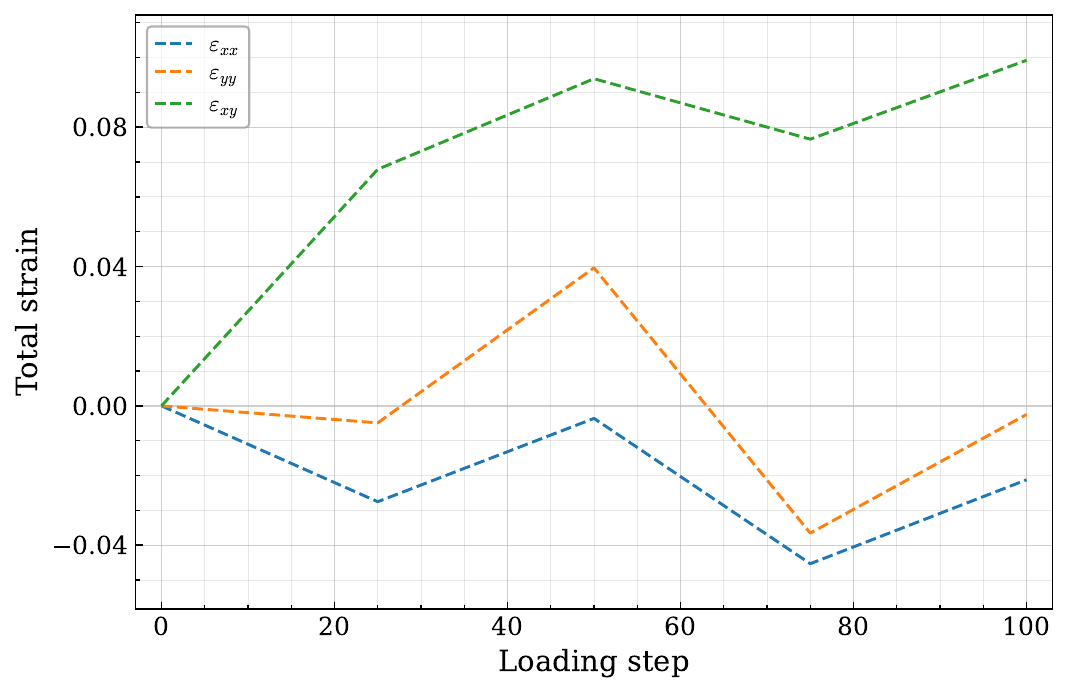}
        \caption{Three-component strain sequence.}
    \end{subfigure}
    \hfill
    \begin{subfigure}[t]{0.38\textwidth}
        \centering
        \includegraphics[width=\textwidth]{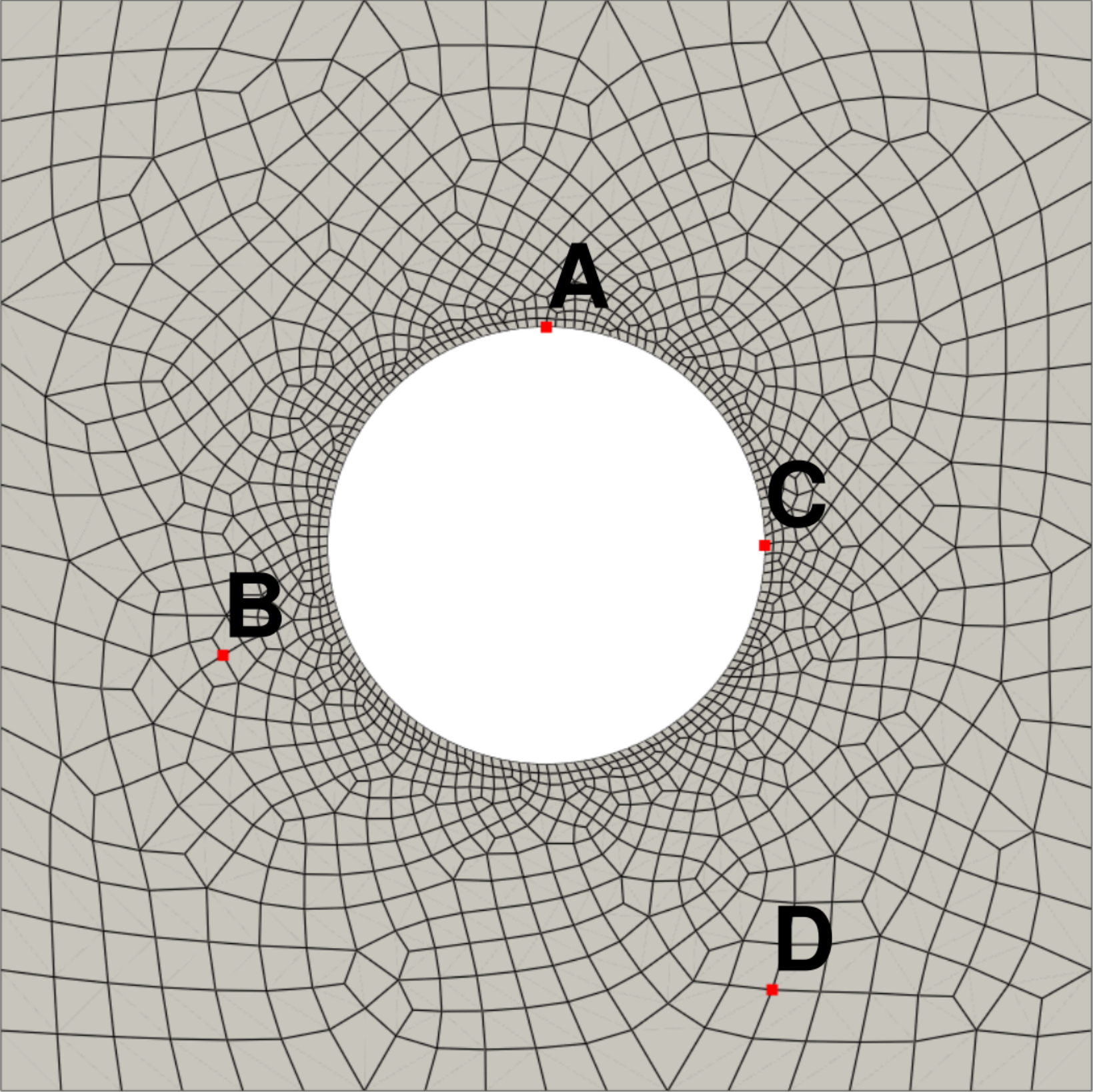}
        \caption{Four representative points on the mesh.}
        \label{fig:mesh_points}
    \end{subfigure}
    \begin{subfigure}[t]{0.98\textwidth}
        \centering
        \includegraphics[width=\textwidth]{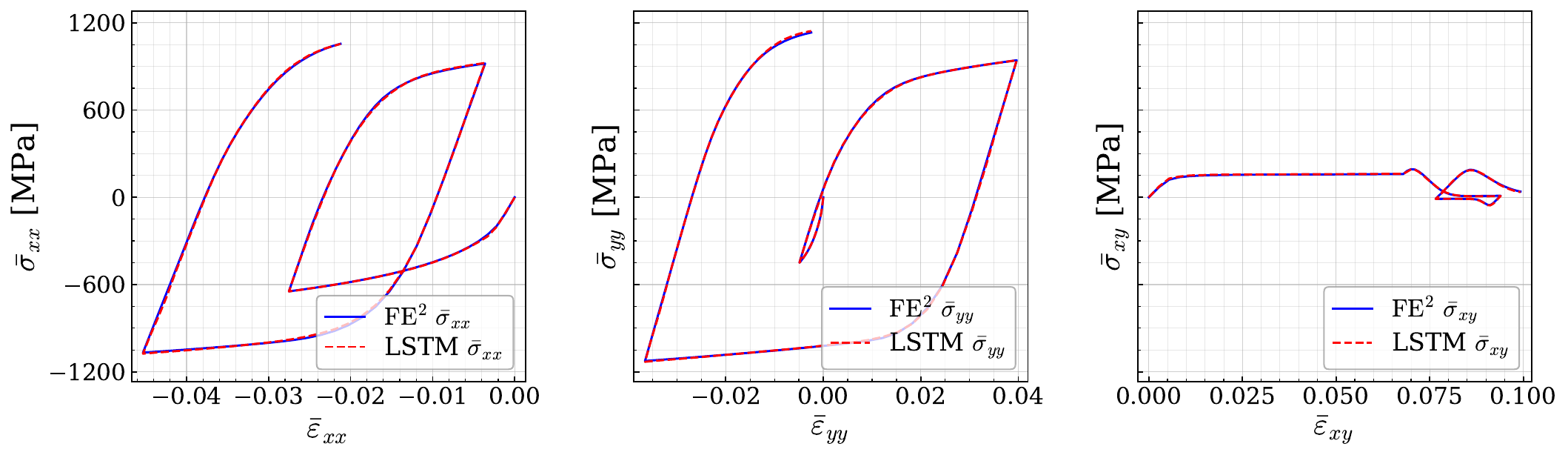}
        \caption{Macroscopic stress-strain response (FE reference vs.\ LSTM prediction).}
        \label{fig:macro_fem_vs_lstm}
    \end{subfigure}
    \vspace{0.3em}
    \centering
    \caption{Microscopic loading case: prescribed strain sequence, representative mesh points, and macroscopic stress-strain response.}
    \label{fig:microscopic_loading}
\end{figure}

The strain sequence and macroscopic stress-strain response are shown in Figure~\ref{fig:microscopic_loading}, where the LSTM prediction tracks the FE reference over the full path (Figure~\ref{fig:macro_fem_vs_lstm}).

\begin{figure}[!htbp]
    \centering
    \includegraphics[width=\textwidth]{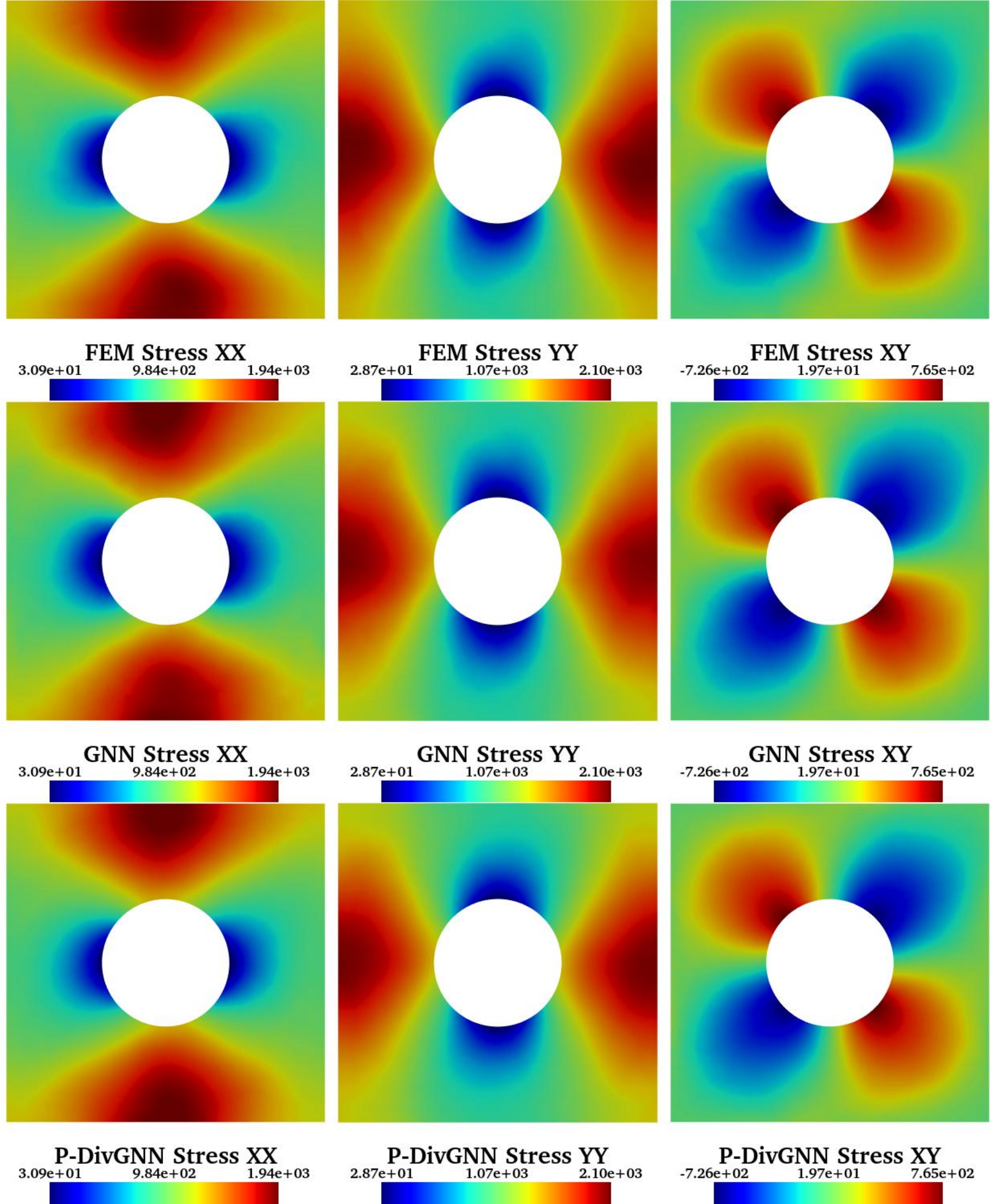}
    \caption{Local stress fields (MPa) at the last loading step: FE (top), GNN (middle), and \textit{\proposedModelName} (bottom). Color scales are determined by the FE solution and shared across rows.}
    \label{fig:stress_field_compare}
\end{figure}

\begin{figure}[!htbp]
    \centering
    \includegraphics[width=\textwidth]{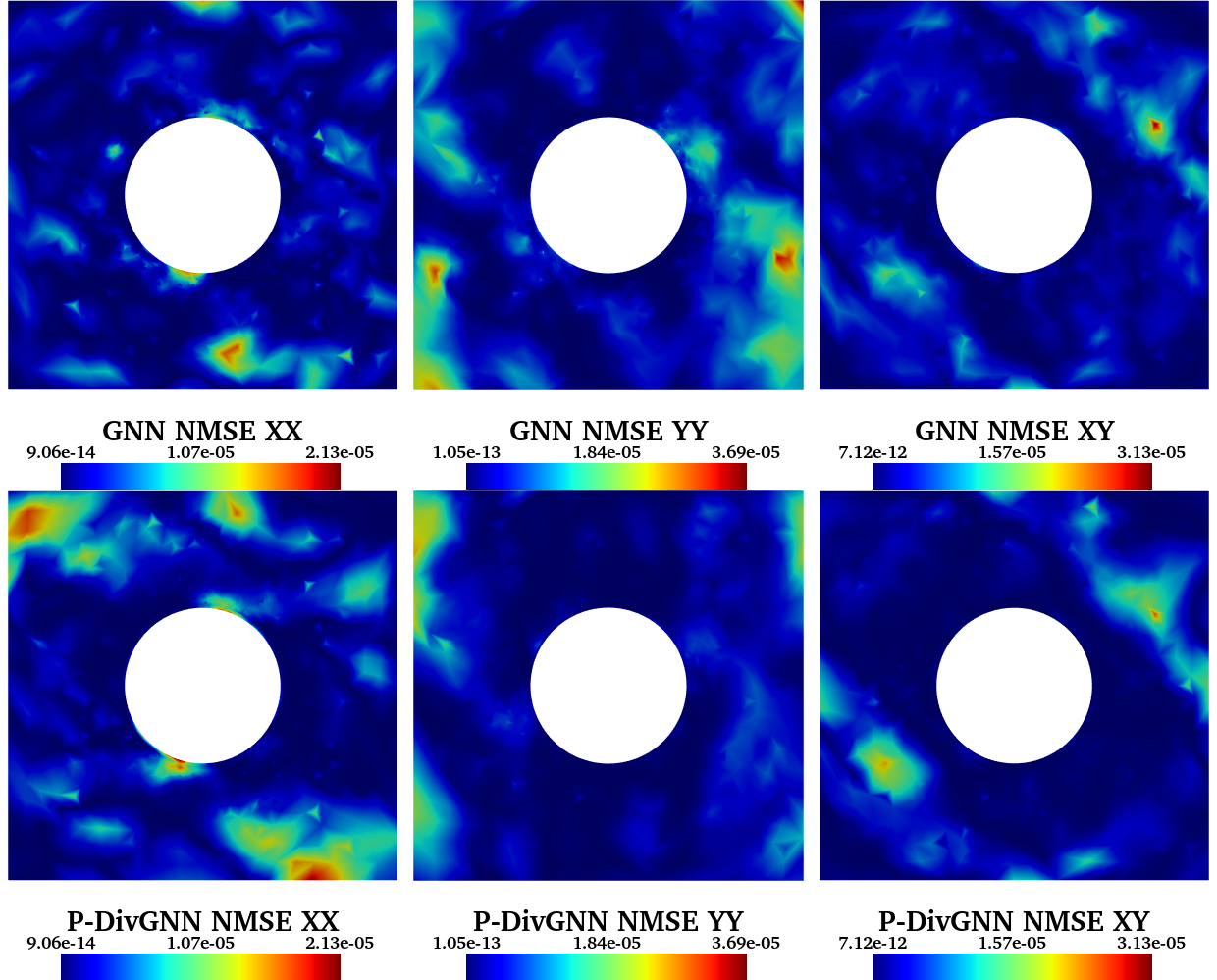}
    \caption{Per-node NMSE of the local stress field at the last loading step, relative to the FE (Quad) solution: GNN (top) and \textit{\proposedModelName} (bottom). Each component ($\sigma_{xx}$, $\sigma_{yy}$, $\sigma_{xy}$) uses its own color scale, shared between the two models.}
    \label{fig:error_fields}
\end{figure}

\begin{figure}[!htbp]
    \centering
    \includegraphics[width=\textwidth]{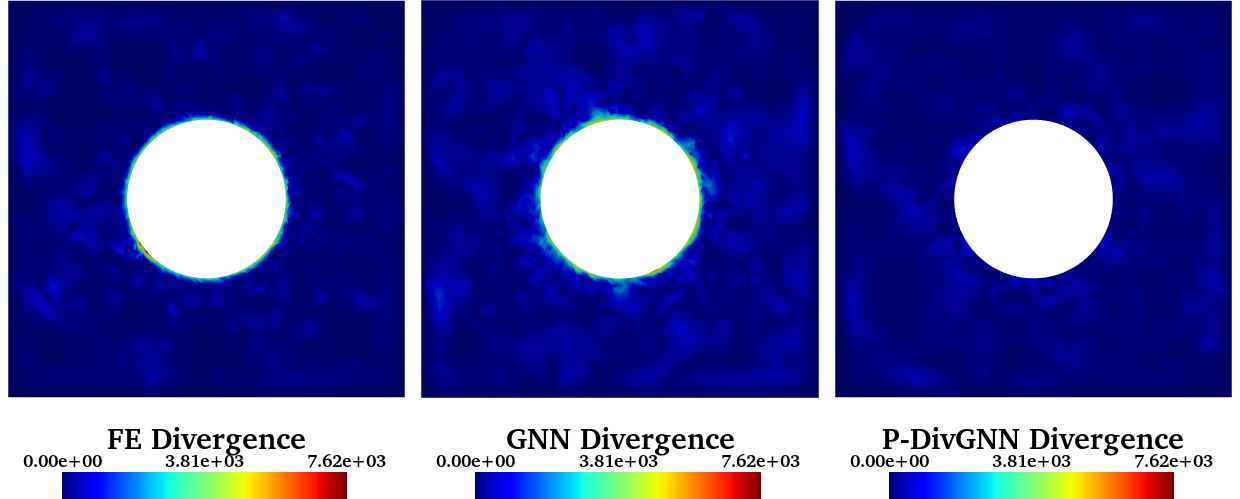}
    \caption{Local stress divergence (MPa per unit length) at the last loading step for FE, GNN, and \textit{\proposedModelName}. \textit{\proposedModelName} reduces the divergence at the hole boundary relative to FE and GNN.}
    \label{fig:divergence_fields}
\end{figure}

Figure~\ref{fig:stress_field_compare} presents the local stress fields at the last loading step for FE, GNN, and \textit{\proposedModelName} on the quadrilateral mesh. Figure~\ref{fig:error_fields} maps the per-node NMSE of each model against the FE (Quad) reference. Both models stay below $0.6\%$ per-component NMSE. The GNN reaches $0.29\%$, $0.53\%$, and $0.36\%$ for $\sigma_{xx}$, $\sigma_{yy}$, and $\sigma_{xy}$; \textit{\proposedModelName} reaches $0.37\%$, $0.29\%$, and $0.25\%$. Figure~\ref{fig:divergence_fields} maps the per-node stress divergence for the three solutions. \textit{\proposedModelName} lowers the divergence at the hole boundary, where FE and the plain GNN both peak. A divergence-based regularization also changes how to read the NMSE. The FE reference satisfies $\textrm{div}(\sigma) = 0$ only weakly: its discrete divergence is small but nonzero ($757$, Table~\ref{tab:unified_metrics}). Since the reference is itself an approximate equilibrium, a lower NMSE against it does not strictly mean a better field once the model also minimizes the divergence.

\begin{figure}[!htbp]
    \centering
    \includegraphics[width=0.95\textwidth]{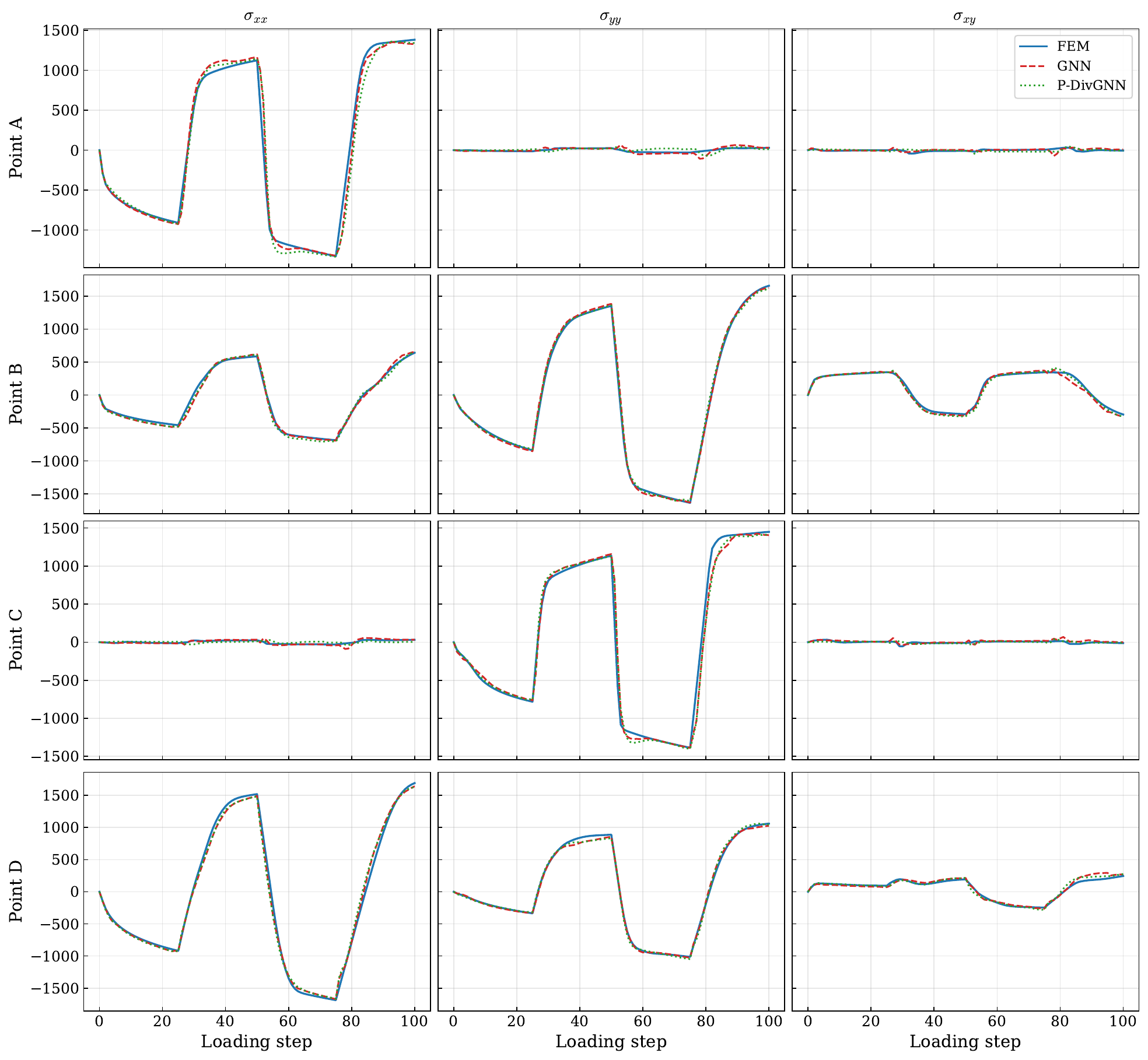}
    \caption{Time evolution of the stress state at four representative points: FE (solid blue), GNN (dashed red), and \textit{\proposedModelName} (dotted green).}
    \label{fig:point_evolution}
\end{figure}

Figure~\ref{fig:point_evolution} reports the time evolution of each stress component at four representative points (Figure~\ref{fig:mesh_points}).
At these sampled locations, the predicted stress histories remain close to the FEA curves for both boundary and interior nodes. We emphasize that the GNN has only been trained on specific snapshots (5 per loading case) that correspond to the random strain values sampled during the dataset generation. In Figure~\ref{fig:point_evolution} we can see that the GNN is accurate even on intermediate local stress fields unseen during training.

Table~\ref{tab:unified_metrics} reports the prediction error metrics and the mean stress divergence, both computed on the full test set ($3{,}000$ graph samples). For the example loading case of the field figures, the mean divergence is $480$ (FE), $893$ (GNN), and $461$ (\textit{\proposedModelName}), with a field NMSE over the path of $7.3\times10^{-3}$ for the GNN and $6.8\times10^{-3}$ for \textit{\proposedModelName}. The divergence values reported here are computed on the raw, unnormalized stress fields and therefore lie on a much larger scale than the NMSE. During training, the relative weighting of Equation~\eqref{eq:relative_weighting} closes this gap by rescaling the divergence contribution to a fixed fraction of the NMSE.

The overall NMSE is essentially unchanged between the two models ($1.70\times10^{-2}$ for the GNN against $1.80\times10^{-2}$ for \textit{\proposedModelName}), while the mean divergence drops from $820$ to $389$, a reduction by more than half. The two metrics measure different
quantities: the NMSE is the distance to the FE reference field, treated here as the ground truth, while the divergence measures how well a
field satisfies local equilibrium, independently of any reference. The equilibrium term therefore acts as a constraint rather than a regularizer: it does not improve both quantities at once, and the accuracy given up is small relative to the reduction in divergence.

The mean divergence of \textit{\proposedModelName} ($389$) also falls below that of the FE solution ($757$). This is not read as the
surrogate being more accurate than the finite element reference: the discrete divergence is a strong, pointwise form of equilibrium that the
FE method does not impose explicitly, since its weak formulation balances forces only in an integral sense. The surrogate settles on a compromise between
agreement with the FE field and satisfaction of this local form, the same behavior reported for the elastic case in
\cite{ref:guevara_ijnme_2025}.

This compromise is most useful away from quadrilateral meshes. On linear triangular meshes the discrete divergence is harder to keep small,
because the interpolation is inconsistent at curved boundaries such as the hole; a surrogate that has learned a low-divergence mapping on
quadrilateral elements and transfers across element types (Section~\ref{sec:cross_mesh}) carries this property to meshes where the finite
element divergence itself grows.

\begin{table}[!htbp]
\centering
\caption{Divergence (MPa per unit length; boundary nodes excluded) and NMSE comparison between GNN, \textit{\proposedModelName}, and FE, both computed on the elasto-plastic test set ($3{,}000$ graph samples).}
\label{tab:unified_metrics}
\begin{tabular}{lcc}
\hline
& \textbf{Divergence} & \textbf{NMSE} \\
\hline
        GNN      & $820$ & $\mathbf{1.70\times10^{-2}}$ \\
        \textit{\proposedModelName}   & $\mathbf{389}$ & $1.80\times10^{-2}$ \\
        FE        & $757$ & -- \\
\hline
\end{tabular}
\end{table}

Table~\ref{tab:nmse_per_component} disaggregates the NMSE by stress component.
The two models are comparable on every component.

\begin{table}[!htbp]
\centering
\caption{Per-component NMSE comparison between GNN and \textit{\proposedModelName} on the full test set.}
\label{tab:nmse_per_component}
\begin{tabular}{lcccc}
\hline
& \textbf{Overall} & $\hat{\sigma}_{xx}$ & $\hat{\sigma}_{yy}$ & $\hat{\sigma}_{xy}$ \\
\hline
GNN & $\mathbf{1.70\times10^{-2}}$ & $\mathbf{1.55\times10^{-2}}$ & $1.79\times10^{-2}$ & $\mathbf{1.75\times10^{-2}}$ \\
\textit{\proposedModelName} & $1.80\times10^{-2}$ & $1.56\times10^{-2}$ & $\mathbf{1.54\times10^{-2}}$ & $2.29\times10^{-2}$ \\
\hline
\end{tabular}
\end{table}

These results show that the regularization strategy from \cite{ref:guevara_ijnme_2025} can also be used in the elasto-plastic regime.
Despite the added complexity of path-dependent behavior and LSTM-based history encoding, the divergence penalty remains effective at reducing equilibrium violations while preserving prediction accuracy.
The relative weighting strategy (Equation~\eqref{eq:relative_weighting}) avoids the need for manual tuning of the penalty weight.

\FloatBarrier
\subsection{Mesh-agnostic reconstruction across element types}\label{sec:cross_mesh}

\begin{figure}[!htbp]
    \centering
    \includegraphics[width=0.7\textwidth]{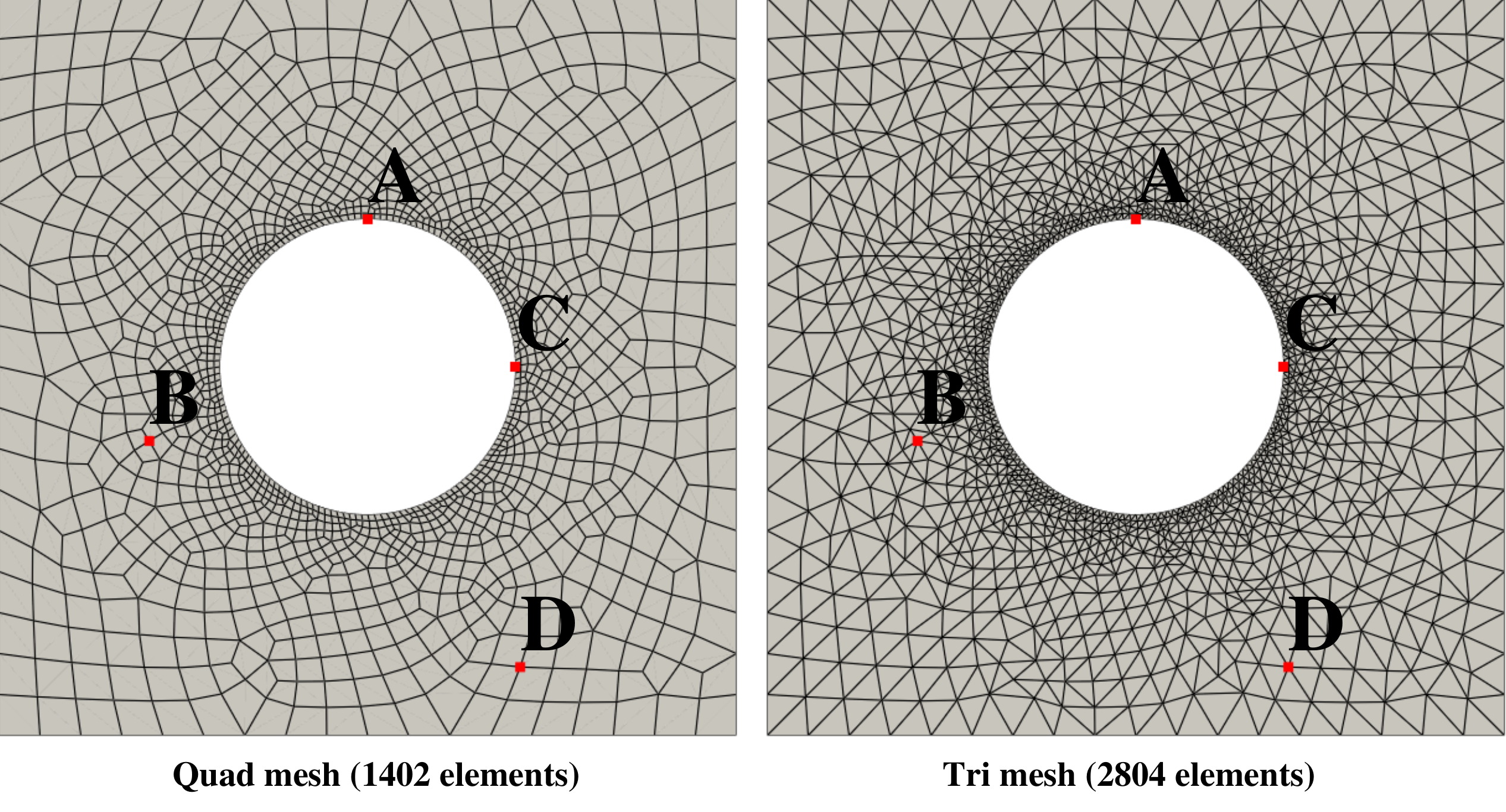}
    \caption{Quad (left) and derived tri (right) meshes used for the cross-mesh generalization experiment. Both meshes share the same node positions; only element connectivity differs.}
    \label{fig:cross_mesh_meshes}
\end{figure}

A notable advantage of the GNN-based surrogate is its independence from element type: the graph is constructed from mesh connectivity, so the learned localization mapping operates on nodes and edges regardless of whether the underlying elements are quadrilaterals or triangles.
To test this, the training quadrilateral mesh is converted into a triangular mesh by splitting each quad element by adding an additional edge between its nodes, preserving all node positions exactly (Figure~\ref{fig:cross_mesh_meshes}).

\begin{figure}[!htbp]
    \centering
    \includegraphics[width=0.8\textwidth]{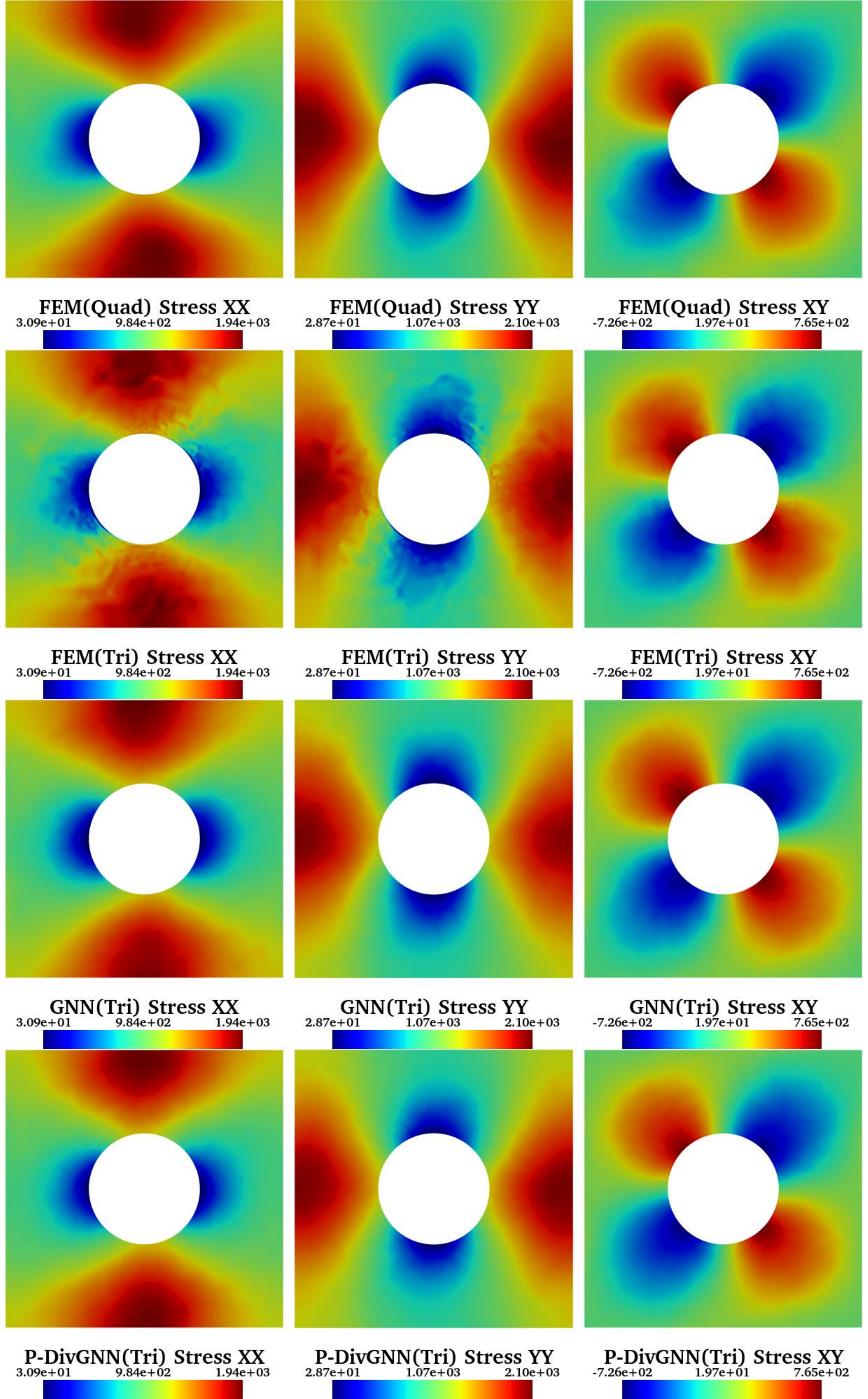}
    \caption{Stress fields (MPa) at the last time step: FE (Quad) reference (a), FE (Tri) (b), GNN(Tri) (c), \textit{\proposedModelName}(Tri) (d). Same color scale throughout.}
    \label{fig:cross_mesh_fields}
\end{figure}

Six simulations are performed: FE, GNN, and \textit{\proposedModelName} on both the quad and tri meshes, with FE on the quad mesh taken as ground truth. Figure~\ref{fig:cross_mesh_fields} compares the predicted stress fields at the last time step. The GNN and \textit{\proposedModelName} fields on the tri mesh remain visually close to the quad FE reference, so the surrogate reconstructs approximately the same field whether the elements are quadrilaterals or triangles. The FE (Tri) field deviates near the hole, as expected for linear triangles under plastic flow.

\begin{table}[!htbp]
\centering
\caption{Overall NMSE relative to FE (Quad) and mean stress divergence (MPa per unit length) for all model-mesh combinations.}
\label{tab:cross_mesh_nmse}
\begin{tabular}{lcccc}
\hline
\textbf{Mesh Type} & \multicolumn{2}{c}{\textbf{Quad}} & \multicolumn{2}{c}{\textbf{Tri}} \\
\cline{2-3} \cline{4-5}
\textbf{Model} & NMSE $\boldsymbol{\sigma}$ & Divergence & NMSE $\boldsymbol{\sigma}$ & Divergence \\
\hline
FE       & --                     & $4.80 \times 10^{2}$         & $9.68 \times 10^{-3}$ & $3.27 \times 10^{3}$ \\
GNN       & $\mathbf{7.31 \times 10^{-3}}$ & $8.93 \times 10^{2}$ & $\mathbf{7.13 \times 10^{-3}}$ & $1.01 \times 10^{3}$ \\
\textit{\proposedModelName}  & $7.34 \times 10^{-3}$  & $\mathbf{4.61 \times 10^{2}}$ & $7.69 \times 10^{-3}$ & $\mathbf{6.92 \times 10^{2}}$ \\
\hline
\end{tabular}
\end{table}

\begin{figure}[!htbp]
    \centering
    \includegraphics[width=0.95\textwidth]{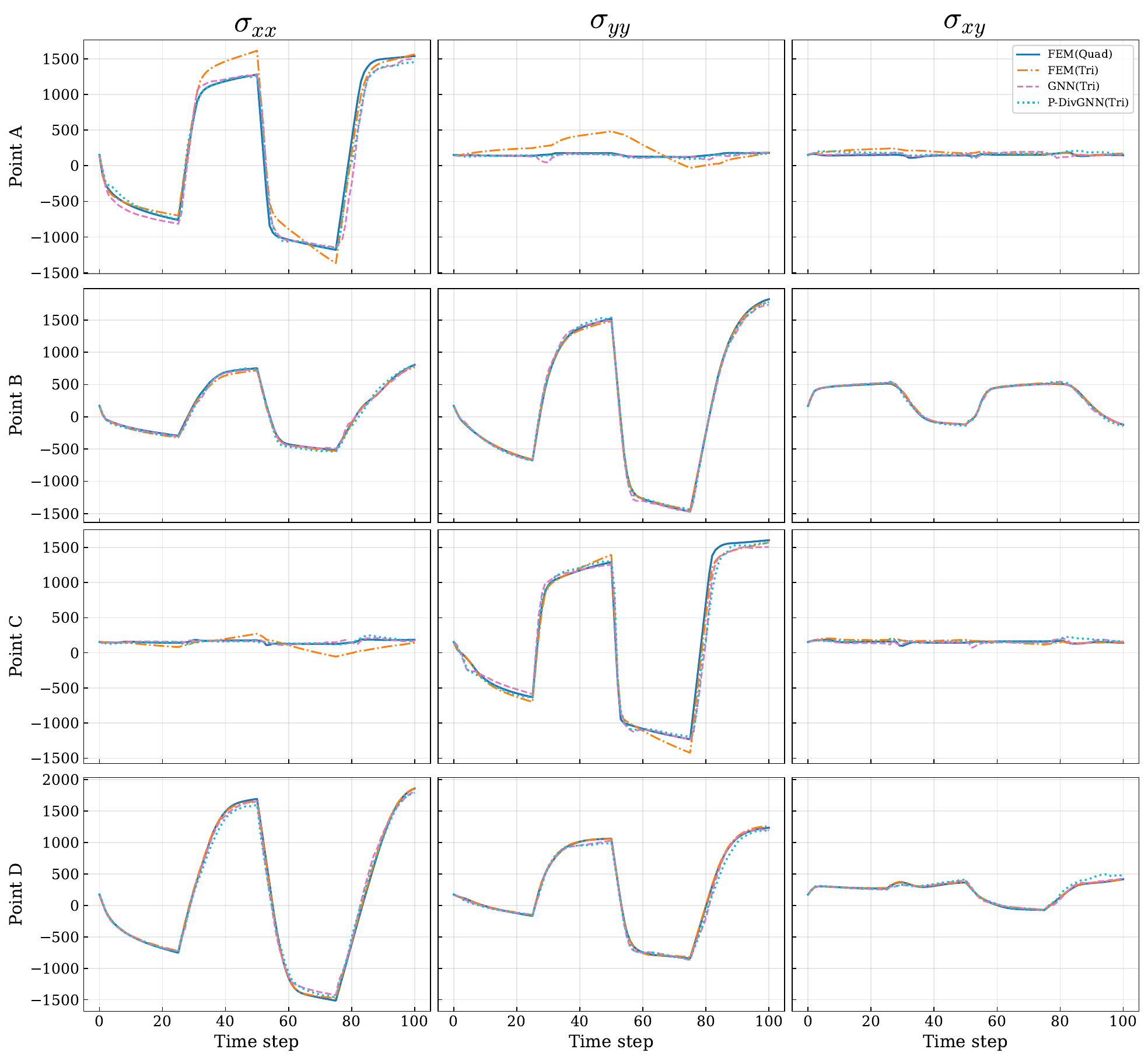}
    \caption{Time evolution of the stress state at four points: FE (Quad) (solid blue), FE (Tri) (dash-dot orange), GNN (Tri) (dashed red), \textit{\proposedModelName}(Tri) (dotted green).}
    \label{fig:cross_mesh_point_evolution}
\end{figure}

The time evolution at four representative points (Figure~\ref{fig:cross_mesh_point_evolution}) shows the FE (Quad), GNN (Tri) and \textit{\proposedModelName} (Tri) traces staying mutually close regardless of element type, with FE (Tri) the only outlier.

\begin{figure}[!htbp]
\centering
\begin{subfigure}[t]{0.48\textwidth}
	\centering
	\includegraphics[width=\textwidth]{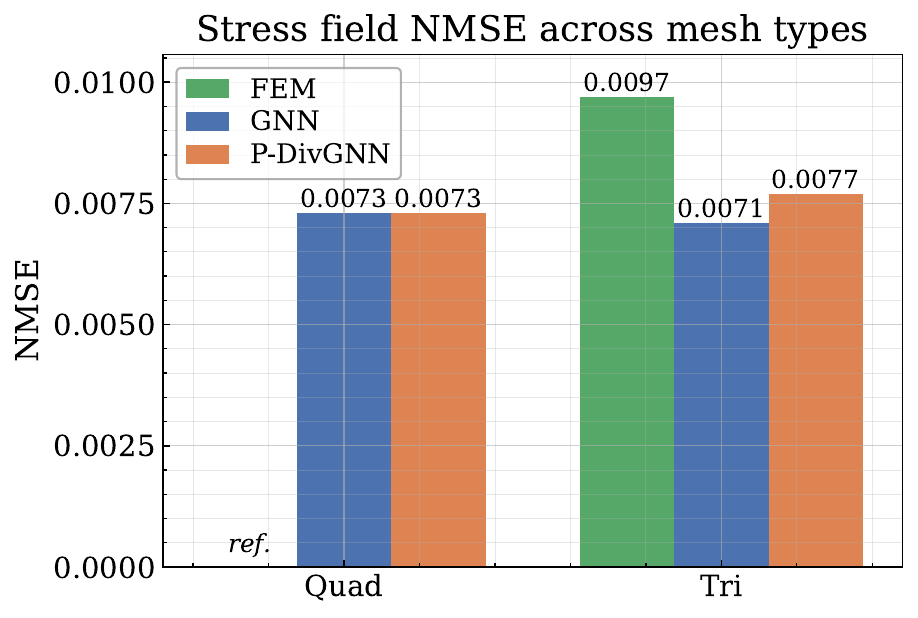}
	\caption{NMSE of the local stress predictions.}
	\label{fig:cross_mesh_metrics}
\end{subfigure}
\hfill
\begin{subfigure}[t]{0.48\textwidth}
	\centering
	\includegraphics[width=\textwidth]{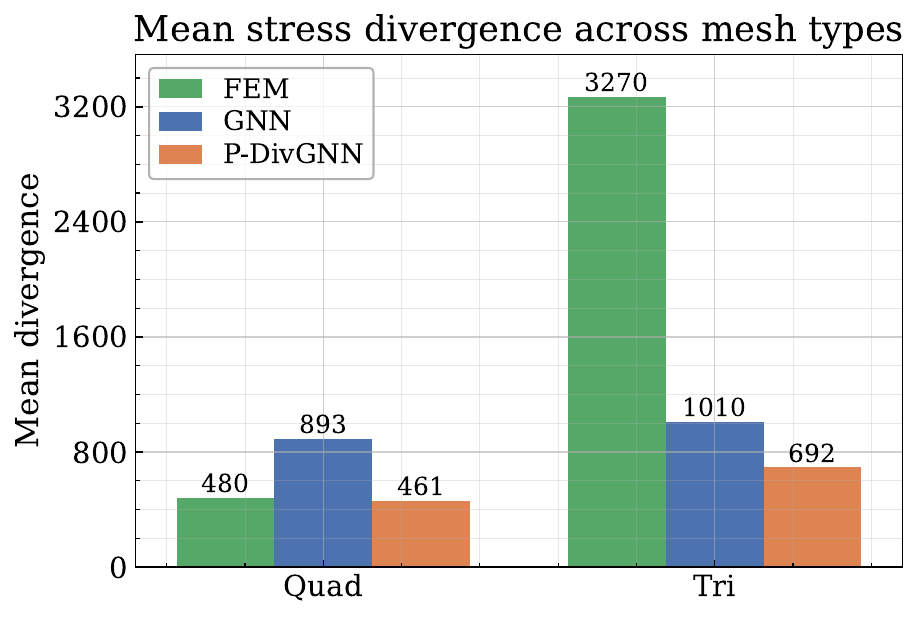}
	\caption{Mean stress divergence.}
	\label{fig:cross_mesh_divergence_bar}
\end{subfigure}
\caption{NMSE and mean stress divergence (MPa per unit length) for FE, GNN, and \textit{\proposedModelName} on the training quad mesh and the unseen triangular mesh.}
\end{figure}

\FloatBarrier
Table~\ref{tab:cross_mesh_nmse} and Figures~\ref{fig:cross_mesh_metrics}--\ref{fig:cross_mesh_divergence_bar} summarize the NMSE and divergence across mesh types.
On the unseen triangular mesh, the GNN and \textit{\proposedModelName} predict nearly the same field as on the quad mesh; their NMSE against the FE (Quad) reference changes little with element type.
However, this ``generalization'' is more precisely a robustness to mesh topology change for the same geometry: both quad and tri meshes share identical node positions, and the GNN was never exposed to triangular element connectivity during training.

Across element types the GNN and \textit{\proposedModelName} divergence stays close to its quad-mesh value, and \textit{\proposedModelName} remains the lowest of the learned models. The larger FE (Tri) divergence is a well-known effect of linear triangles, which stiffen under plastic flow. Using quadratic six-node triangles removes this accuracy gap and converges to the quadrilateral solution.

\begin{figure}[!htbp]
    \centering
    \includegraphics[width=0.85\textwidth]{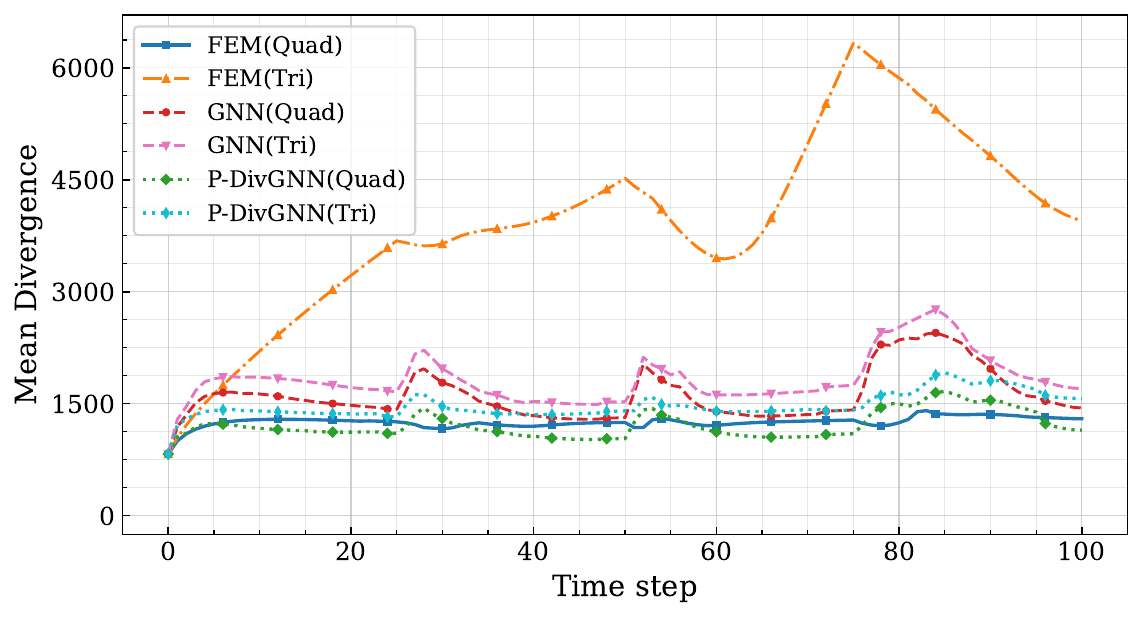}
    \caption{Temporal evolution of the mean stress divergence (MPa per unit length) for all six model-mesh combinations.}
    \label{fig:cross_mesh_divergence}
\end{figure}

Figure~\ref{fig:cross_mesh_divergence} shows the temporal evolution of the mean stress divergence. The FE (Tri) curve separates clearly from the remaining five, while the GNN and \textit{\proposedModelName} curves on the tri mesh stay close to their quad counterparts, confirming that the message-passing mechanism is inherently agnostic to element-level integration.

\FloatBarrier

\subsection{Mesh refinement}\label{sec:mesh_refinement}
\begin{figure}[!htbp]
	\centering
	\includegraphics[width=0.98\textwidth]{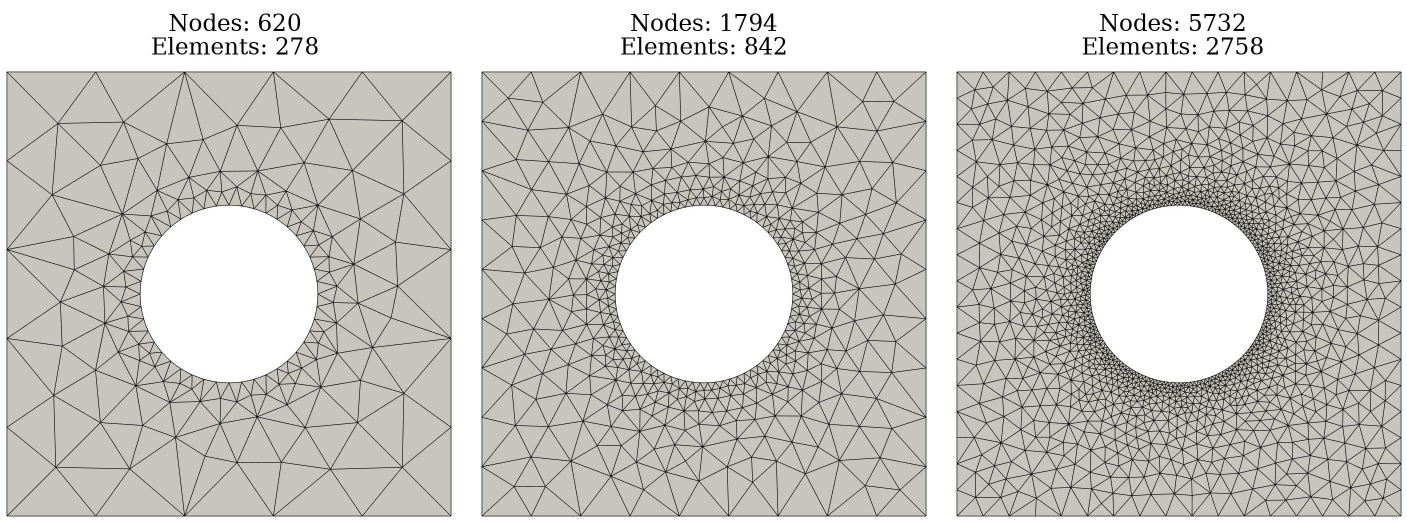}
	\caption{Coarse-to-fine sequence of tri6 meshes (left to right).}
	\label{fig:refinement_mesh_tri}
\end{figure}

To probe the surrogate across mesh resolutions, the evaluation is repeated on a coarse-to-fine sequence of triangular meshes (Figure~\ref{fig:refinement_mesh_tri}), with the loading path and inference setup fixed. The linear constant-strain triangles (CST) of the previous section stiffen under the incompressibility of plastic flow, so the meshes here use quadratic six-node triangles (referred to as tri6 in several finite-element frameworks), which remove that effect and give an accurate finite-element reference compared to linear triangular elements.

\begin{figure}[!htbp]
	\centering
	\includegraphics[width=\textwidth]{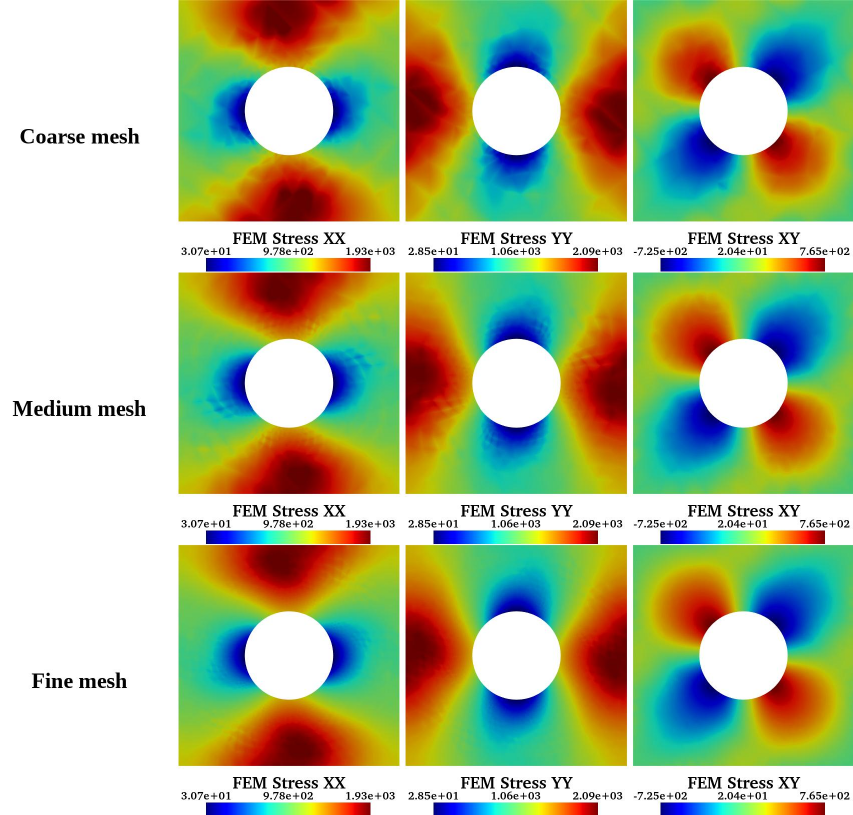}
	\caption{FE stress fields (MPa) across the refinement levels.}
	\label{fig:refinement_fem}
\end{figure}

\begin{figure}[!htbp]
	\centering
	\includegraphics[width=\textwidth]{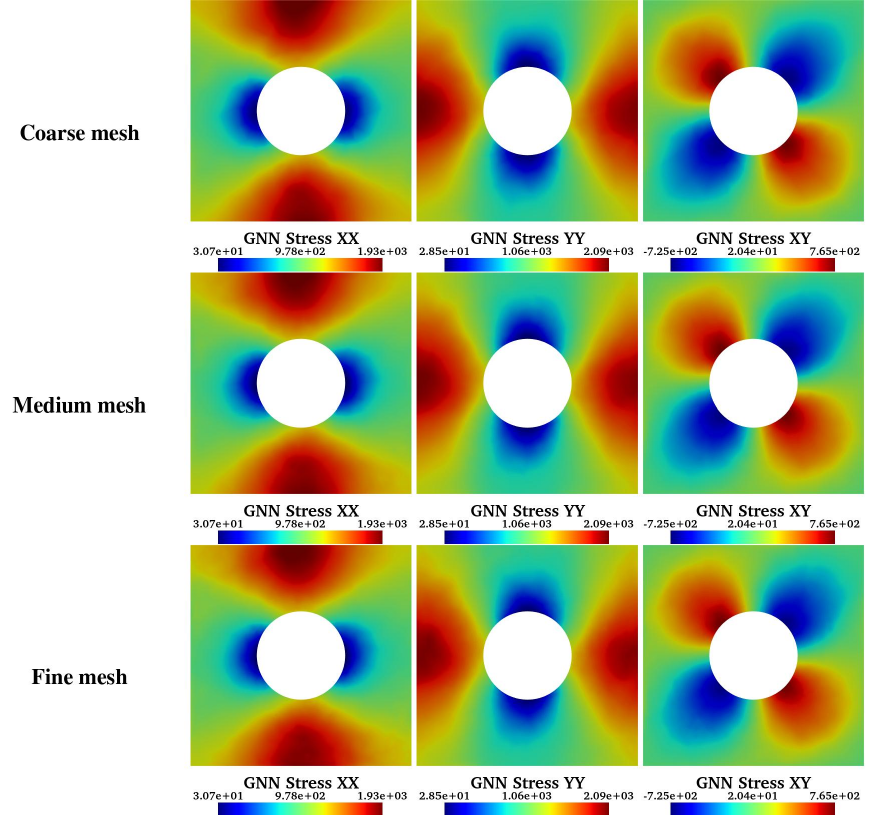}
	\caption{GNN stress predictions (MPa) across the refinement levels.}
	\label{fig:refinement_gnn}
\end{figure}

\begin{figure}[!htbp]
	\centering
	\includegraphics[width=\textwidth]{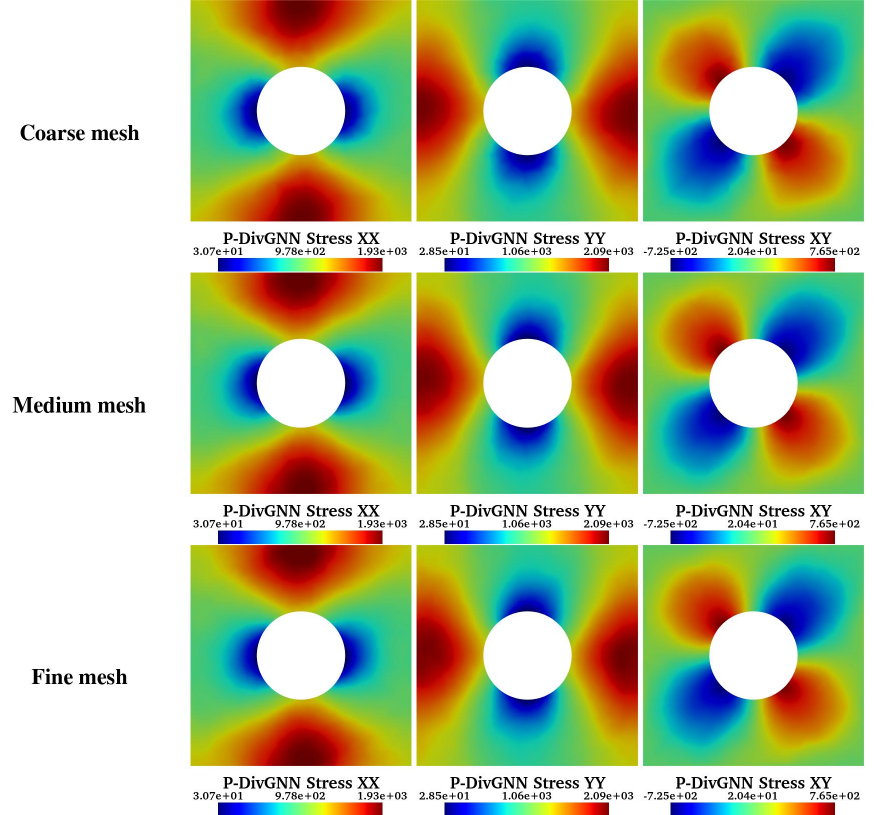}
	\caption{\textit{\proposedModelName} stress predictions (MPa) across the refinement levels.}
	\label{fig:refinement_pdivgnn}
\end{figure}

Figures~\ref{fig:refinement_fem}--\ref{fig:refinement_pdivgnn} compare the FE, GNN and \textit{\proposedModelName} stress fields across the three refinement levels. The quadratic-triangle FE fields sharpen as the mesh refines while staying accurate. The GNN and \textit{\proposedModelName} reproduce the same field quality at every level without retraining, and both move closer to the FE reference as the mesh refines.

\begin{figure}[!htbp]
	\centering
	\includegraphics[width=\textwidth]{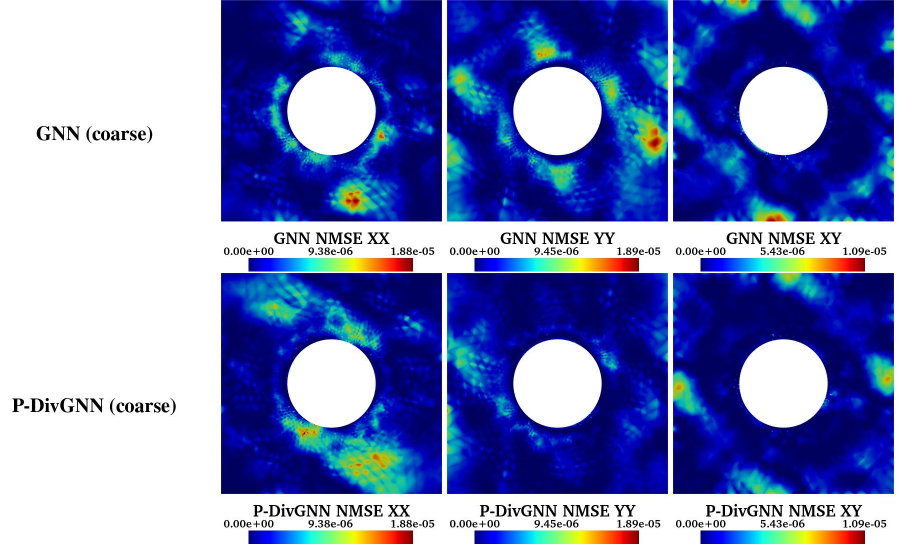}
	\caption{Per-component NMSE of the coarse-mesh GNN and \textit{\proposedModelName} predictions against the fine FE reference.}
	\label{fig:refinement_diff}
\end{figure}

Figure~\ref{fig:refinement_diff} maps the error of the coarse-mesh GNN and \textit{\proposedModelName} predictions against the fine FE reference. Because the coarse and fine meshes do not share the same nodes, the coarse-mesh field is first interpolated onto the fine-mesh nodes with the quadratic triangular shape functions, using the Visualization Toolkit (VTK) \cite{ref:vtk_lib} library. Table~\ref{tab:refinement_nmse} reports the per-level NMSE against the quadratic-triangle FE solution: both surrogates improve as the mesh refines, and \textit{\proposedModelName} has the lower overall error at every level.

\begin{table}[!htbp]
	\centering
	\caption{Per-component and overall NMSE of the surrogates against the FE solution at each refinement level. The lower value of each pair is in bold.}
	\label{tab:refinement_nmse}
    \resizebox{\textwidth}{!}{
	\begin{tabular}{llcccc}
	\toprule
	Level & Model (Tri6) & $\sigma_{xx}$ & $\sigma_{yy}$ & $\sigma_{xy}$ & Overall \\
	\midrule
	\multirow{2}{*}{Coarse} & GNN & $\mathbf{2.39\times10^{-2}}$ & $2.43\times10^{-2}$ & $6.90\times10^{-3}$ & $1.84\times10^{-2}$ \\
	& \textit{\proposedModelName} & $2.63\times10^{-2}$ & $\mathbf{1.53\times10^{-2}}$ & $\mathbf{5.00\times10^{-3}}$ & $\mathbf{1.55\times10^{-2}}$ \\
	\midrule
	\multirow{2}{*}{Medium} & GNN & $\mathbf{1.38\times10^{-2}}$ & $1.49\times10^{-2}$ & $4.20\times10^{-3}$ & $1.10\times10^{-2}$ \\
	& \textit{\proposedModelName} & $1.79\times10^{-2}$ & $\mathbf{7.20\times10^{-3}}$ & $\mathbf{3.20\times10^{-3}}$ & $\mathbf{9.40\times10^{-3}}$ \\
	\midrule
	\multirow{2}{*}{Fine} & GNN & $\mathbf{1.01\times10^{-2}}$ & $1.14\times10^{-2}$ & $3.10\times10^{-3}$ & $8.20\times10^{-3}$ \\
	& \textit{\proposedModelName} & $1.61\times10^{-2}$ & $\mathbf{5.10\times10^{-3}}$ & $\mathbf{2.90\times10^{-3}}$ & $\mathbf{8.00\times10^{-3}}$ \\
	\bottomrule
	\end{tabular}}
\end{table}

This experiment illustrates the mesh-agnostic behavior of the method. \textit{\proposedModelName} lowers the NMSE on the coarse meshes, which are far from the quadrilateral discretization used for training. The graph-based formulation is also convenient when quadrilateral meshing is impractical: the surrogate reconstructs high-fidelity mechanical fields even on coarse, lower-quality triangular meshes.
\FloatBarrier
\subsection{Computational benchmark}\label{sec:benchmark}

\begin{figure}[!htbp]
	\centering
    \includegraphics[width=\textwidth]{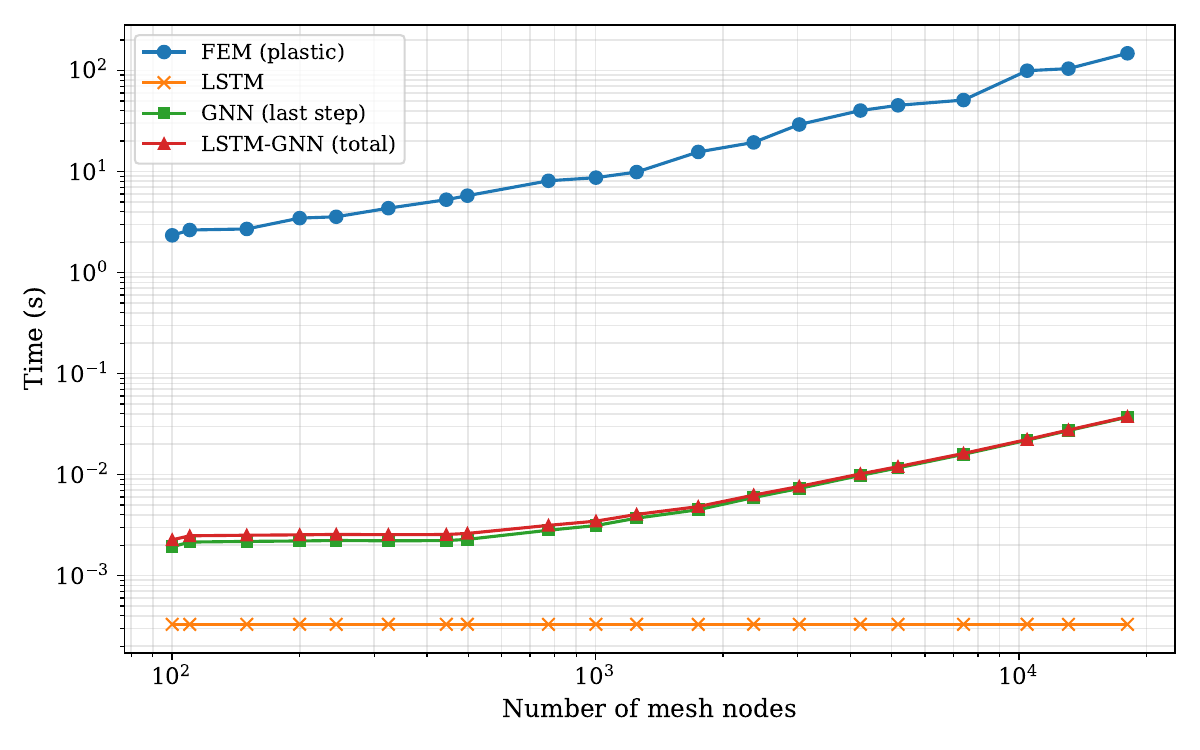}
    \caption{Computational time comparison for FE, LSTM-GNN, and LSTM at the last inference step (double log scale).}
    \label{fig:benchmark}
\end{figure}

Figure~\ref{fig:benchmark} compares the computational time of FE, LSTM-GNN, and LSTM alone, evaluated at the last step (the intended inference usage).
The speedup ranges from $\sim$200$\times$ at 100 nodes to $\sim$3{,}550$\times$ at $56{,}000$ nodes (results reported in Table \ref{tab:benchmark_speedup}).
The GNN inference time scales linearly with the number of nodes, whereas the LSTM time is essentially constant since it operates only on macroscopic quantities.
Across the entire mesh range, the LSTM-GNN reconstructs the response of a single RVE in under $0.2$ seconds, so a complete field prediction is obtained in real time.
Further speedups are expected by exploiting the batching capabilities of the GNN for multi-time step prediction.

These timings compare GPU-based neural network inference (NVIDIA RTX A4000, 16~GB) against CPU-based FE solving (Intel Xeon W5-2455X).
A hardware-fair comparison would require either GPU-accelerated FE or CPU-only neural network inference.
Note that the reported speedups cover only the inference phase; the total training cost is approximately twelve hours for the GNN and ten minutes for the LSTM.
The FE runtime also depends on Newton-Raphson convergence behavior and may vary with loading path difficulty. The computational speedup represents a substantial improvement that can be exploited in multiscale frameworks such as $\textrm{FE}^2$ to efficiently reconstruct the local stress field within a RUC.

\begin{table}[!htbp]
	\centering
\caption{FE and LSTM-GNN computational times (seconds) and speed gain across mesh sizes.}
	\label{tab:benchmark_speedup}
	\begin{tabular}{l c c c}
		\hline
		\textbf{Number of nodes} & \textbf{FE (s)} & \textbf{LSTM-GNN (s)} & \textbf{Speed Gain} \\
		\hline
		100 & 2.379 & 0.012 & $\times$194 \\
		$4{,}196$ & 35.769 & 0.027 & $\times$1{,}303 \\
		$12{,}342$ & 98.149 & 0.058 & $\times$1{,}682 \\
		$28{,}805$ & 262.384 & 0.127 & $\times$2{,}067 \\
		$40{,}494$ & 326.295 & 0.176 & $\times$1{,}853 \\
		$56{,}206$ & 428.071 & 0.121 & $\times$3{,}550 \\
		\hline
	\end{tabular}
\end{table}

\FloatBarrier

\section{Conclusion and perspectives}\label{sec:conclusion}
We coupled an LSTM with a GNN to reconstruct local stress fields in elasto-plastic microstructures under non-proportional loading.
The LSTM encodes the macroscopic stress-strain history into a compact hidden state, which the GNN uses as node-level input for spatial localization at each time step.

Training in two stages reduced the optimization cost and kept both models stable. The LSTM converges in approximately ten minutes, while the GNN, trained on snapshots at main loading steps only, generalizes to intermediate increments not seen during training.

On the test set, the model tracks the macroscopic stress-strain response across elastic-plastic transitions, load reversals, and multiaxial loading.
At the microscale, the predicted stress fields remain close to the FE reference and recover the main heterogeneity near the inner surface in a classical plate with a hole benchmark problem.
A single trained surrogate transfers to triangular meshes and to coarser and finer discretizations without retraining, reproducing the quadrilateral FE field in each case. This independence from element type follows from the message-passing formulation, which operates on connectivity rather than element integration.
The computational speedup reaches three orders of magnitude for last-step inference at mesh sizes exceeding $50{,}000$ nodes, reducing the per-RVE inference to a real-time, sub-second cost.

The relative weighting strategy for the divergence-regularized loss resolves the scale mismatch between the NMSE and divergence terms that often prevents fixed-weight formulations from converging in the elasto-plastic regime. Using scale normalization and a linear warm-up, the divergence term cuts the mean stress divergence by about half while keeping the reconstruction error nearly unchanged. This makes the regularization usable in the history-dependent setting without retuning the loss for each case, thereby extending the approach introduced in \cite{ref:guevara_ijnme_2025}.

PCA and t-SNE of the LSTM hidden states suggest that the network organizes the loading history into a low-dimensional structure. The first principal component appears to track accumulated plastic deformation, while secondary components vary with the current loading segment.
This structure persists on eight-segment paths ($201$ time steps), where the model reaches only $3.0\%$ cumulative error, despite being trained on four-segment sequences.

Notwithstanding these achievements, several limitations should also be acknowledged. The model was trained on a single periodic microstructure geometry; its generalization to unseen geometric configurations remains an open question. The constitutive model is restricted to 2D plane strain with von Mises plasticity and isotropic hardening. Extensions to 3D, more complex hardening laws (kinematic, mixed) or damage-coupled models would require additional investigation. Finally, the benchmark speedups reflect GPU inference vs.\ CPU-based FE solving, and the training cost ought to be amortized over the number of downstream inference queries, for a fairer comparison.

Future directions include extending the framework to multiple microstructure geometries, incorporating more general constitutive behaviors, and investigating 3D periodic microstructures, where the computational savings would be even more significant. For the multi-geometry extension, our recommended strategy is transfer learning of both stages rather than a full dataset and training per geometry. The localization stage is the natural candidate: it operates on the mesh graph and can be initialized from the weights trained on a reference geometry, with the freezing strategy for the encoder, message-passing, and decoder blocks to be selected empirically. The macro-scale recurrent model is the harder part, because its hidden state is specific to the microstructure that produced the homogenized response: we envisage a geometry-informed macro-scale model, conditioned on a descriptor of the microstructure such as the effective stiffness tensor or a learned geometric embedding, trained across a family of geometries so that a new geometry requires only fine-tuning with a reduced dataset. Enriching the model outputs is a further direction, and a possible transfer learning strategy is to train a decoder head for the target output field. The encoder and the message-passing steps describe in a latent space the mechanical state at each node, and the stress field is only one reading of that description. Local strains, displacements and internal variables are within reach of the same architecture.

\section*{Declaration of competing interest}

The authors declare that they have no known competing financial interests or personal relationships that could have appeared to influence the work reported in this paper.

\section*{Data availability}

The source code for reproducing the experiments, generating the database, and training the models is openly available at \url{https://github.com/ricardo0115/plastic-pdivgnn}. The repository also hosts an interactive abstract summarizing the main results, available at \url{https://ricardo0115.github.io/plastic-pdivgnn/}.

\section*{Acknowledgements}

The authors gratefully acknowledge the computational resources provided by the high-performance computing cluster of TU Bergakademie Freiberg for the database generation and training. This research did not receive any specific grant from funding agencies in the public, commercial, or not-for-profit sectors.

\appendix
\numberwithin{table}{section}
\numberwithin{figure}{section}
\renewcommand{\thetable}{\Alph{section}.\arabic{table}}
\renewcommand{\thefigure}{\Alph{section}.\arabic{figure}}

\section{Sensitivity to the commit tolerance}\label{app:tolerance}

The commit tolerance of Algorithm~\ref{alg:committed_state} was varied around its deployed value of $0.1\,\Delta\bar{\varepsilon}_{\mathrm{med}}$, with the model and the test protocol of Section~\ref{sec:stability_tests} otherwise unchanged; Table~\ref{tab:tolerance_sensitivity} reports the sweep.

\begin{table}[!htbp]
	\centering
	\caption{Sensitivity to the commit tolerance. Weighted mean absolute percentage error of the macroscopic prediction when the same validation paths are described with $1\times$, $10\times$ and $100\times$ their native number of steps. The last column is the fraction of the native steps that advance the recurrent state. The deployed configuration is in bold.}
	\label{tab:tolerance_sensitivity}
	\begin{tabular}{lcccc}
		\toprule
		& \multicolumn{3}{c}{wMAPE at refinement} & Committed \\
		\cmidrule(lr){2-4}
		$\delta_{\mathrm{tol}} / \Delta\bar{\varepsilon}_{\mathrm{med}}$ & $1\times$ & $10\times$ & $100\times$ & fraction \\
		\midrule
		$0.05$ & $1.65\%$ & $1.65\%$ & $1.90\%$ & $1.000$ \\
		$\mathbf{0.1}$  & $\mathbf{1.65}\%$ & $\mathbf{1.72}\%$ & $\mathbf{1.90}\%$ & $\mathbf{0.999}$ \\
		$0.25$ & $1.65\%$ & $2.22\%$ & $2.53\%$ & $0.990$ \\
		$0.5$  & $1.83\%$ & $3.39\%$ & $3.75\%$ & $0.943$ \\
		\bottomrule
	\end{tabular}
\end{table}

At the native discretization the four settings are indistinguishable, all between $1.65\%$ and $1.83\%$, so the tolerance does not perturb the nominal operation of the model. Under refinement they separate: $\delta_{\mathrm{tol}} = 0.05$ and $0.1\,\Delta\bar{\varepsilon}_{\mathrm{med}}$ remain flat, whereas $0.25$ and $0.5\,\Delta\bar{\varepsilon}_{\mathrm{med}}$ commit too coarsely and lose accuracy. The deployed value of $0.1\,\Delta\bar{\varepsilon}_{\mathrm{med}}$ therefore lies inside a flat band rather than at a tuned optimum, and the upper end of that band is set by the point at which the rule begins to coarsen the native loading itself.

\section{Sensitivity to the divergence weight and the warm-up length}\label{app:lambda_k}

The relative weight $\lambda_{\mathrm{rel}}$ of the divergence term and the warm-up length $K$ of Equation~\eqref{eq:relative_weighting} were varied around the deployed setting, all other hyperparameters and the training budget being held fixed. Every configuration was trained for $50$ epochs and is reported at its last epoch in Table~\ref{tab:lambda_k}. Accuracy is measured by the validation NMSE and the departure from equilibrium by the relative divergence
\begin{equation}\label{eq:relative_divergence}
R = \frac{\mathcal{L}_{\mathrm{div}}}{\mathcal{L}_{\mathrm{div}}^{\ast}}\ ,
\end{equation}
where $\mathcal{L}_{\mathrm{div}}$ is the divergence penalty of Equation~\eqref{eq:div_term} evaluated on the validation set and $\mathcal{L}_{\mathrm{div}}^{\ast}$ is its value for the deployed configuration at the same epoch, so that $R = 1$ for that configuration by construction. Reporting a ratio rather than an absolute value keeps the comparison independent of the reduction used in Equation~\eqref{eq:div_term}, which is a squared measure and therefore not on the scale of the mean divergence reported in Section~\ref{sec:micro_results}.

\begin{table}[!htbp]
	\centering
	\caption{Effect of the relative divergence weight $\lambda_{\mathrm{rel}}$ and of the warm-up length $K$ on the validation NMSE and on the relative divergence $R$ of Equation~\eqref{eq:relative_divergence}. The deployed configuration is in bold.}
	\label{tab:lambda_k}
	\begin{tabular}{cccc}
		\toprule
		$\lambda_{\mathrm{rel}}$ & $K$ & Validation NMSE & Relative divergence $R$ \\
		\midrule
		$0.01$ & $20$ & $4.95\times10^{-3}$ & $9.96$ \\
		$\mathbf{0.1}$  & $\mathbf{20}$ & $\mathbf{5.37\times10^{-3}}$ & $\mathbf{1.00}$ \\
		$0.5$  & $20$ & $8.32\times10^{-3}$ & $0.088$ \\
		$1.0$  & $20$ & $1.48\times10^{-2}$ & $0.034$ \\
		\midrule
		$0.1$  & $5$  & $5.79\times10^{-3}$ & $0.86$ \\
		$0.1$  & $50$ & $5.13\times10^{-3}$ & $1.41$ \\
		\bottomrule
	\end{tabular}
\end{table}

The relative weight $\lambda_{\mathrm{rel}}$ and the warm-up length $K$ act differently. The relative weight sets the balance between the two terms of the loss, and pushing it to $\lambda_{\mathrm{rel}} = 1.0$ tips that balance too far: the network optimizes the equilibrium residual at the expense of agreement with the finite element field, and the reconstruction error triples while the residual falls to a thirtieth of its deployed value. The warm-up length acts almost entirely on the residual. Between $K = 5$ and $K = 50$ the reconstruction error moves by $13\%$ while the residual moves by a factor of $1.6$: a short warm-up brings the penalty to full strength early and lowers the residual, whereas a warm-up as long as the run itself never lets it reach its nominal weight. The six runs are preliminary, enough to show how the model responds to the two parameters and to support the values used here, but not to map the parameter space; a full hyperparameter study, including how $\lambda_{\mathrm{rel}}$ and $K$ interact with the remaining training settings, is left for future work.

\section{Effect of the dataset size}\label{app:dataset_size}

The size of the training set was varied from $25$ to $5{,}000$ loading paths (Table~\ref{tab:dataset_size}). Both stages were retrained at every rung on the same subset. The validation split is unchanged across rows.

\begin{table}[!htbp]
	\centering
	\caption{Dataset-size ladder. Normalized mean squared error and weighted mean absolute percentage error of the macroscopic stress predicted by the LSTM and of the local stress field reconstructed by the GNN, both on the full validation split. The last row is the training set used throughout the paper.}
	\label{tab:dataset_size}
	\begin{tabular}{rcccc}
		\toprule
		& \multicolumn{2}{c}{Macroscopic} & \multicolumn{2}{c}{Local} \\
		\cmidrule(lr){2-3} \cmidrule(lr){4-5}
		Training paths & NMSE & wMAPE & NMSE & wMAPE \\
		\midrule
		$25$      & $1.77\times10^{-1}$ & $33.93\%$ & $6.77\times10^{-2}$ & $22.93\%$ \\
		$50$      & $1.10\times10^{-1}$ & $26.17\%$ & $3.44\times10^{-2}$ & $15.41\%$ \\
		$100$     & $3.18\times10^{-2}$ & $14.52\%$ & $2.32\times10^{-2}$ & $12.44\%$ \\
		$250$     & $9.39\times10^{-3}$ & $7.73\%$  & $1.53\times10^{-2}$ & $9.53\%$ \\
		$500$     & $4.22\times10^{-3}$ & $5.07\%$  & $1.34\times10^{-2}$ & $8.78\%$ \\
		$1{,}000$ & $1.90\times10^{-3}$ & $3.22\%$  & $9.38\times10^{-3}$ & $6.93\%$ \\
		$2{,}500$ & $9.51\times10^{-4}$ & $2.25\%$  & $5.53\times10^{-3}$ & $5.35\%$ \\
		$5{,}000$ & $5.78\times10^{-4}$ & $1.72\%$  & $4.28\times10^{-3}$ & $4.68\%$ \\
		\midrule
		$7{,}000$ (full) & $5.57\times10^{-4}$ & $1.65\%$ & $3.90\times10^{-3}$ & $4.42\%$ \\
		\bottomrule
	\end{tabular}
\end{table}

The step from $5{,}000$ paths to the full training set improves the local wMAPE by less than $6\%$, so three quarters of the data already recovers most of the accuracy of the full set. Both stages improve monotonically with no plateau up to $2{,}500$ paths, and the local error falls below $10\%$ between $100$ and $250$ paths. Below $100$ paths the reconstruction is qualitatively wrong rather than merely inaccurate. The database size is therefore the binding constraint on accuracy in the regime studied here, ahead of the capacity of either network, which bears directly on the cost of transferring the method to a new microstructure: the expense lies in the finite element database, not in the training itself.

\section{Recurrent cell and hidden state size}\label{app:cell}

The macro-scale stage was retrained with different recurrent cells and hidden state widths, under the recipe of Section~\ref{sec:lstm_training} and evaluated on the native validation discretization; Table~\ref{tab:cell_ablation} reports the comparison.

\begin{table}[!htbp]
	\centering
	\caption{Recurrent-cell ablation at the macro-scale stage. Parameter counts are those of the recurrent model alone. The deployed configuration is in bold.}
	\label{tab:cell_ablation}
	\begin{tabular}{lrc}
		\toprule
		Cell & Parameters & Best validation wMAPE \\
		\midrule
		LSTM, $16$ units            & $3{,}571$   & $3.87\%$ \\
		LSTM, $32$ units            & $13{,}283$  & $1.80\%$ \\
		GRU, $64$ units             & $38{,}403$  & $1.31\%$ \\
		\textbf{LSTM, 64 units (deployed)} & $\mathbf{51{,}139}$  & $\mathbf{1.19}\%$ \\
		LSTM, $128$ units           & $200{,}579$ & $0.94\%$ \\
		LSTM, $256$ units           & $794{,}371$ & $0.78\%$ \\
		\bottomrule
	\end{tabular}
\end{table}

The LSTM follows the architecture validated in~\cite{ref:these_aymen} for mean stress prediction, and the ablation does not overturn that choice. The gated recurrent unit reaches a comparable error at equal width with a quarter fewer parameters, so it is a parameter-efficient alternative. Below $64$ units the error rises steeply, more than tripling at $16$ units, whereas above $64$ units the returns are gradual, about $20\%$ each time the number of parameters is multiplied by four. The size of the recurrent cell is not what limits the pipeline in any case: the benchmark of Section~\ref{sec:benchmark} shows the inference time scaling with the number of mesh nodes through the GNN, while the macro-scale contribution stays essentially constant, so both speed and memory are set by the localization stage. Should memory become the binding constraint, a narrower cell is where to look, at $32$ units for a quarter of the parameters and $1.80\%$ against $1.19\%$, and the saving compounds because the state vector is broadcast to every mesh node as an input feature of the GNN.

\bibliographystyle{elsarticle-num-names}
\bibliography{biblio}

\end{document}